\documentclass[journal]{IEEEtran}
\IEEEoverridecommandlockouts
\usepackage[utf8]{inputenc}

\usepackage{amsmath,amssymb,amsthm}
\usepackage{enumerate}
\usepackage{stfloats}
\usepackage{comment}
\usepackage{siunitx}
\usepackage{mathtools}
\usepackage{accents,latexsym,cancel}
\usepackage{cite}

\usepackage[dvipsnames]{xcolor}
\definecolor{myPink}{RGB}{255,105,183}

\usepackage[T1]{fontenc}
\usepackage{graphics} 
\usepackage{epsfig} 
\usepackage[mathscr]{euscript}
\usepackage{algorithm}
\usepackage[noend]{algpseudocode}
\usepackage{bbm}
\makeatletter
\def\BState{\State\hskip-\ALG@thistlm}
\makeatother

\usepackage{tikz}
\usetikzlibrary{arrows,shapes,chains,matrix,positioning,scopes,patterns,calc,
decorations.markings,
decorations.pathmorphing,
}

\usepackage{pgfplots}
\pgfplotsset{compat=1.3}
\usepgflibrary{shapes}

\newtheorem{theorem}{Theorem}
\newtheorem{lemma}[theorem]{Lemma}
\newtheorem{condition}[theorem]{Condition}
\newtheorem{proposition}[theorem]{Proposition}
\newtheorem{definition}[theorem]{Definition}

\newtheorem{remark}[theorem]{Remark}
\newtheorem{corollary}[theorem]{Corollary}

\renewcommand{\epsilon}{\varepsilon}

\newcommand{\RNum}[1]{\uppercase\expandafter{\romannumeral #1\relax}}

\newcommand{\bv}{\ensuremath{\mathbf{b}}}

\newcommand{\dv}{\ensuremath{\mathbf{d}}}
\newcommand{\Dv}{\ensuremath{\mathbf{D}}}
\newcommand{\ev}{\ensuremath{\mathbf{e}}}

\newcommand{\lv}{\ensuremath{\boldsymbol{\ell}}}
\newcommand{\Lv}{\ensuremath{\mathbf{L}}}
\newcommand{\mv}{\ensuremath{\mathbf{m}}}
\newcommand{\Mv}{\ensuremath{\mathbf{M}}}
\newcommand{\nv}{\ensuremath{\mathbf{n}}}
\newcommand{\Nv}{\ensuremath{\mathbf{N}}}

\newcommand{\rv}{\ensuremath{\mathbf{r}}}
\newcommand{\Rv}{\ensuremath{\mathbf{R}}}
\newcommand{\sv}{\ensuremath{\mathbf{s}}}
\newcommand{\Sv}{\ensuremath{\mathbf{S}}}
\newcommand{\uv}{\ensuremath{\mathbf{u}}}
\newcommand{\vv}{\ensuremath{\mathbf{v}}}

\newcommand{\wv}{\ensuremath{\mathbf{w}}}
\newcommand{\xv}{\ensuremath{\mathbf{x}}}

\newcommand{\yv}{\ensuremath{\mathbf{y}}}

\newcommand{\zv}{\ensuremath{\mathbf{z}}}

\newcommand{\etav}{\ensuremath{\boldsymbol{\eta}}}
\newcommand{\zetav}{\ensuremath{\boldsymbol{\zeta}}}

\newcommand{\zerov}{\ensuremath{\boldsymbol{0}}}

\newcommand{\Am}{\ensuremath{\mathbf{A}}}

\newcommand{\IDm}{\ensuremath{\mathbf{I}}}

\def\Pr{\mathrm{Pr}}

\DeclareMathAlphabet{\mcl}{OMS}{cmsy}{m}{n}

\newlength\tikzwidth
\newlength\tikzheight

\definecolor{mycolor1}{rgb}{0.63529,0.07843,0.18431}%
\definecolor{mycolor2}{rgb}{0.00000,0.44706,0.74118}%
\definecolor{mycolor3}{rgb}{0.00000,0.49804,0.00000}%
\definecolor{mycolor4}{rgb}{0.87059,0.49020,0.00000}%
\definecolor{mycolor5}{rgb}{0.00000,0.44700,0.74100}%
\definecolor{mycolor6}{rgb}{0.74902,0.00000,0.74902}%

\def\fig_path{./Figures}
\IEEEoverridecommandlockouts

\title{Sparse Regression LDPC Codes}

\author{Jamison R. Ebert, \textit{Student Member, IEEE},
Jean-Francois Chamberland, \textit{Senior Member, IEEE}, \\
Krishna R. Narayanan, \textit{Fellow, IEEE}
\thanks{
This material is based upon work supported, in part, by the National Science Foundation (NSF) under Grants CCF-2131106 \& CNS-2148354, and by Qualcomm Technologies, Inc., through their University Relations Program.
This work was presented, in part, at the 2023 Information Theory Applications (ITA) workshop and the 2023 IEEE International Symposium on Information Theory (ISIT). 

Jamison R. Ebert, Jean-Francois Chamberland, and Krishna R. Narayanan are with the Department of Electrical and\ Computer Engineering, Texas A\&M University, College Station, TX 77843, USA (emails: \{jrebert, chmbrlnd, krn\}@tamu.edu).
\textit{Corresponding author: J.-F. Chamberland.}
}
}

\begin{document}

\maketitle

\begin{abstract}
This article introduces a novel concatenated coding scheme called sparse regression LDPC (SR-LDPC) codes.
An SR-LDPC code consists of an outer non-binary LDPC code and an inner sparse regression code (SPARC) whose respective field size and section sizes are equal.
For such codes, an efficient decoding algorithm is proposed based on approximate message passing (AMP) that dynamically shares soft information between inner and outer decoders. 
This dynamic exchange of information is facilitated by a denoiser that runs belief propagation (BP) on the factor graph of the outer LDPC code within each AMP iteration. 
It is shown that this denoiser falls within the class of non-separable pseudo-Lipschitz denoising functions and thus that state evolution holds for the proposed AMP-BP algorithm.
Leveraging the rich structure of SR-LDPC codes, this article proposes an efficient low-dimensional approximate state evolution recursion that can be used for efficient hyperparameter tuning, thus paving the way for future work on optimal code design. 
Finally, numerical simulations demonstrate that SR-LDPC codes outperform contemporary codes over the AWGN channel for parameters of practical interest. 
SR-LDPC codes are shown to be viable means to obtain shaping gains over the AWGN channel.
\end{abstract}


\section{Introduction}
\label{section:introduction}

Low-density parity check (LDPC) codes have been studied extensively over the past several decades~\cite{gallager1962low,mackay1999good,luby2001improved,richardson2001capacity,chung2001analysis,davey1998low,bennatan2006design,richardson2008modern} and are known to be capacity approaching.  
Furthermore, under certain conditions, encoded messages can be recovered efficiently using iterative belief propagation (BP) decoding.
Since the complexity per iteration of BP decoding grows linearly with the block length, this paradigm offers a pragmatic solution for decoding codes with long block lengths~\cite{costello2014spatially}.
Moreover, some spatially coupled LDPC constructions feature capacity approaching iterative decoding thresholds while also avoiding the pitfall of error floors~\cite{felstrom1999time,lentmaier2010iterative,kudekar2013spatially,yedla2014simple,kumar2014threshold,andriyanova2016threshold}.
However, systems operating at shorter block lengths may not be conducive to the application of spatial coupling.
In such situations, non-binary LDPC codes have been leveraged as means to provide adequate performance~\cite{davey1998low,bennatan2006design,declercq2007decoding,voicila2010low,chang2012non}.

In a seemingly unrelated research direction, Joseph and Barron introduce the concept of a sparse regression code (SPARC)~\cite{joseph2012least,joseph2013fast,venkataramanan2019sparse} which establishes a connection between code design and sparse recovery in high dimensions.
SPARC codewords consist of sparse linear combinations of the columns of a design matrix; thus, the problem of SPARC decoding is equivalent to that of noisy support recovery for which many low complexity frameworks have been studied in the literature.
Most notably, Barbier \textit{et al.} introduce an approximate message passing (AMP) decoder for SPARCs in~\cite{barbier2014amp}.
It is shown that SPARCS with AMP decoding achieves the asymptotic single-user additive white Gaussian noise (AWGN) channel capacity under an appropriately chosen power allocation~\cite{barbier2017approximate, rush2017capacity}.
Concurrently, there have also been efforts to maximize the finite block-length performance of SPARCs~\cite{greig2017techniques,cao2021using}.

A popular strategy for improving the performance of codes in practical settings is to adopt a concatenated structure.
For example, Greig and Venkataramanan combine a binary LDPC code with a SPARC and propose the following decoding algorithm~\cite{greig2017techniques}. 
First, AMP is run to decode the SPARC; then, the factor graph of the LDPC code is initialized using the soft outputs of AMP; BP is subsequently run to decode the LDPC code; and finally, AMP is run once more to decode the SPARC after removing the contribution of confidently decoded sections.
This approach is shown to provide significant performance benefits over uncoded SPARCs for finite block lengths.
Similarly in \cite{cao2021using}, Cao and Vontobel concatenate a SPARC and a cyclic redundancy check (CRC) code for the complex AWGN channel.
In this scheme, the outer CRC code serves as an error detection mechanism to determine the true codeword among the multiple candidate paths retained from the static outputs of AMP. 
Interestingly, the schemes discussed in both~\cite{greig2017techniques} and ~\cite{cao2021using} produce a steep waterfall in error performance, a phenomenon that is not achieved by the standalone AMP decoder when operating over short block-lengths.
From a theoretic perspective, Liang \textit{et al.} show that, with a carefully designed outer code, a compressed-coding scheme can asymptotically achieve the single-user Gaussian capacity provided that the state evolution for AMP remains accurate in the presence of the outer code~\cite{liang2020compressed}.
These results suggest that concatenated schemes involving SPARCs are promising codes for the AWGN channel.

Concatenated structures with SPARC-like inner codes have also been proposed in the context of unsourced random access~\cite{fengler2021sparcs,amalladinne2021unsourced,ebert2022coded}.
In~\cite{amalladinne2021unsourced}, Amalladinne \textit{et al.} demonstrate that, under AMP decoding, the structure of a judiciously designed outer code can be integrated into the composite iterative recovery algorithm for the inner code via a dynamic denoising function.
Surprisingly, despite being mentioned by Liu \textit{et al.}\ in~\cite{liu2021capacity} as a possible future research direction, such an approach has not been considered for the single-user scenario.
Therefore, the purpose of this article is to address this deficiency. 

\subsection{Main Contributions}

In this article, a novel concatenated coding scheme is introduced consisting of an outer non-binary LDPC code and an inner sparse regression code, where the field size of the outer code equals the section size of the inner code. 
An efficient decoding algorithm based on AMP is presented that allows for information to be dynamically shared between inner and outer decoders. 
This dynamic exchange of information is facilitated by a denoising function that performs BP on the factor graph of the outer LDPC code during each AMP iteration. 
It is shown that the proposed dynamic denoiser falls into the framework of non-separable pseudo-Lipschitz denoising functions and therefore that the state evolution formalism for AMP holds in the presence of the outer code.
Leveraging the rich mathematical structure inherent in both the code design and the decoding algorithm, an approximate state evolution recursion is proposed for efficient hyperparameter tuning and code optimization. 
Finally, the proposed code, referred to as a \textit{Sparse Regression LDPC (SR-LDPC)}\cite{ebert2023srldpc} code, is shown to outperform other SPARC and LDPC code constructions over the AWGN channel for parameters of practical interest. 
Numerical results suggest that SR-LDPC codes may be leveraged as effective means to obtain shaping gain over the AWGN channel. 

\subsection{Organization}
The remainder of this article is organized as follows. 
Section~\ref{section:system_model} describes the channel model and introduces SR-LDPC encoding and decoding. 
Section~\ref{section:bp_denoiser} describes the design of the dynamic denoiser and investigates some of its properties.
Section~\ref{section:state_evolution} utilizes the structure of SR-LDPC codes and the proposed decoding algorithm to develop a low-dimensional approximate state evolution recursion for code optimization. 
Then, Section~\ref{section:simulation_results} presents numerical simulation results highlighting the benefits of SR-LDPC codes.
Finally, Section~\ref{section:conclusion} offers concluding remarks. 
Derivations and proofs for the theorems contained throughout this article may be found in Appendices~\ref{appendix:properties_bp_denosier} and~\ref{appendix:state_evolution}. 

\section{System Model} 
\label{section:system_model}

We consider a memoryless point-to-point AWGN channel where both the transmitter and the receiver are equipped with a single antenna.
In this model, the received signal $\yv \in \mathbb{R}^n$ is given by
\begin{equation} \label{equation:ChannelModel}
\yv = \xv + \zv,
\end{equation}
where $\xv \in \mathbb{R}^n$ is the transmitted signal, $\zv\sim\mathcal{N}\left(\zerov, \sigma^2\IDm\right)$ represents AWGN, and $n$ denotes the number of channel uses or, equivalently, the number of (real) degrees of freedom available.
The signal-to-noise ratio (SNR) is expressed as
\begin{equation} \label{equation:SNR}
\frac{E_b}{N_0} = \frac{\mathbb{E} \left[ \| \xv \|^2 \right]}{2 B \sigma^2},
\end{equation}
where $B$ denotes the number of information bits conveyed in $\xv$. 
The set of codewords is subject to an average power constraint which, without loss of generality, can be set to one (i.e., $\mathbb{E}\left[\|\xv\|^2\right] = 1$) with the understanding that a given SNR may be achieved by adjusting the noise variance.
As mentioned above, we wish to study a coding architecture composed of a sparse regression inner code~\cite{joseph2012least,joseph2013fast,venkataramanan2019sparse}, and a non-binary LDPC outer code~\cite{davey1998low,bennatan2006design,richardson2008modern}.
We elaborate on the encoding process and the decoding scheme below.

\subsection{SR-LDPC Encoding}
\label{subsection:Encoding}

The proposed encoding process features a sequence of three distinct steps: $q$-ary LDPC encoding, indexing of LDPC symbols, and inner CS encoding.
In the first step, the information bits are encoded into a $q$-ary LDPC codeword via well-established operations~\cite{davey1998low,bennatan2006design,richardson2008modern}.
The second step transforms the $q$-ary LDPC codeword into a suitable sparse vector.
The last step is the matrix multiplication emblematic of a sparse regression code; the output is sometimes referred to as a large random matrix system~\cite{joseph2012least,joseph2013fast,venkataramanan2019sparse,liu2021capacity}.
We summarize the notions pertaining to this encoding process below while concurrently introducing necessary notation.

\subsubsection{$q$-ary LDPC Encoding}
The LDPC encoder takes a binary sequence $\wv \in \mathbb{F}_{2}^B$ as its input and maps it to a $q$-ary codeword $\vv \in \mathbb{F}_{q}^L$, where $q$ denotes the size of the Galois field~\cite{davey1998low,bennatan2006design}.
Note that for $\mathbb{F}_q$ to be a field, $q$ must be a power of a prime. 
Throughout this article, we assume $q$ is of the form $q = 2^m$ for some $m \in \mathbb{N}, m > 1$.
We represent the resultant codeword in concatenated form as
\begin{equation} \label{equation:CodewordLDPC2}
\vv = \left( v_1, v_2, \ldots, v_L \right),
\end{equation}
where the $\ell$th element $v_{\ell}$ lies in finite field $\mathbb{F}_{q}$ and $L$ is the length of the resulting codeword.

\begin{remark} \label{remark:NotationOverload}
There exists a bijection $\Phi: \mathbb{F}_q \rightarrow [q]$ between the elements of $\mathbb{F}_q$ and the integers $[q] = \{0, 1, \ldots, q-1\}$, where the integer $0$ represents the zero element of $\mathbb{F}_q$ and the integer $1$ represents the unity element of $\mathbb{F}_q$ \cite{bennatan2006design}.
Throughout this article, we adopt such an arbitrary, but fixed bijection.
We exploit this relation by employing the same variable for a field element $g \in \mathbb{F}_q$ and for its corresponding integer $\Phi\left(g\right) \in [q]$. 
This slight abuse of notation greatly simplifies the exposition of SR-LDPC codes.
Furthermore, its use should not lead to confusion because one can unambiguously infer from context whether $g$ refers to the field element or to its integer representation. 
\end{remark}

\subsubsection{LDPC Symbol Indexing}
With Remark~\ref{remark:NotationOverload} in mind, it becomes straightforward to explain the second step of the encoding process.
Coded symbol sparsification/indexing consists of mapping $v_{\ell} \in \mathbb{F}_q$ to standard basis vector $\ev_{v_{\ell}} \in \mathbb{R}^{q}$ and subsequently stacking the $L$ basis vectors together.
We seize this opportunity to reinforce the notion that entry $\ell$ of $\vv$ is an element of $\mathbb{F}_q$, but $v_{\ell}$ in $\ev_{v_{\ell}}$ refers to an integer in $[q]$ under our overloaded notation.
With that, the output of the indexing process becomes
\begin{equation} \label{equation:Indexing}
\sv = \begin{bmatrix} \ev_{v_1} \\ \vdots \\ \ev_{v_L} \end{bmatrix} ,
\end{equation}
where $\sv$ is an $L$-sparse vector of length $qL$.
Vector $\sv$ has a structure akin to that of a sparse regression code prior to multiplication by a large random matrix.
This structured sparsity can be exploited during decoding.

\subsubsection{Inner CS Encoding}
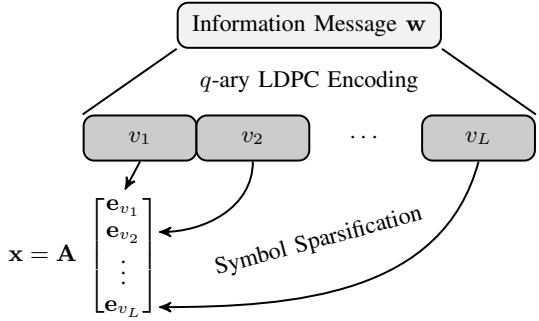
\begin{figure}
\centering
\begin{tikzpicture}[
  font=\small, >=stealth', line width = 0.75pt,
  fragment/.style={rectangle, minimum height=6mm, minimum width=35mm, draw=black, fill=gray!10, rounded corners},
  block/.style={rectangle, minimum height=6mm, minimum width=15mm, draw=black, fill=gray!40, rounded corners}
]

\node[fragment] (fragment) at (0,2.5) {Information Message $\wv$};

\foreach \j in {1,2} {
    \node[block] (block\j) at (1.5*\j-3.75,1) {$v_{\j}$};
}
\node at (0.75,1) {$\cdots$};
\node[block] (block4) at (2.25,1) {$v_{L}$};

\draw (fragment.south west) to (block1.north west);
\draw (fragment.south east) to (block4.north east);
\node (encoding) at (0,1.75) {$q$-ary LDPC Encoding};

\node[rotate=15] (index) at (0.125,-0.375) {Symbol Sparsification};

\draw[->] (block1.south) to (-2.45,0.3125);
\draw[->] (block2.south) to [out=-90,in=0] (-2,-0.25);
\node (encoding) at (-3,-0.5625) {$\xv = \Am \; \begin{bmatrix} \ev_{v_1} \\ \ev_{v_2} \\ \vdots \\ \ev_{v_L} \end{bmatrix} $};
\draw[->] (block4.south) to [out=-105,in=0] (-2,-1.25);

\end{tikzpicture}
\caption{This notional diagram depicts the encoding process for an SR-LDPC code.
    Information message $\wv$ is first outer encoded using an LDPC code over $\mathbb{F}_q$. 
    Every LDPC-encoded field element is subsequently converted into a one-sparse basis vector.
    The collection of one-sparse vectors are then stacked into a SPARC-like sequence.
    The resulting $L$-sparse vector is pre-multiplied by matrix $\Am$.
    The outcome of this process is the signal $\xv$.}
\label{figure:EncodingProcess}
\end{figure}

The last phase of the encoding process consists in pre-multiplying vector $\sv$ by matrix $\Am$ to obtain $\xv = \Am \sv$, where $\Am \in \mathbb{R}^{n \times qL},~n \ll qL$ and $\Am_{i,j}\sim\mathcal{N}\left(0, \frac{1}{n}\right)$.
Equation \eqref{equation:ChannelModel} may thus be rewritten as:
\begin{equation} \label{equation:UpdatedChannelModel}
\yv = \Am \sv + \zv .
\end{equation}
The overall encoding process is depicted in Fig.~\ref{figure:EncodingProcess}.
We are now ready to discuss the decoding process for SR-LDPC codes.

\subsection{SR-LDPC Decoding}
\label{subsection:SRLDPCdecoding}

Paralleling the development of AMP for sparse regression codes~\cite{rush2017capacity} and drawing inspiration from concatenated AMP systems~\cite{liang2020compressed,liu2021capacity}, we wish to create an iterative process to recover state vector $\sv$ from $\yv$ using AMP.
However, a notable distinction between our system model and previously published articles is the presence of a $q$-ary LDPC outer code.
Thus, we wish to create an AMP decoder that simultaneously takes advantage of the structured sparsity in $\sv$ and the parity structure embedded in the LDPC outer code.
This can be accomplished by incorporating message passing on the factor graph of the LDPC code into the AMP denoiser.
A similar approach proposed by Amalladinne \textit{et al.}\ can be found in~\cite{amalladinne2022unsourced}, where the intended application is unsourced random access.
While the two strategies are conceptually similar, the denoiser we wish to utilize below differs from the one employed in \cite{amalladinne2022unsourced} because, in the problem at hand, only one codeword is present within $\yv$.
This distinction simplifies the structure of the code and enables us to leverage a denoiser that more closely parallels traditional message passing algorithms for $q$-ary LDPC codes.

Our AMP composite algorithm is as follows,
\begin{align}
\zv^{(t)} &= \yv - \Am \sv^{(t)} + \frac{\zv^{(t-1)}}{n} \operatorname{div} \etav_{t-1} \left( \rv^{(t-1)} \right) \label{equation:AMP-Residual} \\
\rv^{(t)} &= \Am^{\mathrm{T}} \zv^{(t)} + \sv^{(t)} \label{equation:EffectiveObservation} \\
\sv^{(t+1)} &= \etav_{t} \left( \rv^{(t)} \right), \label{equation:AMP-Denoising}
\end{align}
where the superscript $t$ denotes the iteration count.
The algorithm is initialized with conditions $\rv^{(0)} = \sv^{(0)} = \zerov$ and $\zv^{(0)} = \yv$.
Furthermore, every quantity with a negative iteration count is equal to the zero vector.

Equation \eqref{equation:AMP-Residual} computes the \textit{residual} error under the current state estimate $\sv^{(t)}$ enhanced with an Onsager correction term.
This residual error is used to compute an \textit{effective observation} in \eqref{equation:EffectiveObservation}, which is passed through a denoiser to produce a revised state estimate in \eqref{equation:AMP-Denoising}. 
The denoising functions $\left( \etav_t (\cdot) \right)_{t \geq 0}$ seek to exploit the structure of $\sv$ to promote AMP's convergence to the true state. 
\begin{figure*}[t]
\centering
\begin{tikzpicture}
  [
  font=\small, >=stealth', line width=1pt,
  check/.style={rectangle, minimum height=2.5mm, minimum width=2.5mm, draw=black},
  varnode/.style={circle, minimum size=2mm, draw=black},
  mmse/.style={rectangle, minimum height=7.5mm, minimum width=25mm, rounded corners, draw=black},
  quantity/.style={rectangle, minimum height=8mm, minimum width=8mm, rounded corners, draw=black},
  multiply/.style={trapezium, trapezium angle=75, draw=black, minimum width=10mm, minimum height=8mm, rounded corners}
  ]

%
\node (received) at (-7,4.0625) {Received Signal};
\node[quantity] (signal) at (-7,3.375) {$\mathbf{y}$};
\filldraw[black] (-9,4) circle (1mm);
\draw[->, rounded corners] (-9, 4) -- (-9, 3.375) -- node[below] {Input} (signal);

%
\node[quantity] (residual) at (-5.5,2) {$\mathbf{z}^{(t)}$};
\node[circle, minimum width=8mm, draw=black] (bigsum) at (-7,2) {$\sum$}
  edge[<-] (signal)
  edge[->] (residual);
\draw[->, rounded corners] (-7,-1) -- (-7,-1.75) -- node[above] {Output} (-5.5,-1.75);
\node[multiply] (Amatrix) at (-7,0.5) {$\Am$}
  edge[->] node[right] {$-$} (bigsum)
  edge[<-] (-7,-1);
\filldraw[black] (-7,-1) circle (1mm);


\foreach \g in {1} {
    %
    \node (observation) at (-2,3.0625) {Effective Observation};
    \draw[draw=black, densely dotted, rounded corners, fill=white] (-4.375,1.25) rectangle (0.375,2.75);
    \node[multiply, shape border rotate=270] (dual) at (-3.5,2) {$\Am^\intercal$}
      edge[<-] (residual);
    \node[circle, minimum width=8mm, draw=black] (sum) at (-2,2) {$\sum$}
      edge[<-] (dual);
    \node[quantity] (rv\g) at (-0.5,2) {$\rv^{(t)}$}
      edge[<-] (sum)
      edge[->] (1,2);

    %
    \node (denoiser) at (2.5,4.0625) {Denoiser};
    \draw[draw=black, densely dotted, rounded corners, fill=white] (1,0.25) rectangle (4,3.75);
    \node[mmse] (mmse) at (2.5,0.875){Dynamic BP};
    \draw[draw=black, rounded corners] (1.25,2) rectangle (3.75,3.5);
    \foreach \s in {1,2,3,4,5} {
      \node[varnode] (var-\s) at (1+0.5*\s,2.25) {}
        edge (mmse);
    }
    \foreach \c in {1,2,3} {
      \node[check] (check-\c) at (1+0.75*\c,3.25) {};
    }
    \draw (var-5) -- (check-3.south);
    \draw (var-4) -- (check-3.south);
    \draw (var-3) -- (check-3.south);
    \draw (var-2) -- (check-2.south);
    \draw (var-5) -- (check-2.south);
    \draw (var-4) -- (check-1.south);
    \draw (var-2) -- (check-1.south);
    \draw (var-1) -- (check-1.south);
    
    %
    \node[quantity] (state) at (-2,-1) {$\sv^{(t)}$}
      edge[->] (sum)
      edge[->] (-6.875,-1);
    \node[quantity] (state-delay) at (2.5,-1) {Delay}
      edge[<-] (2.5,0.25)
      edge[->] (state);
}


%
\draw[draw=black, densely dotted, rounded corners] (-10,-1.75) rectangle (-8.0, 1.25);
\node[draw=none,rotate=90] (onsagerlabel) at (-10.25, -0.25) {Onsager Term};
\node[quantity] (onsagerdelay) at (-5.5,0.5) {Delay}
  edge[<-](residual);
\node[quantity] (div) at (-9,-1) {$\frac{1}{n} \mathrm{div} (\cdot)$}
  edge[<-,dashed] (-7,-1);
\node[circle, minimum width=8mm, draw=black] (times) at (-9,0.5) {$\times$}
  edge[<-,dashed] (div);
\draw[dashed,rounded corners] (onsagerdelay) -- (-5.5,-0.5) -- (-6.875,-0.5);
\draw[->,dashed,rounded corners] (-7.125,-0.5) -- (-8,-0.5) -- (times);
\draw[dashed] (-6.875,-0.5) arc[radius=0.125, start angle=0, end angle=180];
\draw[->,dashed,rounded corners] (times) -- (-9,2) -- (bigsum);

\end{tikzpicture}
\caption{This diagram depicts the operation of the dynamic AMP-BP decoder.
The input comes in the form of observation $\yv$ at the top left.
During every AMP iteration, the algorithm computes the residual $\zv$, which incorporates the effect of the Onsager term.
This vector is then turned into an effective observation, which acts as the input to the BP denoiser.
After message passing, a state estimate vector is produced for every section.
This, in turn, yields the updated global estimate via concatenation.
The computation of the Onsager term, which is intrinsic to AMP, is highlighted on the left.
The iterative process repeats itself until convergence is achieved, at which point the state estimate $\hat{\sv}$ is taken as the output of the algorithm.
}
\label{figure:DecodingProcess}
\end{figure*}
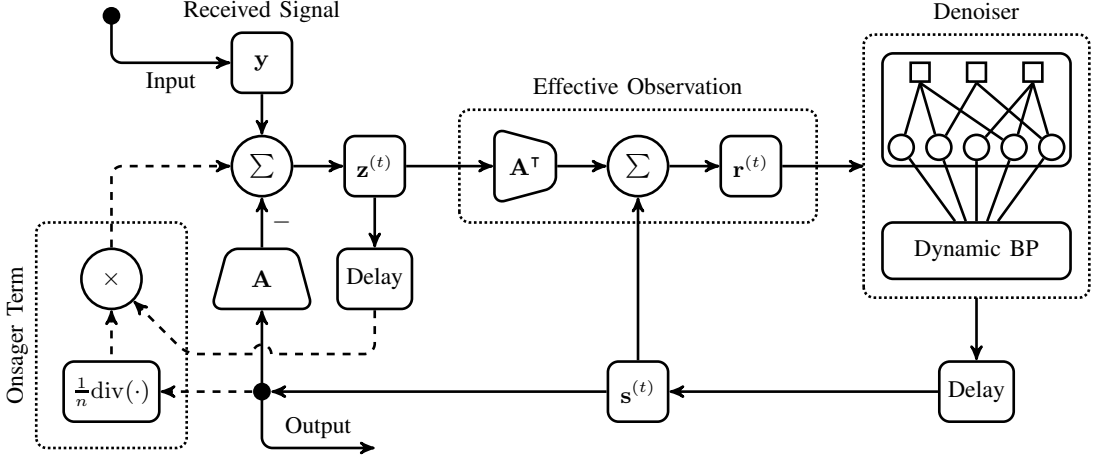

Generally speaking, one may want to employ the Bayes-optimal denoiser, which is the conditional expectation of $\sv$ given observation $\rv^{(t)}$,
Unfortunately, this approach is computationally intractable in the context of SR-LDPC codes because it entails summing over all possible codewords.
As an alternative, we know that BP can be applied to $q$-ary LDPC codes in a computationally efficient manner.
Furthermore, at any point during BP, a belief on individual LDPC symbols can be formed based on incoming messages from neighboring factor nodes, including the information afforded by the local observation.
Thus, we can potentially run a few rounds of BP as a means to get an estimate for the state vector by leveraging the connection between sections of $\sv$ and LDPC symbols.
Such an approach will simultaneously exploit the block sparsity and parity structure in $\sv$. 
In this sense, iterative message passing can act as a foundation for pragmatic denoising functions.
We elaborate on this connection below and, concurrently, we review pertinent notions of BP applied to $q$-ary LDPC codes.
As a final observation, we note that a standalone BP decoder can be employed after the AMP iterative process has terminated to further improve performance.

\section{BP Denoiser}
\label{section:bp_denoiser}

In this section, we introduce the denoising function we wish to employ within AMP.
To begin, we emphasize that $\rv$ admits a sectionized representation akin to that of the state vector $\sv$ in \eqref{equation:Indexing}.
That is, we can view both the state estimate and the effective observation as a concatenation of $L$ vectors, each of length $q$.
Mathematically, we have
\begin{xalignat*}{2}
\rv &= \begin{bmatrix} \rv_1 \\ \vdots \\ \rv_L \end{bmatrix} &
\hat{\sv} &= \begin{bmatrix} \hat{\sv}_1 \\ \vdots \\ \hat{\sv}_L \end{bmatrix} .
\end{xalignat*}
This point is crucially important because the denoiser is constructed in a block-wise fashion.
As a side note, we neglect the superscript $(t)$, which denotes the iteration count, for most of the discussion below to lighten notation; instead, we employ the hat symbol to distinguish the estimate $\hat{\sv}$ from the true state vector $\sv$.
To help keep track of variables, Fig.~\ref{figure:DecodingProcess} illustrates several of the key quantities we employ throughout.

Each section $\rv_{\ell}$ in $\rv$ acts as a vector observation about the value of $\sv_{\ell} \in \left\{ \ev_g : g \in [q] \right\}$.
An astounding and enabling property of AMP is that, under certain technical conditions, the effective observation $\rv$ is asymptotically distributed as $\sv + \tau \boldsymbol{\zeta}$, where $\boldsymbol{\zeta}$ is a random vector with independent $\mathcal{N}(0,1)$ components and $\tau$ is a deterministic quantity.
This fact hinges on the presence of the Onsager term in \eqref{equation:AMP-Residual} and on some smoothness conditions for the denoising functions.
While we delay the treatment of these technical conditions until Section~\ref{subsection:properties_bp_denoiser}, we take advantage of the Gaussian distribution in our discussion below.
For the time being, we posit this property and formally introduce it as a condition.

\begin{condition} \label{condition:AsymptoticCharacterization}
The effective observation $\rv^{(t)}$ is asymptotically distributed as $\sv + \tau_t \boldsymbol{\zeta}_t$, where $\boldsymbol{\zeta}_t$ is a random vector with independent $\mathcal{N}(0,1)$ components and $\tau_t$ is specified by a set of deterministic equations.
This asymptotic characterization takes place in the dimensions of the system, as opposed to time or iteration count.
\end{condition}

We describe below our rationale behind the denoising function assuming Condition~\ref{condition:AsymptoticCharacterization} holds; we eventually provide a rigorous foundation for this condition, but this can only be done once the structure of the denoiser is established.
Consider the effective observation restricted to section~$\ell$.
Under Condition~\ref{condition:AsymptoticCharacterization}, the distribution of random observation vector $\Rv_{\ell}$ given section $\Sv_{\ell} = \ev_g$ is given by
\begin{equation*}
f_{\Rv_{\ell} | \Sv_{\ell}} \left( \rv_{\ell} | \ev_g \right)
= \frac{1}{(2 \pi)^{\frac{q}{2}} \tau^q }
\exp \left( - \frac{\left\| \rv_{\ell} - \ev_g \right\|^2}{2 \tau^2} \right) .
\end{equation*}
It may be beneficial to think of the inner AMP loop as being equivalent to accessing a Gaussian vector channel $L$ times, with every channel use being attached to one LDPC symbol in a manner akin to pulse position modulation (PPM).
Under a uniform input distribution, the conditional distribution of $\Sv_{\ell}$ becomes
\begin{equation} \label{equation:ConditionalProbability}
\begin{split}
\boldsymbol{\alpha}_{\ell} (g)
&= \Pr \left( \Sv_{\ell} = \ev_g \middle| \Rv_{\ell} = \rv_{\ell} \right)
= \Pr \left( V_{\ell} = g \middle| \Rv_{\ell} = \rv_{\ell} \right) \\
&= \frac{f_{\Rv_{\ell} | V_{\ell}} \left( \rv_{\ell} | g \right)}
{\sum_{h \in \mathbb{F}_q} f_{\Rv_{\ell} | V_{\ell}} \left( \rv_{\ell} | h \right)}
= \frac{e^{- \frac{\left\| \rv_{\ell} - \ev_g \right\|^2}{2 \tau^2}}}
{\sum_{h \in \mathbb{F}_q} e^{- \frac{\left\| \rv_{\ell} - \ev_h \right\|^2}{2 \tau^2}}} \\
&= \frac{e^{\frac{\rv_{\ell}(g)}{\tau^2}}}
{\sum_{h \in \mathbb{F}_q} e^{\frac{\rv_{\ell}(h)}{\tau^2}}} .
\end{split}
\end{equation}

A possible estimate for $\Sv_{\ell}$ can be formed by taking its conditional expectation, given observation $\Rv_{\ell} = \rv_{\ell}$, with
\begin{equation} \label{equation:LocalMMSE}
\mathbb{E} \left[ \Sv_{\ell} \middle| \Rv_{\ell} = \rv_{\ell} \right]
= \sum_{g \in \mathbb{F}_q} \ev_g \Pr \left( \Sv_{\ell} = \ev_g \middle| \Rv_{\ell} = \rv_{\ell} \right) .
\end{equation}
A variant of this approach can be found in \cite{rush2017capacity} for a system without an outer code.
It is also employed in \cite{fengler2019sparcs} in the context of unsourced random access.
Yet, this approach overlooks the redundancy found in the outer code for the system under consideration.
Ideally, we would like to take advantage of the outer code with the more precise MMSE estimate of the form $\mathbb{E} \left[ \Sv_{\ell} \middle| \Rv = \rv \right]$.
Unfortunately, as mentioned above, computing this conditional expectation is far too complex to be implemented in practice.
A viable alternative that trades off performance and complexity is to perform BP on the factor graph of the outer LDPC code.
Implicitly, this approach computes an estimate for every $\Sv_{\ell}$ based on the observations contained within the corresponding computation tree of the code, up to a certain depth~\cite{wiberg1995codes,richardson2008modern}.

Frameworks to perform BP on factor graphs are well-established~\cite{kschischang2001factorgraph}; thus, we assume some familiarity with such iterative procedures. 
For the $q$-ary LDPC portion of the article, we borrow definitions and concepts from Bennatan and Burshtein~\cite{bennatan2006design}, who offer a compelling exposition of $q$-ary LDPC codes.
We proceed by first considering the nuances of non-binary LDPC factor graphs, then presenting a BP algorithm, then proposing a dynamic BP denoiser for SR-LDPC codes, and finally by considering the properties of the proposed denoiser. 

\subsection{Non-Binary LDPC Graphs}
\label{subsection:nonbinary_ldpc_graphs}

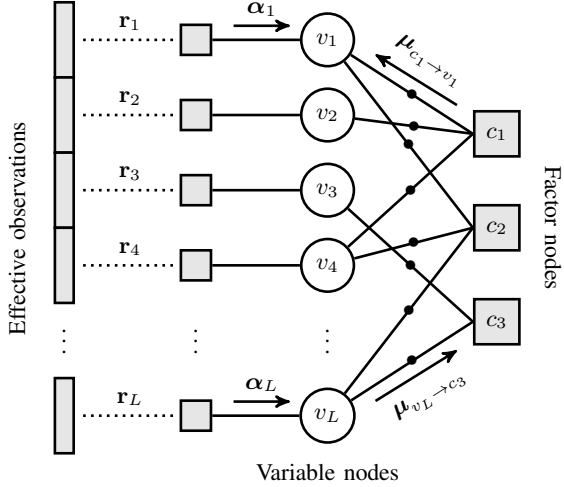
\begin{figure}[t]
  \centering
  \begin{tikzpicture}
  [
  font=\small, >=stealth', line width=1pt, draw=black,
  check/.style={rectangle, minimum height=6mm, minimum width=6mm, draw=black, fill=gray!20},
  trivialcheck/.style={rectangle, minimum height=4mm, minimum width=4mm, draw=black, fill=gray!20},
  section/.style={circle, minimum size=7mm, draw=black}
  ]

\foreach \l in {1,2,3,4} {
  \node[section] (v\l) at (0,3.5-\l) {$v_{\l}$};
}
\node[section] (vL) at (0,-2.5) {$v_{L}$};
\node at (0,-1.4) {$\vdots$};

\foreach \t in {1,2,3,4} {
  \node[trivialcheck] (t\t) at (-1.75,3.5-\t) {}
    edge (v\t);
}
\node[trivialcheck] (tL) at (-1.75,-2.5) {}
  edge (vL);
\node at (-1.75,-1.4) {$\vdots$};

\foreach \r in {1,2,3,4} {
  \draw[fill=gray!20] (-3.625,4-\r) rectangle (-3.375,3-\r);
  \draw[dotted] (-3.25,3.5-\r) -- node[above] {$\mathbf{r}_{\r}$} (-2,3.5-\r);
}
\draw[fill=gray!20] (-3.625,-2) rectangle (-3.375,-3);
\draw[dotted] (-3.25,-2.5) -- node[above] {$\mathbf{r}_{L}$} (-2,-2.5);
\node at (-3.5,-1.4) {$\vdots$};

\node[check] (a1) at (2.25,1.25) {$c_1$};
\node[check] (a2) at (2.25,0) {$c_2$};
\node[check] (a3) at (2.25,-1.25) {$c_3$};

\node[rotate=-90] (check) at (3,0) {Factor nodes};
\node (variable) at (0,-3.25) {Variable nodes};
\node[rotate=90] (observation) at (-4.125,0) {Effective observations};

\draw (v1) -- node {$\bullet$} (a1.west);
\draw (v2) -- node {$\bullet$} (a1.west);
\draw (v4) -- node {$\bullet$} (a1.west);
\draw (v1) -- node {$\bullet$} (a2.west);
\draw (v4) -- node {$\bullet$} (a2.west);
\draw (vL) -- node {$\bullet$} (a2.west);
\draw (v3) -- node {$\bullet$} (a3.west);
\draw (vL) -- node {$\bullet$} (a3.west);

\draw[shorten <=0.75cm,shorten >=0.25cm,<-] (v1)++(0,0.2cm) -- node[above,yshift=-0.1cm,xshift=0.25cm,rotate=-32.5] {$\boldsymbol{\mu}_{c_1 \to v_1}$} ([yshift=0.25cm]a1.west);
\draw[shorten <=0.75cm,shorten >=0.25cm,->] (vL)++(0,-0.2cm) -- node[below,yshift=0.1cm,xshift=0.25cm,rotate=32.5] {$\boldsymbol{\mu}_{v_L \to c_3}$} ([yshift=-0.25cm]a3.west);
\draw[->] (-1.25,2.6875) -- node[above] {$\boldsymbol{\alpha}_1$} (-0.5,2.6875);
\draw[->] (-1.25,-2.3125) -- node[above] {$\boldsymbol{\alpha}_L$} (-0.5,-2.3125);

\end{tikzpicture}
  \caption{
  This illustration shows the augmented factor graph for the denoising function with the variable nodes, the parity check constraints, and the extra factors associated with local observations.
  The effective observation vector is sectionized in a way that matches variable nodes.}
  \label{figure:FactorGraph}
\end{figure}

The factor graph for an $\mathbb{F}_q$ LDPC code features $L$ variable (left) nodes, which correspond to the symbols of the codewords, and $L (1 - R)$ check (right) nodes enforcing parity constraints, where $R$ is the \emph{design rate} of the LDPC code~\cite{davey1998low,bennatan2006design,richardson2008modern}.
An important distinction between binary and non-binary LDPC codes is that a factor graph for a non-binary LDPC code typically includes edge labels, which take values in $\mathbb{F}_q \setminus \{ 0 \}$.
Fig.~\ref{figure:FactorGraph} offers a notional factor graph for a non-binary LDPC code, where the edge labels are represented as dots along the graph edges.
A vector $\vv \in \mathbb{F}_q^L$ is a valid codeword if it satisfies the parity equations
\begin{equation} \label{equation:ParityConstraint}
\sum_{v_{\ell} \in N(c_p)} \omega_{\ell, p} \otimes v_{\ell} = 0 \qquad \forall p \in [L (1 - R)],
\end{equation}
where $N(c_p)$ is the collection of variable nodes adjacent to parity check~$c_p$, i.e., neighbors on the factor graph.
The summation and the multiplication operator $\otimes$ in \eqref{equation:ParityConstraint} take place over finite field $\mathbb{F}_q$.
Parameter $\omega_{\ell, p}$ represents the label or weight assigned with the edge connecting variable node $v_{\ell}$ and check node $c_p$.
Adopting common factor graph concepts~\cite{kschischang2001factorgraph,loeliger2004introduction}, we denote the graph neighbors of variable node $v_{\ell}$ by $N (v_\ell)$.
The factor associated with $c_p$ and derived from parity equation \eqref{equation:ParityConstraint} can be expressed as an indicator function
\begin{equation} \label{equation:LocalFactors}
\mathcal{G}_{p} ( \vv_{p} ) = \mathbf{1} \left( \sum_{v_\ell \in N(c_p)} \omega_{\ell, p} \otimes v_{\ell} = 0 \right)
\end{equation}
where $\vv_{p} = \left( v_{\ell} \in N(c_p) \right)$ is a shorthand notation for the restriction of $\vv$ to entries associated with graph neighborhood $N(c_p)$.
With these definitions, the factor function associated with our LDPC code assumes the product decomposition given by
\begin{equation} \label{equation:FactorFunction}
\begin{split}
\mathcal{G} ( \vv )
= \prod_{p \in [L(1-R)]} \mathcal{G}_{p} ( \vv_{p} ) .
\end{split}
\end{equation}
Succinctly, $\mathcal{G} ( \vv )$ is an indicator function that assesses whether its argument is a valid codeword.

To create a suitable LDPC code, one can first construct a Tanner graph~\cite{tanner1981recursive} according to established techniques~\cite{luby2001efficient,richardson2008modern} and then assign labels to edges, possibly randomly and independently from a uniform distribution over $\mathbb{F}_q \setminus \{ 0 \}$.

\subsection{Belief Propagation}
\label{subsection:belief_propagation}

We view messages for a non-binary LDPC code as multi-dimensional belief vectors over $\mathbb{F}_q$.
Messages from variable nodes to check nodes are denoted as $\boldsymbol{\mu}_{v \to c}$, and messages in the reverse direction are represented as $\boldsymbol{\mu}_{c \to v}$.
Formally, a message going from check node $c_p$ to variable node $v_{\ell} \in N(c_p)$ is computed component-wise through the equation
\begin{equation} \label{equation:BP-Check2Variable}
\boldsymbol{\mu}_{c_p \to v_{\ell}} (g)
= \sum_{\vv_{p}: v_{\ell} = g} \mathcal{G}_{p} \left( \vv_{p} \right)
\prod_{v_j \in N(c_p) \setminus v_{\ell}} \boldsymbol{\mu}_{v_j \to c_p} (g_j) .
\end{equation}
While \eqref{equation:BP-Check2Variable} is shown in compact form, the actual summation operation is cumbersome.
Finding the set of summands entails identifying sequences of the form $\left( g_j \in \mathbb{F}_q: v_j \in N(c_p) \setminus v_{\ell} \right)$ that fulfill local condition \eqref{equation:ParityConstraint} or, equivalently,
\begin{equation} \label{equation:LocalCondition}
\sum_{v_j \in N(c_p) \setminus v_{\ell}} \omega_{j, p} \otimes g_j
= - \omega_{\ell, p} \otimes g .
\end{equation}
Likewise, a belief vector passed from variable node $v_{\ell}$ to check node $c_p$, where $p \in N(v_{\ell})$, is calculated component-wise via
\begin{equation} \label{equation:BP-Variable2Check}
\boldsymbol{\mu}_{v_{\ell} \rightarrow c_p} (g)
\propto \boldsymbol{\alpha}_{\ell} (g) \prod_{c_{\xi} \in N(v_{\ell}) \setminus c_p} \boldsymbol{\mu}_{c_{\xi} \to v_{\ell}}(g) .
\end{equation}
The `$\propto$' symbol indicates that the positive measure should be normalized before being sent out as a message.
Vector $\boldsymbol{\alpha}_{\ell}$ in \eqref{equation:BP-Variable2Check} can be viewed as a collection of beliefs based on local observations, as in \eqref{equation:ConditionalProbability}.
That is, entry $\boldsymbol{\alpha}_{\ell}(g)$ captures the posterior probability that symbol $g$ is the true field element within section~$\ell$, given the local observation.
Other BP messages are initialized with $\boldsymbol{\mu}_{v \to c} = \mathbf{1}$ and $\boldsymbol{\mu}_{c \to v} = \mathbf{1}$.
These message passing operations appear in Fig.~\ref{figure:FactorGraph}.
The traditional parallel sum-product algorithm iterates between \eqref{equation:BP-Check2Variable} and \eqref{equation:BP-Variable2Check}, alternating between updated rightbound messages and leftbound messages.

One of the key advantages of indexing vectors using field elements in $\mathbb{F}_q$, as pointed out in \cite{bennatan2006design}, is the ensuing ability to define pertinent operators on these vectors.
Paralleling existing literature, we consider two operators.

\begin{definition}[Vector $+g$ Operator \cite{bennatan2006design}]
For field element $g \in \mathbb{F}_q$, the vector $+g$ operator acting on $\bv \in \mathbb{R}^{q}$ and denoted by $\bv^{+g}$ is defined as
\begin{equation*}
\begin{split}
\bv^{+g} &= \left( b_g, b_{g \oplus 1}, \ldots, b_{g \oplus (q-1)} \right) \\
&= \left( b_{h \oplus g} : h \in \mathbb{F}_q \right),
\end{split}
\end{equation*}
where subscript addition $\oplus$ is performed in $\mathbb{F}_q$.
\end{definition}

\begin{definition}[Vector $\times g$ Operator \cite{bennatan2006design}] \label{defintion:TimesOperator}
For field element $g \in \mathbb{F}_g \setminus \{ 0 \}$, we define the vector $\times g$ operator acting on $\bv \in \mathbb{R}^{q}$ and denoted by $\bv^{\times g}$ by
\begin{equation*}
\begin{split}
\bv^{\times g} &= \left( b_0, b_g, b_{2 \otimes g}, \ldots, b_{(q-1) \otimes g} \right) \\
&= \left( b_{h \otimes g} : h \in \mathbb{F}_q \right)
\end{split}
\end{equation*}
where subscript product $\otimes$ takes place in $\mathbb{F}_q$.
\end{definition}

\noindent We emphasize that the $+g$ and $\times g$ operators introduced above are reversible, with
\begin{xalignat*}{2}
\left( \bv^{+g} \right)^{- g}
&= \bv & g &\in \mathbb{F}_q \\
\left( \bv^{\times g} \right)^{\times g^{-1}} &= \bv
& g &\in \mathbb{F}_q \setminus \{ 0 \} .
\end{xalignat*}
These operations essentially permute the entries of $\bv$ in a structured fashion that naturally meshes with field actions.
These operations are especially meaningful in the computation of BP messages for non-binary LDPC codes, as factor nodes impose constraints that are easily expressible within the Galois field $\mathbb{F}_q$.
Vector operators then become a convenient way to track the distribution of belief vectors during message passing.
Specifically, under these operations, we can rewrite \eqref{equation:BP-Check2Variable} in a concise manner:
\begin{equation} \label{equation:FactorOperationField1}
\boldsymbol{\mu}_{c_p \to v_{\ell}}
= \left( \bigodot_{v_j \in N(c_p) \setminus v_{\ell}} \left( \boldsymbol{\mu}_{v_j \to c_p} \right)^{\times \omega_{j,p}^{-1}} \right)^{\times \left(- \omega_{\ell,p} \right)}
\end{equation}
where $\omega_{j,p}$ is the label on the edge between variable node $v_j$ and factor node $c_p$~\cite{bennatan2006design}.
Here, the operator $\odot$ denotes the $\mathbb{F}_q$-convolution between two vectors,
\begin{equation*}
\left[ \boldsymbol{\mu} \odot \boldsymbol{\nu} \right]_{g}
= \sum_{h \in \mathbb{F}_q} \mu_h \cdot \nu_{g - h}
\qquad g \in \mathbb{F}_q .
\end{equation*}
The exposition can be simplified further if we absorb the edge labels within the messages themselves.
Specifically, we adopt the definitions
\begin{align}
\overline{\boldsymbol{\mu}}_{v_j \to c_p} &= \left( \boldsymbol{\mu}_{v_j \to c_p} \right)^{\times \omega_{j,p}^{-1}}
\label{equation:AbsorbedMessageVar2Check} \\
\overline{\boldsymbol{\mu}}_{c_p \to v_{\ell}} &= \left( \boldsymbol{\mu}_{c_p \to v_{\ell}} \right)^{\times \left(- \omega_{\ell,p}^{-1} \right)} .
\label{equation:AbsorbedMessageCheck2Var}
\end{align}
Then, \eqref{equation:FactorOperationField1} morphs into the simpler expression
\begin{equation} \label{equation:BP-Check2Variable-Convolution}
\overline{\boldsymbol{\mu}}_{c_p \to v_{\ell}}
= \bigodot_{v_j \in N(c_p) \setminus v_{\ell}} \overline{\boldsymbol{\mu}}_{v_j \to c_p} .
\end{equation}
This equation highlights the role of the $\mathbb{F}_q$-convolution within BP for non-binary LDPC codes.
The message from variable node $v_{\ell}$ to check node $c_p$ found in \eqref{equation:BP-Variable2Check} also admits a more compact form.
For $p \in N(v_{\ell})$, the traditional outgoing message from a variable node can be written as
\begin{equation} \label{equation:VariableNodeEstimate}
\boldsymbol{\mu}_{v_{\ell} \to c_p}
= \frac{ \boldsymbol{\alpha}_{\ell} \circ \left( \operatorname*{\bigcirc}_{c_{\xi} \in N(v_{\ell}) \setminus c_p} \boldsymbol{\mu}_{c_{\xi} \to v_{\ell}} \right) }
{ \left\| \boldsymbol{\alpha}_{\ell} \circ \left( \operatorname*{\bigcirc}_{c_{\xi} \in N(v_{\ell}) \setminus c_p} \boldsymbol{\mu}_{c_{\xi} \to v_{\ell}} \right) \right\|_1 }
\end{equation}
where $\circ$ represents the Hadamard product.

A natural estimate for the distribution associated with variable node~$v_{\ell}$, including intrinsic information, is
\begin{equation} \label{equation:BP-Estimate}
\begin{split}
\hat{\Sv}_{\ell}
&= \frac{ \boldsymbol{\alpha}_{\ell} \circ \left( \operatorname*{\bigcirc}_{c_p \in N(v_{\ell})} \boldsymbol{\mu}_{c_p \to v_{\ell}} \right) }
{ \left\| \boldsymbol{\alpha}_{\ell} \circ \left( \operatorname*{\bigcirc}_{c_p \in N(v_{\ell})} \boldsymbol{\mu}_{c_p \to v_{\ell}} \right) \right\|_1 } .
\end{split}
\end{equation}
As we will see shortly, \eqref{equation:BP-Estimate} is the output of our proposed denoiser.

\begin{remark}
\label{remark:fwht}
In our construction, $q = 2^m, m \geq 1$ because indexing is derived from sequences of bits.
This invites the application of fast techniques to implement message passing over the corresponding factor graph.
Specifically, the fast Walsh-Hadamard transform (FWHT) can be utilized to rapidly and efficiently compute \eqref{equation:BP-Check2Variable-Convolution}, with
\begin{equation*}
\overline{\boldsymbol{\mu}}_{c_p \to v_{\ell}}
\propto \operatorname{fwht}^{-1} \left( \prod_{v_j \in N(c_p) \setminus v_{\ell}} \operatorname{fwht} \left( \overline{\boldsymbol{\mu}}_{v_j \to c_p} \right) \right) .
\end{equation*}
This technique is especially meaningful given that it may be desirable to maintain large sections and, hence, a large alphabet size for sparse regression codewords.
Alternatively, one could adopt a different finite field convolution or a ring structure amenable to the circular convolution to create local factor functions conducive to the fast Fourier transform \cite{davey1998low,song2003reduced,goupil2007fft}.
\end{remark}

With these tools in mind, we are ready to formally define our proposed denoiser.

\subsection{BP Denoiser}
\label{subsection:BP-denoiser}

Conceptually, one can initiate the state of the LDPC factor graph using the effective observation $\rv$, run a few rounds of BP, and then form an estimate for the state based on \eqref{equation:BP-Estimate}.
As mentioned before, in the absence of BP iterations, local estimates reduce to the conditional expectation $\mathbb{E} \left[ \Sv_{\ell} \middle| \Rv_{\ell} = \rv_{\ell} \right]$ found in \eqref{equation:LocalMMSE}.
Yet, as more iterations of the BP algorithm are performed, the estimate for $\Sv_{\ell}$ can be refined based on the computation tree of the outer code, up to a certain depth. 

\begin{definition}[BP Denoiser] \label{definition:BPDenoiser}
Let $N_t$ denote the number of BP iterations to perform during AMP iteration $t$. 
The BP denoiser:
\begin{enumerate}
    \item initializes the LDPC factor graph with estimates $\boldsymbol{\alpha}_{\ell}$ computed from $\rv_{\ell}$ for $\ell \in [L]$ according to \eqref{equation:ConditionalProbability};
    \item computes and passes variable to check messages (see \eqref{equation:VariableNodeEstimate}) and check to variable messages (see \eqref{equation:AbsorbedMessageVar2Check}, \eqref{equation:BP-Check2Variable-Convolution}, \eqref{equation:AbsorbedMessageCheck2Var} and Remark~\ref{remark:fwht}) along the edges of the factor graph in an alternating fashion $N_t$ times;
    \item computes updated state estimates according to \eqref{equation:BP-Estimate}.
\end{enumerate}
The output of this denoiser can then be passed to the AMP composite algorithm for the computation of the next residual, enhanced with the Onsager term.
\end{definition}

To the reader familiar with the iterative decoding of LDPC codes, it may seem more natural to construct an estimate for the distribution associated with variable node~$v_{\ell}$ based on extrinsic information, i.e.,
\begin{equation} \label{equation:BP-Estimate-Extrinsic}
\begin{split}
\hat{\Sv}_{\ell}^{\mathrm{ext}}
&= \frac{ \operatorname*{\bigcirc}_{c_p \in N(v_{\ell})} \boldsymbol{\mu}_{c_p \to v_{\ell}} }
{ \left\| \operatorname*{\bigcirc}_{c_p \in N(v_{\ell})} \boldsymbol{\mu}_{c_p \to v_{\ell}} \right\|_1 } .
\end{split}
\end{equation}
One may be tempted to argue that $\boldsymbol{\alpha}_{\ell}$ should not be used when passing a message back to the left-most nodes in Fig.~\ref{figure:FactorGraph}.
However, the presence of the Onsager term in \eqref{equation:AMP-Residual} serves to break first-order dependencies and, hence, one need not worry about the presence of $\boldsymbol{\alpha}_{\ell}$ in \eqref{equation:BP-Estimate} as part of the iterative process.

When AMP is used with the BP denoiser as presented in this paper, the algorithm is referred to as the \textit{AMP-BP} algorithm. 

\subsection{Properties of BP Denoiser}
\label{subsection:properties_bp_denoiser}

As mentioned previously, the proposed BP denoiser relies on Condition~\ref{condition:AsymptoticCharacterization}, which states that the effective observation $\rv$ is asymptotically distributed as the true state vector embedded in i.i.d. Gaussian noise, or 
\begin{equation}
    \rv^{(t)} \sim \sv + \tau_t\zetav_t,
\end{equation}
where $\tau_t$ is a deterministic quantity and $\zetav_t$ has i.i.d. $\mathcal{N}(0, 1)$ components.
If the proposed non-separable BP denoiser satisfies the conditions set forth by Berthier et al. in~\cite{berthier2020state}, then Condition~\ref{condition:AsymptoticCharacterization}, and more generally, the state evolution of AMP, are guaranteed to hold. 
The required conditions include the sensing matrix $\Am$ having i.i.d. Gaussian entries with mean zero and variance $1/n$ and the denoiser being pseudo-Lipschitz of a certain order. 
The requirement on the sensing matrix is satisfied for SR-LDPC codes by construction; however, as will be shown, whether the BP denoiser satisfies the requirement of being pseudo-Lipschitz depends on the following condition. 

\begin{condition}[Sub-Girth BP] \label{condition:Sub-Girth-BP}
The BP denoiser is said to possess the \emph{Sub-Girth BP} condition when fewer message passing iterations are performed on the factor graph of the LDPC code than the shortest cycle of this same graph, per AMP denoising step.
\end{condition}

This condition is reasonable because, in contrast to the traditional technique of performing many BP iterations at once, we are primarily interested in repeatedly performing a few BP iterations at a time as the BP algorithm is run within each AMP iteration. 
Though this condition is sufficient for the theory to hold, in practice, one may be able to violate this condition and still obtain reasonable performance.
With this condition in mind, we obtain the following result. 

\begin{theorem} [BP Denoiser is Lipschitz Continuous] \label{theorem:bp_denoiser_lipschitz}
Under Condition~\ref{condition:Sub-Girth-BP}, the BP denoiser presented in Definition~\ref{definition:BPDenoiser} is Lipschitz continuous.
\end{theorem}

Note that the proof for this and all subsequent theorems in this section may be found in Appendix~\ref{appendix:properties_bp_denosier}. 

Given Theorem~\ref{theorem:bp_denoiser_lipschitz}, it can be shown that the proposed BP denoiser falls within the framework of non-separable pseudo-Lipschitz denoising functions functions~\cite{berthier2020state} and that, under Condition~\ref{condition:Sub-Girth-BP}, state evolution holds for the AMP-BP algorithm.
This endows the algorithm with a significant amount of mathematical structure that will be exploited in Section~\ref{section:state_evolution} to obtain a computationally efficient recursion for hyperparameter tuning and code optimization. 

The final step in completing our AMP-BP algorithm is computing the Onsager correction term, which we provide in Proposition~\ref{proposition:onsager_correction_term}. 

\begin{proposition} \label{proposition:onsager_correction_term}
The Onsager correction term associated with the BP denoiser is given by
\begin{equation} \label{equation:BP-OnsagerCorrection}
\begin{split}
&\frac{\zv^{(t-1)}}{n} \operatorname{div} \etav_{t-1} \left( \rv^{(t-1)} \right)\\
&= \frac{\zv^{(t-1)}}{n\tau^2} \left( \left\| \etav_{t-1} \left( \rv^{(t-1)} \right) \right\|_1 - \left\| \etav_{t-1} \left( \rv^{(t-1)} \right) \right\|_2^2 \right) .
\end{split}
\end{equation}
\end{proposition}

Note that this term has a particularly simple form that is amenable to efficient computation. 
Having established these results, we are now ready to consider the state evolution of the SR-LDPC decoder. 

\section{State Evolution}
\label{section:state_evolution}

State evolution is a mathematical formalism that seeks to characterize the performance of AMP as a function of its iteration count~$t$. 
It is an asymptotic tool, rooted in the analysis of large systems, that captures performance through a Gaussian approximation.
Under suitable regularity conditions, this asymptotic approach is valid in that random vectors in the approximate Gaussian model converge in distribution to their counterparts in the original system~\cite{bayati2011dynamics,bayati2015universality}, a property that greatly simplifies mathematical analysis.
The foundation for state evolution in the current setting is the AMP framework for non-separable, pseudo-Lipschitz denoising functions put forth by Berthier, Montanari, and Nguyen in~\cite{berthier2020state}.
Recall that, under Condition~\ref{condition:AsymptoticCharacterization}, the effective observation is distributed as the true state embedded in zero-mean i.i.d. Gaussian noise with variance $\tau_t^2$. 
Using the state evolution formalism, the value of $\tau_t^2$ at iteration $t$ can be computed through the following recursion: 
\begin{equation}
    \label{equation:state_evolution}
    \begin{split}
        \tau_0^2 &= \lim_{n \rightarrow \infty} \frac{\| \yv \|_2^2}{n} \\
        \tau_{t+1}^2 &= \sigma^2 + \lim_{n \rightarrow \infty} \frac{1}{n} \mathbb{E} \left[ \left\| \boldsymbol{\eta}_{t} \left( \sv + \tau_t \boldsymbol{\zeta}_{t} \right) - \sv \right\|_2^2 \right], \\
    \end{split}
\end{equation}
where $\sigma^2$ is the variance of the AWGN channel and $n$ is the number of channel uses employed. 

Note that $\mathbb{E}\left[\tau_0^2\right] = \sigma^2 + \frac{L}{n}$ and thus $\tau_0^2$ may be approximated directly from the channel noise variance and the system parameters, notably without the need for high-dimensional computations.
Furthermore, the expected mean-squared-error (MSE) of the input to the denoiser can be computed as a function only of $\tau_t^2$ and the system parameters. 
It follows that, if we can track how the expected MSE changes as the effective observation passes through the denoiser, then we could obtain a low-dimensional recursion for predicting the performance of a given SR-LDPC code. 
Such a recursion could be used for hyperparameter tuning and code optimization without the need for extensive high-dimensional Monte-Carlo simulation campaigns.  

The goal of this section is to develop such an algorithm for approximating the state evolution \eqref{equation:state_evolution} as a function only of the channel noise variance and system parameters. 
To accomplish this goal, we study the rich structure of SR-LDPC codes and the properties of the BP messages passed during SR-LDPC decoding.
The remainder of this section is organized as follows. 
In Section~\ref{subsection:geometric_uniformity}, we show that SR-LDPC codes are geometrically uniform and thus that the probability of error is independent of which codeword is sent. 
Assuming the all-zero codeword is sent, we further show that the factor graph's edge labels have no effect on the distribution of BP messages and thus can be neglected. 
Continuing our study, in Section~\ref{subsection:properties_bp_messages}, we show that the expected MSE of a graph message can be computed directly from the expected $L_2$-norm of that same message. 
Furthermore, we study the effects of $\mathbb{F}_q$ convolution on the distribution of BP messages.
In Section~\ref{subsection:computing_state_evolution}, we show that the expected MSE at the output of a check node can be computed based on the expected MSEs of messages sent from graph neighbors, and likewise, that the expected MSE at the output of a variable node can be approximated given that same information. 
In this section, we combine these results to obtain an MSE message passing algorithm that tracks how the expected MSE changes as the effective observation is passed through the BP denoiser. 
Using this MSE message passing algorithm, we define an approximate state evolution recursion that, we claim, can be used for hyperparameter tuning and code optimziation. 
Finally, in Section~\ref{subsection:performance_of_state_evolution}, we investigate the performance of our proposed algorithm and discuss its limitations. 

We note that the proofs associated with all propositions, corollaries, lemmas, and theorems from this section are contained in Appendix~\ref{appendix:state_evolution}. 

\subsection{Geometric Uniformity of Indexed $\mathbb{F}_{q}$ LDPC Codes}
\label{subsection:geometric_uniformity}

We begin by examining the symmetry properties of indexed $\mathbb{F}_{q}$ LDPC codewords and show that SR-LDPC codewords are geometrically uniform. 
The notion of geometric uniformity, as presented in \cite{forney1991geometrically}, is of great value because it guarantees that the error probability over a Gaussian channel does not depend on which codeword is transmitted.
In particular, the sets of distances (distance profile) from any codeword to all other codewords are all the same.

First, note that the signal constellation produced by mapping a field element $g \in \mathbb{F}_q$ to vector element $\ev_g \in \mathbb{R}^q$ is invariant under coordinate permutations.
That is, suppose $\Pi : \mathbb{R}^q \to \mathbb{R}^q$ is a permutation matrix. then the following set equality (trivially) holds,
\begin{equation} \label{equation:PermutationInvariant}
\left\{ \ev_g : g \in \mathbb{F}_q \right\}
= \left\{ \Pi \ev_g : g \in \mathbb{F}_q \right\} .
\end{equation}
Furthermore, it is known that every coordinate permutation operator is an \emph{isometry}, with
\begin{equation} \label{equation:PermutationIsometry}
\begin{split}
\left\| \Pi \ev_g - \Pi \ev_h \right\|^2
&= \left\| \ev_g - \ev_h \right\|^2\quad \forall g, h \in \mathbb{F}_q, \\
\end{split}
\end{equation}
where $\pi: [q] \to [q]$ is the permutation function corresponding to matrix $\Pi$.
It follows that the sets in \eqref{equation:PermutationInvariant} are \emph{geometrically congruent} under any permutation operator.

We can extend these observations to state vectors in $\mathbb{R}^{qL}$.
Consider a set of permutation matrices on $\mathbb{R}^{q \times q}$, which we denote by $\Pi^{(1)}, \ldots, \Pi^{(L)}$.
Define the block diagonal permutation matrix $\boldsymbol{\Pi} = \operatorname{diag} \left( \Pi^{(1)}, \ldots, \Pi^{(L)} \right)$.
Given that we can write
\begin{equation*}
\left\| \sv - \sv' \right\|^2
= \sum_{\ell=1}^L \left\| \sv_{\ell} - \sv_{\ell}' \right\|^2 ,
\end{equation*}
we deduce that the original codebook $\mathcal{S} \subset \mathbb{R}^{qL}$ and any section-wise permutation $\boldsymbol{\Pi}$ thereof must also be \emph{geometrically congruent}.

\begin{definition}[Geometric Uniformity \cite{forney1991geometrically}]
A signal set $\mathcal{S}$ is \emph{geometrically uniform} if, given any two points $\sv$ and $\sv'$ in $\mathcal{S}$, there exists an isometry that transforms $\sv$ to $\sv'$ while leaving $\mathcal{S}$ invariant.
\end{definition}

Since any section-wise permutation $\boldsymbol{\Pi}$ acting on $\mathcal{S}$ produces a symmetry of $\mathcal{S}$, it becomes straightforward to show that this set is geometrically uniform.

\begin{proposition}
\label{proposition:geometric_uniformity}
Let $\mathcal{S}$ be the codebook produced by combining the $\mathbb{F}_q$ LDPC outer code and the indexing step.
Then, the set $\mathcal{S}$ is geometrically uniform.
\end{proposition}

This result should not be too surprising to the reader familiar with LDPC codes, vector indexing, and sparse regression codes.
Nevertheless, this is important because it permits an analysis of the system under the all-zero codeword.
We elaborate on the section symmetry in the following proposition.

\begin{proposition} \label{proposition:SectionPermuationInvariance}
Let $\Pi$ be any permutation on the entries of vectors in $\mathbb{R}^q$.
The distribution of $\Rv_{\ell}$ conditioned on the input $\Sv_{\ell}$ is permutation invariant in the sense that
\begin{equation} \label{equation:SectionPermuationInvariance}
f_{\Rv_{\ell} | \Sv_{\ell}} \left( \rv_{\ell} | \ev_g \right)
= f_{\Rv_{\ell} | \mathbf{S}_{\ell}} \left( \Pi \rv_{\ell} \middle| \ev_{\pi(g)} \right)
\end{equation}
for any $\rv_{\ell} \in \mathbb{R}^q$.
Above, $\pi(\cdot)$ is a representation of $\Pi$ where $\pi(g)$ denotes the permutation of the integer position of $g$ under the bijection of Remark~\ref{remark:NotationOverload}.
In other words, $\ev_{\pi(g)} = \Pi \ev_g$ for any $g \in \mathbb{F}_q$.
\end{proposition}

While there are only $q$ permutations induced through field mapping of the form $g \mapsto g \oplus u$, determined by choosing $u \in \mathbb{F}_q$, the mathematical statement holds for all $q!$ possible permutations of the vector indices.
Thus, this attribute forms a strong notion of statistical symmetry that is related to
a symmetry property of binary LDPC codes~\cite{richardson2001design} and non-binary LDPC codes over finite fields~\cite{bennatan2006design}.
As mentioned in the latter article, the capacity-achieving distribution for constrained channels with such statistical symmetry is uniform over the $q$ possible inputs.
Fortunately, the marginal input distribution to the Gaussian vector channel corresponding to section~$\ell$ under SR-LDPC encoding is indeed uniform over the admissible inputs.

\begin{corollary}
\label{corollary:edge_label_effect_on_distribution}
Suppose that vector $\ev_0$ is the input to the Gaussian vector channel of Condition~\ref{condition:AsymptoticCharacterization}.
Then, observation vectors $\Rv_{\ell}$ and $\Rv_{\ell}^{\times \omega}$, where $\omega \in \mathbb{F}_q \setminus \{ 0 \}$, have identical distributions.
\end{corollary}

This corollary is pertinent because, as discussed above, system analysis for a geometrically uniform codebook can be performed assuming that the all-zero codeword has been transmitted.
Under such circumstances, the action of edge label $\omega_{\ell, p}$ does not affect the distribution of the rightbound messages $\boldsymbol{\mu}_{v_{\ell} \to c_p}$. 
This complexity reduction also extends to the distribution of $\boldsymbol{\mu}_{c_p \to v_{\ell}}$.
Hence, the effects of the edge labels can be disregarded when studying the statistical properties of the BP denoiser.

\subsection{Statistical Properties of BP Messages}
\label{subsection:properties_bp_messages}

We now explore certain statistical properties of the BP messages that are passed during SR-LDPC decoding.
Throughout the remainder of this section, we assume that the all-zero codeword has been sent (i.e., $v_\ell = 0~\forall\ell \in [L]$) and we assume that Condition~\ref{condition:AsymptoticCharacterization} holds.
For our purpose, it is necessary to entertain the notions of likelihood-vector random variables and probability-vector random variables.
Consider the following definitions that seek to capture these notions.

\begin{definition} \label{permuation-symmetric}
A \emph{likelihood-vector random variable} is defined as a random vector $\Lv = \left( L_0, L_1, \ldots, L_{q-1} \right)$ in $\mathbb{R}^q$ that takes on values from the set of likelihood vectors, whose entries are non-negative.
Furthermore, a likelihood-vector random variable $\Lv$ is called \emph{group-symmetric} if
\begin{equation} \label{equation:GroupSymmetricVectorRV}
f_{\Lv} \left( \lv \right)
= f_{\Lv} \left( \lv^{\times g} \right)
\end{equation}
for any field element $g \in \mathbb{F}_q \setminus \{ 0 \}$.
Similarly, $\Lv$ is said to be \emph{permutation-symmetric} if
\begin{equation} \label{equation:PermutationSymmetricVectorRV}
f_{\Lv} \left( \lv \right)
= f_{\Lv} \left( \Pi_0 \lv \right)
\end{equation}
for any permutation matrix $\Pi_0$ that preserves the location of the zeroth entry in its argument.
Such a random vector is qualified as \emph{dominant} if, in addition to \emph{symmetry}, the mean of the zeroth element $\mathbb{E} [L_0]$ is greater than or equal to the expected value of any other entry.
\end{definition}

\begin{remark} \label{remark:BulletExpectation}
We emphasize that, if a likelihood-vector random variable is symmetric, then the expected value of all its components, except for the zeroth entry, are equal.
We can therefore unambiguously adopt the uniform notation $\mathbb{E} [L_{\bullet}]$, where $\bullet$ can be any field element $g \in \mathbb{F}_q \setminus \{ 0 \}$.
\end{remark}

A likelihood-vector random variable that takes on values in $\mathbb{R}^q$ can be normalized to produce a probability-vector random variable on $\mathbb{F}_q$.
Such vector random variables are defined below.
Both notions are important in analyzing the performance of sparse regression LDPC codes.

\begin{definition} \label{definition:SymmetricProbVectorRV}
A \emph{probability-vector random variable} is a random vector $\Dv = \left( D_0, D_1. \ldots, D_{q-1} \right)$ in $\mathbb{R}^q$ that takes on values in the probability simplex.
Such a probability-vector random variable $\Dv$ is called \emph{group-symmetric} if
\begin{equation} \label{equation:GroupSymmetricProbVectorRV}
f_{\Dv} \left( \dv \right)
= f_{\Dv} \left( \dv^{\times g} \right)
\end{equation}
for any field element $g \in \mathbb{F}_q \setminus \{ 0 \}$.
Moreover, $\Dv$ is said to be \emph{permutation-symmetric} if
\begin{equation} \label{equation:PermutationSymmetricProbVectorRV}
f_{\Dv} \left( \dv \right)
= f_{\Dv} \left( \Pi_0 \dv \right)
\end{equation}
for any permutation matrix $\Pi_0$ that preserves the location of the zeroth entry in its argument.
Such a random vector is \emph{dominant} if, in addition to \emph{symmetry}, the mean of the zeroth element $\mathbb{E} [D_0]$ is greater than or equal to the expected value of any other entry.
\end{definition}

The most important probability-vector random variables for the problem at hand are normalized likelihood-vector random variables.
We adopt the notation
\begin{equation}
\bar{\Lv} = \frac{\Lv}{\left\| \Lv \right\|_1}
\end{equation}
for the normalized version of a likelihood-vector random variable.
Under Condition~\ref{condition:AsymptoticCharacterization}, the components of likelihood vectors associated with the effective observation are derived from the Gaussian distribution.
This fact, which we use extensively throughout, acts as a motivation for the next definition.

\begin{definition} \label{definition:SymmetricGaussianLikelihoodRV}
We define a \emph{permutation-symmetric Gaussian likelihood-vector random variable} as a dominant permutation-symmetric likelihood-vector random variable 
that is component-wise equal to
\begin{equation} \label{equation:GaussianSymmetricVectorRV}
L_g = f_{\Rv_{\ell} | \Sv_{\ell}} \left( \Rv_{\ell} | \ev_g \right)
\end{equation}
where $\Rv_{\ell}$ has distribution $f_{\Rv_{\ell} | \Sv_{\ell}} \left( \cdot | \ev_0 \right)$, as defined in \eqref{equation:ConditionalGaussianDistribution}.
When a permutation-symmetric Gaussian likelihood-vector random variable $\Lv$ is normalized, we call the resulting vector $\bar{\Lv}$ a \emph{permutation-symmetric Gaussian probability-vector random variable}.
\end{definition}

Part of the motivation for introducing these definitions is rooted in the operations that take place on the factor graph of the $\mathbb{F}_q$ LDPC outer code during belief propagation.
One benefit of working with the likelihood-vector, as opposed to the probability vector, is the fact that vector components in \eqref{equation:GaussianSymmetricVectorRV} are independent, with the joint distribution assuming a product form.
We turn to the effects of the $\mathbb{F}_q$-convolution on random likelihood vectors and show that this operation preserves certain properties.

\begin{lemma} \label{lemma:fq_conv_dominant_gs_lv_rv}
The $\mathbb{F}_q$-convolution of a finite set of independent dominant group-symmetric likelihood-vector random variables produces a dominant group-symmetric likelihood-vector random variable.
\end{lemma}

Another interesting property of the $\mathbb{F}_q$-convolution of likelihood vectors pertains to the one-norm of the output.
This result is analogous to the one-norm relation for the regular convolution; it is included below for the sake of completeness.

\begin{lemma} \label{lemma:ConvolutionNormL1}
Let $\big\{ \lv^{(p)} \big\}$ be a collection of likelihood vectors and define
\begin{equation*}
\nv = \bigodot_{p \in [n]} \lv^{(p)}
\end{equation*}
where $n$ is a natural number.
Then, the one-norm of the output of the $\mathbb{F}_q$-convolution is equal to the product of the one-norms of the input vectors,
\begin{equation*}
    \left\| \nv \right\|_1 = \prod_{p \in [n]} \left\| \lv^{(p)} \right\|_{1}.
\end{equation*}
\end{lemma}

\begin{corollary} \label{corollary:OneNormConvolution}
Let $\left\{ \Lv^{(p)} \right\}$ be a collection of independent likelihood-vector random variables and define
\begin{equation*}
\Nv = \bigodot_{p \in [n]} \Lv^{(p)}
\end{equation*}
where $n$ is a natural number.
Then, it necessarily holds that
\begin{equation*}
\mathbb{E} \left[ \left\| \Nv \right\|_1 \right]
= \prod_{p \in [n]} \mathbb{E} \left[ \left\| \Lv^{(p)} \right\|_1 \right].
\end{equation*}
\end{corollary}

With the properties identified above, we can characterize the expectation of the $\mathbb{F}_q$-convolution of certain collections of likelihood-vector random variables.
This is meaningful in that we can then track the mean behavior of certain BP messages passed on the factor graph of the outer LDPC code.

\begin{proposition} \label{proposition:ConvolutionGroupSymmetricLikelihoodRV}
Suppose $\left\{ \Lv^{(p)} \right\}$ forms a collection of independent dominant group-symmetric likelihood-vector random variables.
For any natural number $n$, the expectation of $\Nv = \bigodot_{p \in [n]} \Lv^{(p)}$ is governed by
\begin{align}
\begin{split} \label{equation:ConvolutionGroupSymmetricLikelihoodRVMeanN0}
\mathbb{E} [N_0]
&= \frac{1}{q} \prod_{p \in [n]} \left( \mathbb{E} \left[ L_0^{(p)} \right] + (q-1) \mathbb{E} \left[ L_{\bullet}^{(p)} \right] \right) \\
&\qquad + \left( 1 - \frac{1}{q} \right)  \prod_{p \in [n]} \left( \mathbb{E} \left[ L_0^{(p)} \right] - \mathbb{E} \left[ L_{\bullet}^{(p)} \right] \right)
\end{split} \\
\begin{split} \label{equation:ConvolutionGroupSymmetricLikelihoodRVMeanNBullet}
\mathbb{E} [N_{\bullet}]
&= \frac{1}{q} \prod_{p \in [n]} \left( \mathbb{E} \left[ L_0^{(p)} \right] + (q-1) \mathbb{E} \left[ L_{\bullet}^{(p)} \right] \right) \\
&\qquad - \frac{1}{q} \prod_{p \in [n]} \left( \mathbb{E} \left[ L_0^{(p)} \right] - \mathbb{E} \left[ L_{\bullet}^{(p)} \right] \right) .
\end{split}
\end{align}
\end{proposition}

We can extend these findings to probability vectors of the form $\bar{\Lv} = \Lv / \left\| \Lv \right\|_1$, where $\Lv$ is a dominant group-symmetric likelihood-vector random variable.

\begin{corollary} \label{corollary:ConvolutionGroupSymmetricProbabilityRV}
Suppose $\left\{ \bar{\Lv}^{(p)} \right\}$ is a collection of independent dominant group-symmetric probability-vector random variables.
Then, for any natural number $n$, the expectation of $\bar{\Nv} = \bigodot_{p \in [n]} \bar{\Lv}^{(p)}$ is governed by
\begin{align}
\label{equation:ConvolutionGroupSymmetricProbabilityRVMeanN0}
\mathbb{E} \left[ \bar{N}_0 \right]
&= \frac{1}{q} + \left( \frac{q}{q-1} \right)^{n-1} \prod_{p \in [n]} \left( \mathbb{E} \left[ \bar{L}_0^{(p)} \right] - \frac{1}{q} \right) \\
\label{equation:ConvolutionGroupSymmetricProbabilityRVMeanNBullet}
\mathbb{E} \left[ \bar{N}_{\bullet} \right]
&= \frac{1}{q} - \frac{1}{q} \left( \frac{q}{q-1} \right)^n \prod_{p \in [n]} \left( \mathbb{E} \left[ \bar{L}_0^{(p)} \right] - \frac{1}{q} \right) .
\end{align}
\end{corollary}

An important application of Corollary~\ref{corollary:ConvolutionGroupSymmetricProbabilityRV} for our analysis is the situation where $\left\{ \bar{\Lv}^{(p)} \right\}$ is a collection of independent permutation-symmetric Gaussian probability-vector random variables, each with parameter $\tau$.
Our last set of results on the statistical properties of likelihood vectors pertains to the two-norm of permutation-symmetric Gaussian probability-vector random variables.

\begin{lemma} \label{lemma:SymmetricGaussianProbabilityRV}
Suppose $\bar{\Lv}$ is a dominant permutation-symmetric Gaussian probability-vector random variable with standard deviation parameter $\tau$.
The expected two-norm of $\bar{\Lv}$ is related to $\mathbb{E} \left[ \bar{L}_0 \right]$ through the equation
\begin{equation} \label{equation:SymmetricGaussianProbabilityRV}
\mathbb{E} \left[ \left\| \bar{\Lv} \right\|_2^2 \right]
= \mathbb{E} \left[ \bar{L}_0 \right] .
\end{equation}
\end{lemma}

Inspecting the proof of Lemma~\ref{lemma:SymmetricGaussianProbabilityRV}, one notices that \eqref{equation:SymmetricGaussianProbabilityRV} hinges on the property $\mathbb{E} \left[ \bar{L}_{\bullet}^2 \right] = \mathbb{E} \left[ \bar{L}_0 \bar{L}_{\bullet} \right]$.
This relation arises naturally for dominant permutation-symmetric Gaussian probability-vector random variables, yet it may occur more generally.
For instance, this property may be preserved under certain factor graph operations such as the $\mathbb{F}_q$-convolution.
Before studying this property more thoroughly, we give it a formal name.

\begin{definition} \label{definition:BalancedProbabilityVector}
We say that a dominant permutation-symmetric probability-vector random variable $\bar{\Lv}$ is \emph{balanced} if
\begin{equation*}
\mathbb{E} \left[ \bar{L}_{g}^2 \right] = \mathbb{E} \left[ \bar{L}_0 \bar{L}_{g} \right]
\end{equation*}
for all $g \in \mathbb{F}_q$.
\end{definition}

\begin{proposition} \label{proposition:BalancedProbabilityRV}
Let $\bar{\Lv}$ be a dominant permutation-symmetric probability-vector random variable.
Then, $\bar{\Lv}$ is \emph{balanced} if and only if
\begin{equation} \label{equation:BalancedProbabilityRV}
\mathbb{E} \left[ \left\| \bar{\Lv} \right\|_2^2 \right]
= \mathbb{E} \left[ \bar{L}_0 \right] .
\end{equation}
\end{proposition}

This structure leads to a corollary that will become very important when computing the mean-squared-error (MSE) of graph messages. 

\begin{corollary}
\label{corollary:mse_of_structured_rv}
Let $\Bar{\Lv}$ be a balanced dominant permutation-symmetric probability-vector random variable. 
Then,
\begin{equation}
    \mathbb{E}\left[\|\Bar{\Lv} - \ev_0\|_2^2\right] = 1 - \mathbb{E}\left[\|\Bar{\Lv}\|_2^2\right].
\end{equation}
\end{corollary}

Thus, when a graph message is a balanced dominant permutation-symmetric probability-vector random variable, the expected $L_2$-norm of that message is sufficient to compute the expected MSE of that same message.
Understanding the close connection between the $L_2$ norm and the expected MSE, we now seek to compute the expected $L_2$ norm of the output of the $\mathbb{F}_q$ convolution operator. 

\begin{theorem} \label{theorem:ConvolutionGaussianSymmetricMSE}
Suppose $\left\{ \bar{\Lv}^{(p)} \right\}$ is a collection of independent, balanced dominant permutation-symmetric probability-vector random variables.
For any natural number $n$, the two-norm of $\bar{\Nv} = \bigodot_{p \in [n]} \bar{\Lv}^{(p)}$ is given by
\begin{equation} \label{equation:ConvolutionGaussianSymmetricMSE}
\begin{split}
&\mathbb{E} \left[ \left\| \bar{\Nv} \right\|_2^2 \right]
= \mathbb{E} \left[ \bar{N}_0 \right] \\
&= \frac{1}{q} + \left( \frac{q}{q-1} \right)^{n-1} \prod_{i \in [n]} \left( \mathbb{E} \left[ \left\| \bar{\Lv}^{(p)} \right\|_2^2 \right]
- \frac{1}{q} \right).
\end{split}
\end{equation}
Furthermore, $\bar{\Nv}$ is itself a balanced dominant permutation-symmetric probability-vector random variable.
\end{theorem}

Recall that our overarching goal is to develop an efficient algorithm for computing $\mathbb{E} \left[ \left\| \boldsymbol{\eta}_{t} \left( \sv + \tau_t \boldsymbol{\zeta}_{t} \right) - \sv \right\|_2^2 \right]$, where the denoising function $\etav_t\left(\cdot\right)$ is the BP denoiser of Section~\ref{section:bp_denoiser}. 
Our strategy for computing this expectation is to develop expressions for approximating the expected MSE of BP graph messages as a function of the expected MSEs of their inputs.
Equipped with such expressions, we can compute the expected MSE at the output of the BP denoiser by passing MSE messages between variable and check nodes on the LDPC factor graph. 
After a fixed number of iterations, we can compute the expected output MSE and use that value to compute $\tau_{t+1}^2$ in the state evolution iteration. 
In the next section, we elaborate on this strategy and define  message passing rules for such an MSE message passing algorithm. 

\subsection{Computing the State Evolution}
\label{subsection:computing_state_evolution}

We begin with the first round of message passing in which variable to check node messages consist only of local observations. 
Under the all-zero codeword assumption, the messages arriving at any check node constitute a set of independent, balanced, dominant, permutation-symmetric Gaussian probability-vector random variables. 
Thus, Theorem~\ref{theorem:ConvolutionGaussianSymmetricMSE} may be employed to obtain an exact expression for the expected two-norm of the resultant check to variable BP messages.
This is consequential as it offers a means to calculate the MSE of a section estimate based on incoming BP messages. 

\begin{proposition}
\label{proposition:check_to_var_mse}
Let $\{\boldsymbol{\mu}_{v_{j} \rightarrow c_p} : v_{j} \in N(c_p) \setminus v_{\ell} \}$ constitute a set of independent, balanced, dominant, permutation-symmetric, Gaussian probability-vector random variables. 
Then, the MSE associated with leftbound message message $\boldsymbol{\mu}_{c_p \to v_{\ell}}$
is equal to
\begin{equation*}
\begin{split}
&\mathbb{E} \left[ \left\| \Sv_{\ell} - 
\boldsymbol{\mu}_{c_p \to v_{\ell}} \right\|_2^2 \right] =   \\
&\frac{q-1}{q} - \left( \frac{q}{q-1} \right)^{n-1} \prod_{\substack{v_j \in N(c_p) \\ v_j \neq  v_{\ell}}} \left( \mathbb{E} \left[ \left\| \boldsymbol{\mu}_{v_j \rightarrow c_p} \right\|_2^2 \right] - \frac{1}{q} \right) \\
\end{split}
\end{equation*}
where $n = |N(c_{p})|-1$.
\end{proposition}

We now turn our attention to the MSE of variable to check messages. 
The optimal BP message from variable node $v_{\ell}$ to check node $c_p$, defined in equation \eqref{equation:VariableNodeEstimate}, is created as the Hadamard product of leftbound messages.
Unfortunately, an exact characterization of the MSE associated with this operation remains elusive to the authors.
It is challenging to calculate the MSE of this estimator due, in part, to the normalization step that appears in its construction.
Furthermore, while it can be shown that  $\boldsymbol{\mu}_{v_{\ell} \to c_p}$ is a dominant permutation symmetric probability-vector random variable if its inputs are similarly structured, it remains unclear whether this BP message is also balanced.

For the sake of tractability, we propose introducing a mild approximation for the incoming check to variable node messages.
Specifically, note that under Condition~\ref{condition:AsymptoticCharacterization}, there exists a bijective function $\Psi$ such that
\begin{equation}
    \mathbb{E}\left[\boldsymbol{\alpha}(0)\right] = \Psi\left(\tau_t^2\right).
\end{equation}
Using this bijection, we propose approximating the messages $\{\boldsymbol{\mu}_{c_{p} \to v_{\ell}}\}$ as having been generated from a Gaussian model
\begin{equation}
\label{eq:leftbound_approximation}
\boldsymbol{\mu}_{c_{p} \to v_{\ell}}(g) \approx \tilde{\boldsymbol{\mu}}_{c_{p} \to v_{\ell}} (g) \sim \frac{e^{\frac{\rv_{p, \ell}(g)}{\tau_{p, \ell}^2}}}
{\sum_{h \in \mathbb{F}_q} e^{\frac{\rv_{p, \ell}(h)}{\tau_{p, \ell}^2}}},
\end{equation}
where $\rv_{p, \ell} = \sv_{p, \ell} + \nv_{p, \ell}$ and 
\begin{equation}
\label{eq:taupl2}
\nv_{p, \ell} \sim \mathcal{N} \left(0, \tau^2_{p, \ell}\right) = \mathcal{N}\left(0, \Psi^{-1}\left(\mathbb{E} \left[ \boldsymbol{\mu}_{c_{p} \to v_{\ell}} (0) \right]\right)\right).
\end{equation}
Under this approximation, it becomes straightforward to track the MSE of the rightbound messages.

\begin{proposition}
\label{proposition:var_to_chk_approximation}
Let $\{\Tilde{\boldsymbol{\mu}}_{c_{\xi} \rightarrow v_{\ell}} : c_{\xi} \in N(v_{\ell}) \setminus c_p \}$ be a set of independent, balanced, dominant, permutation-symmetric, Gaussian probability-vector random variables having the distribution provided in~\eqref{eq:leftbound_approximation}. 
Then, the MSE associated with the rightbound message $\Tilde{\boldsymbol{\mu}}_{v_{\ell} \rightarrow c_{p}}$ is given by
\begin{equation}
    \mathbb{E}\left[\left\|\Sv_{\ell} - \tilde{\boldsymbol{\mu}}_{v_{\ell} \rightarrow c_{p}} \right\|_2^2 \right] = 1 - \Psi\left(\Tilde{\tau}^2_{\ell, p}\right).
\end{equation}
Here, the effective noise variance $\Tilde{\tau}_{\ell, p}^2$ is given by
\begin{equation}
\Tilde{\tau}_{\ell, p}^2 = \frac{1}{\frac{1}{\tau^2} + \sum_{c_{\xi} \in N(v_{\ell})\setminus c_p} \frac{1}{\tau^2_{\xi, \ell}}},
\end{equation}
where $\tau^2$ denotes the effective noise variance of the effective observation from AMP and $\tau_{\xi, \ell}^2 = \Psi^{-1}\left(\mathbb{E}\left[\tilde{\boldsymbol{\mu}}_{c_{\xi} \rightarrow v_{\ell}}(0)\right]\right).$
\end{proposition}

Note that a similar result may be obtained for the output MSE of the denoiser by including all messages from neighboring check nodes as well as the local observation in the computation of $\Tilde{\tau}_{\ell, p}^2$.
Note that the approximate variable to check node message computed in Proposition~\ref{proposition:var_to_chk_approximation} is also a balanced, dominant, permutation-symmetric, Gaussian probability-vector random variable. 
Furthermore, if Condition~\ref{condition:Sub-Girth-BP} is satisfied, the variable to check node messages received at a given check node will be independent. 
Thus, the MSE of the next round's check to variable messages can also be computed using Proposition~\ref{proposition:check_to_var_mse}, thus setting the stage for an iterative procedure. 
With these facts in mind, we are now ready to present a low-dimensional recursion for approximating the state evolution from \eqref{equation:state_evolution}. 

\begin{definition} (Approximate State Evolution)
\label{definition:approximate_state_evolution}
Let $T$ denote the number of AMP iterations to model, let $\{N_t : t \in [T]\}$ denote the number of BP rounds to perform during each AMP iteration $t \in [T]$, and let $\sigma^2$ denote the channel noise variance. 
The approximate state evolution algorithm proceeds by
\begin{enumerate}
    \item Approximate $\tau_0^2 \approx \mathbb{E}\left[\tau_0^2\right] = \sigma^2 + \frac{L}{n}$
    \item For each AMP iteration $t \in [T]$:
    \begin{enumerate}
        \item Initialize all graph messages to be $\frac{1}{q}$
        \item For all $\ell \in [L]$, set  $\boldsymbol{\mu}_{c_0 \rightarrow v_{\ell}} = \mathbb{E}\left[\boldsymbol{\alpha}_{\ell}(0)\right] = \Psi\left(\tau_t^2\right)$
        \item For each BP iteration $\varsigma \in [N_t]$:
        \begin{enumerate}
            \item  Approximate expected MSE of variable to check node messages according to Proposition~\ref{proposition:var_to_chk_approximation}
            \item Compute expected MSE of check to variable node messages according to Proposition~\ref{proposition:check_to_var_mse}
        \end{enumerate}
        \item Approximate expected output MSE using Proposition~\ref{proposition:var_to_chk_approximation}, except that no check nodes are excluded in the computation of $\Tilde{\tau}_{\ell, p}^2$
        \item Compute $\tau_{t+1}^2 = \sigma^2 + \frac{1}{n}\sum_{\ell \in [L]} \mathbb{E}\left[\|\Sv_{\ell} - \hat{\Sv}_{\ell}\|_2^2 \right]$
    \end{enumerate}
\end{enumerate}
\end{definition}

Note that each graph message in this MSE message-passing algorithm is a scalar; thus, this approximate state evolution algorithm is of a dimensionality that is several orders of magnitude lower than the full SR-LDPC decoding algorithm. 
As will be shown in the next section, this property makes the approximate state evolution algorithm an attractive solution for hyperparameter tuning and code optimization tasks. 

\subsection{Performance of the State Evolution}
\label{subsection:performance_of_state_evolution}

\begin{figure}
    \centering
    \input{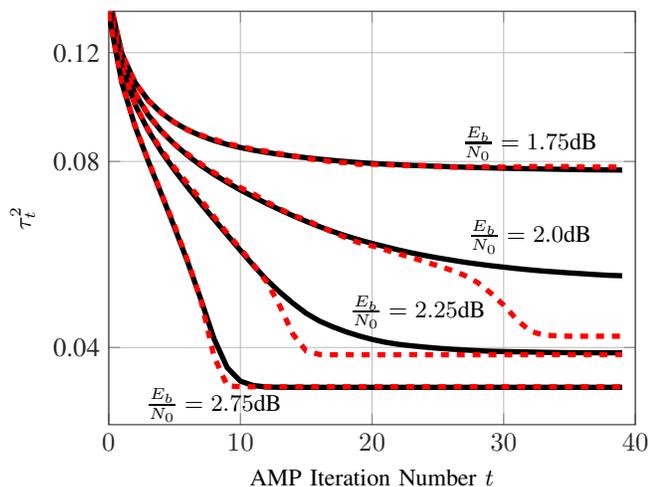}
    \caption{This figure compares the true $\tau_t^2$ values with the estimated $\hat{\tau}_t^2$ values predicted by the approximate state evolution algorithm defined in Definition~\ref{definition:approximate_state_evolution}. 
    The approximate state evolution algorithm is very accurate for low and high SNRs but provides reduced insight when AMP is at the edge of its convergence region. }
    \label{fig:se_vs_truth}
\end{figure}

Having defined a low-dimension approximate state evolution algorithm, we now seek to characterize its performance and identify its limitations.
Before doing so, we note that a popular method for approximating the true value of $\tau_t^2$ during AMP decoding is to use the following relation:
\begin{equation}
\label{eq:tau2_approximation}
\tau_t^2 \approx \frac{\|\zv^{(t)}\|_2^2}{n}.
\end{equation}
As a benchmark, we thus run the full SR-LDPC decoder for a given code many times and average the $\tau_t^2$ values computed from \eqref{eq:tau2_approximation}.
Using this benchmark, we then compare the true $\tau_t^2$ values with those predicted by the approximate state evolution algorithm for a specific SR-LDPC codes. 
The details of this code will not be given in this section, but will be addressed in detail in Section~\ref{section:simulation_results}. 
Figure~\ref{fig:se_vs_truth} highlights the results of this experiment.

From Fig.~\ref{fig:se_vs_truth}, we see that the $\hat{\tau}_t^2$ values predicted by the approximate state evolution algorithm are very accurate for low and high SNRs. 
When the $E_b/N_0 = 1.75$~dB, the SR-LDPC decoding algorithm fails to converge in the sense that $\tau_T^2 > \sigma^2$, or there is a non-zero fixed-point MSE. 
Clearly, the approximate state evolution algorithm identifies this fixed point MSE very well.
Conversely, when the $E_b/N_0 = 2.75$~dB, the SR-LDPC decoding algorithm converges in the sense that $\tau_T^2 = \sigma^2$ and thus the final expected MSE is zero. 
However, when the SNR is such that the decoding algorithm is operating on the edge of its convergence region, the predictions provided by the approximate state evolution algorithm tend to be overconfident. 
In the example provided, at $E_b/N_0 = 2.0$~dB, the approximate state evolution algorithm predicts a $\tau_T^2$ value that is lower than that observed in practice. 
Despite this caveat, the proposed approximate state evolution algorithm may be useful for code optimization as it enables the rapid comparison of different hyperparameter configurations (e.g. field size $q$, $\{N_t : t \in [T]\}$, outer LDPC code, etc) for a variety of SNRs.
Of course, exact performance for the hyperparameter selected under approximate state evolution can be validated through full Monte-Carlo simulations of the corresponding SR-LDPC code. 

\begin{figure}[t!]
    \centering
    \begin{tikzpicture}

\definecolor{customred}{rgb}{0.63529,0.07843,0.18431} 
\definecolor{customblue}{rgb}{0.00000,0.44706,0.74118} 
\definecolor{customgreen}{rgb}{0.00000,0.49804,0.00000} 

\begin{semilogyaxis}[
    font=\small,
    width=7cm,
    height=5.5cm,
    scale only axis,
    every outer x axis line/.append style={white!15!black},
    every x tick label/.append style={font=\color{white!15!black}},
    xmin=0.7,
    xmax=1.0,
    xtick = {0.7, 0.75, ..., 1.0},
    xlabel={$R_{\mathrm{LDPC}}$},
    xmajorgrids,
    every outer y axis line/.append style={white!15!black},
    every y tick label/.append style={font=\color{white!15!black}},
    ymin=0.00001,
    ymax=0.2,
    ytick = {0.00001, 0.0001, 0.001, 0.01, 0.1}, 
    ylabel={$\tau_{T}^2 - \sigma^2$},
    ymajorgrids,
    yminorgrids,
    legend style={at={(0,0)},anchor=south west, draw=black,fill=white,legend cell align=left}
]

\addplot [
    color=black,
    solid,
    line width=2.0pt,
    mark size=1.4pt,
    mark=o,
    mark options={solid}
]
table[row sep=crcr]{
    0.67 0.11769992179296572 \\
    0.68 0.11352183690836695 \\
    0.69 0.11034428585818826 \\
    0.71 0.10607691094562582 \\
    0.72 0.10211654158589972 \\
    0.74 0.09757451690674336 \\
    0.75 0.09326460076192664 \\
    0.77 0.08818820353894172 \\
    0.78 0.08351460513872544 \\
    0.80 0.07702427431120636 \\
    0.82 0.07104484992026926 \\
    0.84 0.06099712684183889 \\
    0.86 0.047856686055457315 \\
    0.88 0.027695320050536168 \\
    0.90 0.013070294880985163 \\
    0.92 0.002156190751706165 \\
    0.94 0.0003530571023319523 \\
    0.96 0.000016971413467126162 \\
    0.98 0.00020563862794686222 \\
    0.9959 0.0024863784084699844 \\
};
\addlegendentry{Full SR-LDPC Decoder};

\addplot [
    color=red,
    dashed,
    line width=2.0pt,
    mark size=1.4pt,
    mark=triangle,
    mark options={solid}
]
table[row sep=crcr]{
    0.67 0.11723758820401937 \\
    0.68 0.113484537400768 \\
    0.69 0.11024609188779108 \\
    0.71 0.10574047968278799 \\
    0.72 0.10136816833254647 \\
    0.74 0.09773558757816565 \\
    0.75 0.09389556641253502 \\
    0.77 0.0888506342035178 \\
    0.78 0.08399164655315206 \\
    0.80 0.0783941735576375 \\
    0.82 0.07123613452548884 \\
    0.84 0.0629831997775364 \\
    0.86 0.052030567233903716 \\
    0.88 0.008496461562405383 \\
    0.90 0.000025325350476418373 \\
    0.92 0.000024795307863964555 \\
    0.94 0.00002321055810341338 \\
    0.96 0.000050889595239834995 \\
    0.98 0.00018052946563017896 \\
    0.9959 0.0023264923139284938 \\
};
\addlegendentry{Approx. State Evolution};

\end{semilogyaxis}
\end{tikzpicture}
    \vspace{-4.5mm}
    \caption{This figure demonstrates the utility of the approximate state evolution algorithm for hyperparameter optimization by comparing $\tau_T^2 - \sigma^2$ for SR-LDPC codes with a constant overall rate but a varying inner LDPC code rate. The approximate state evolution tool may be used as a coarse optimization tool, possibly followed by fine tuning through a local rate search. }
    \label{fig:ldpc_rate_tuning}
\end{figure}
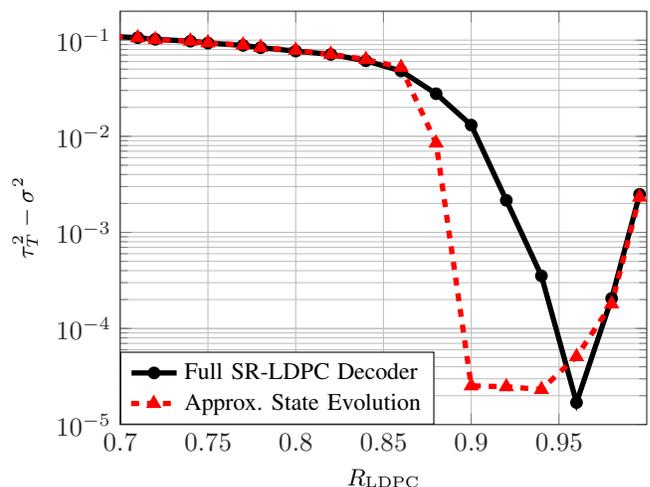

As an example of using the approximate state evolution algorithm for hyperparameter tuning, consider the task of choosing the outer LDPC code rate $R_{\textrm{LDPC}}$ while keeping the number of information bits and the number of channel uses fixed. 
In Fig.~\ref{fig:ldpc_rate_tuning}, we compare the performance of various SR-LDPC codes that are identical in every way except for their choices of outer codes.
The SNR of this comparison is chosen to be $E_b/N_0 = 2.5$~dB, the number of AMP iterations is set at $T = 20$, and the difference $\hat{\tau}_T^2 - \sigma^2$ is plotted in Fig.~\ref{fig:ldpc_rate_tuning} alongside the associated benchmark values. 
We choose to plot $\hat{\tau}_T^2 - \sigma^2$ because the $\sigma^2$ values are slightly different under each code and thus this difference provides a clearer comparison of the residual MSE. 
From Fig.~\ref{fig:ldpc_rate_tuning}, we see that the optimal $R_{\mathrm{LDPC}} \approx 0.96$, which is relatively high. 
The intuition behind this phenomenon is that when $R_{\textrm{LDPC}}$ is low, the undersampling ratio of AMP, or the ratio of the number of measurements to the dimensionality of the sparse vector $\delta = \frac{n}{qL}$, is so small that AMP may not be operating in its convergence region~\cite{donoho2009message}.
Conversely, when $R_{\mathrm{LDPC}}$ is high, the LDPC code has minimal error-correcting capabilities. 
The rate $R_{\mathrm{LDPC}} \approx 0.96$ thus appears to offer the best tradeoff between competing design criteria. 
As before, we see that the approximate state evolution algorithm tends to be overconfident when operating on the edge of AMP's convergence region. 
Nevertheless, it is clear that the approximate state evolution algorithm significantly narrows down the search space for the optimal $R_{\mathrm{LDPC}}$ and thus can be used as a coarse optimization tool, where more precise optimization may be done via full Monte-Carlo (MC) simulations. 

\section{Simulation Results}
\label{section:simulation_results}

In this section, we investigate the performance of SR-LDPC codes\footnote{The source code used to generate these results is available online at https://github.com/EngProjects/mMTC/tree/code.}.
Specifically, we simulate a randomly generated SR-LDPC code that encodes $5888$ information bits into $7350$ coded symbols. 
We define the rate of the SR-LDPC code to be the number of information bits over the number of channel uses; thus, we have that $R_{\mathrm{SRLDPC}} \approx 0.80$. 
The non-binary LDPC code employed is a $(766, 736)$ code of rate $R_{\mathrm{LDPC}} \approx 0.96$ over GF($256$) whose edges are generated via progressive edge growth (PEG) and whose weights are chosen uniformly at random from the elements of $\mathbb{F}_{256}\setminus 0$. 
These parameters were chosen to facilitate comparisons with similar codes in the literature. 
Throughout this section, we run $25$ AMP iterations using the BP denoiser followed by $100$ final BP iterations.
If at any point during the decoding process a valid codeword is obtained, the decoding process is terminated. 

Recall that the BP denoiser is parameterized by $\{N_t : t \in [T]\}$, or the schedule of the number of BP iterations to perform per AMP iteration.
In this section, we will investigate various schedules for the BP denoiser.
To begin our study, we consider the BP-N denoiser, which is defined as the BP denoiser in which $N_t = t + 1$ for every AMP iteration $t = 0, 1, 2, \ldots T-1$. 

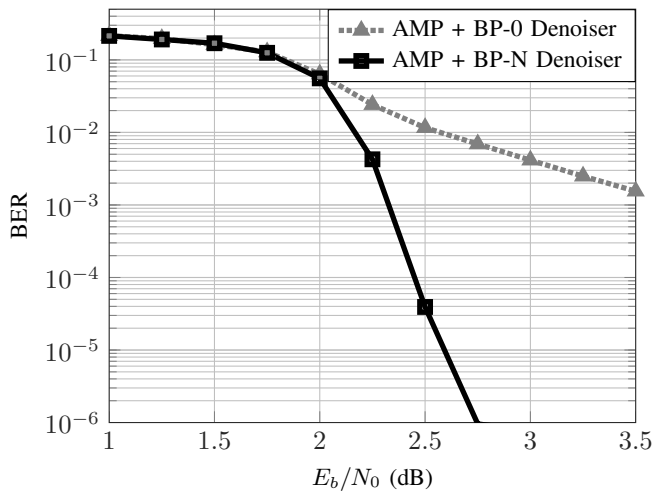
\begin{figure}[t!]
    \centering
    \begin{tikzpicture}

\definecolor{customred}{rgb}{0.63529,0.07843,0.18431} 
\definecolor{customblue}{rgb}{0.00000,0.44706,0.74118} 
\definecolor{customgreen}{rgb}{0.00000,0.49804,0.00000} 

\begin{semilogyaxis}[
    font=\small,
    width=7cm,
    height=5.5cm,
    scale only axis,
    every outer x axis line/.append style={white!15!black},
    every x tick label/.append style={font=\color{white!15!black}},
    xmin=1,
    xmax=3.5,
    xtick = {1.0, 1.5, 2.0, 2.5, 3.0, 3.5},
    xlabel={$E_b/N_0$ (dB)},
    xmajorgrids,
    every outer y axis line/.append style={white!15!black},
    every y tick label/.append style={font=\color{white!15!black}},
    ymin=0.000001,
    ymax=0.5,
    ytick = {0.0000001, 0.000001, 0.00001, 0.0001, 0.001, 0.01, 0.1, 1.0},
    ylabel={BER},
    ymajorgrids,
    yminorgrids,
    legend style={at={(1,1)},anchor=north east, draw=black,fill=white,legend cell align=left}
]

\addplot [
    color=gray,
    densely dotted,
    line width=2.0pt,
    mark size=2.0pt,
    mark=triangle,
    mark options={solid}
]
table[row sep=crcr]{
    1.0 0.215650 \\
    1.25 0.196280 \\
    1.5 0.162240 \\
    1.75 0.128410 \\
    2.0 0.063713 \\
    2.25 0.024155 \\
    2.5 0.011623 \\
    2.75 0.006919 \\
    3.0 0.004145 \\
    3.25 0.002493 \\
    3.5 0.001548 \\
};
\addlegendentry{AMP + BP-0 Denoiser};

\addplot [
    color=black,
    solid,
    line width=2.0pt,
    mark size=2.0pt,
    mark=square,
    mark options={solid}
]
table[row sep=crcr]{
    1.0 0.214110 \\
    1.25 0.191220 \\
    1.5 0.169673 \\
    1.75 0.125142 \\
    2.0 0.055629 \\
    2.25 0.004260 \\
    2.5 0.000039268 \\ 
    2.75 0.00000095788 \\ 
    3.00 0.000000771 \\ 
    3.25 0.0000003634 \\
    3.50 0.00000036005 \\
};
\addlegendentry{AMP + BP-N Denoiser};

\end{semilogyaxis}
\end{tikzpicture}
    \caption{This figure compares the BER performance of AMP with the BP-N denoiser and AMP with the BP-0 denoiser, where no BP iterations are performed per AMP iteration. The AMP-BP decoding process provides a steep BER waterfall region, outperforming the alternate version.}
    \label{fig:bpn_vs_bp0}
\end{figure}

Before comparing the performance of our SR-LDPC code to other error-correcting codes, we first seek to evaluate the performance of AMP with the dynamic denoiser from Definition~\ref{definition:BPDenoiser}. 
To do this, we introduce a second denoiser, which we refer to as the BP-$0$ denoiser, as this denoiser performs zero rounds of BP per AMP iteration, or $N_t = 0~\forall t$. 
Note that, under this denoiser, no information is shared between inner and outer decoders. 
We then compute and compare the BER performance of our randomly-generated SR-LDPC code under AMP + BP-N decoding and AMP + BP-$0$ decoding. 
In both cases, we run $100$ rounds of BP after the AMP-BP process has terminated.  
In essence, this experiment compares the performance of jointly decoding the inner and outer codes vs decoding the inner, then outer codes in disjoint succession.  
Fig.~\ref{fig:bpn_vs_bp0} compares the bit error rate (BER) performance of the SR-LDPC code with the BP-N denosier and the BP-0 denoiser.
Clearly, AMP + BP-N decoding endows the SR-LDPC code with a steep waterfall in BER, a phenomenon not seen in AMP + BP-$0$ decoding. 
We thus conclude that the proposed joint decoder is superior to a disjoint decoder in terms of error performance.

The intuition behind the BP-N denoiser becomes clear when one considers how the effective noise variance $\tau_t^2$ of $\rv^{(t)}$ is ideally decreasing with AMP iteration count $t$, up to a certain point. 
During the first few AMP iterations, the outer factor graph is initialized with very noisy local observations so consequently, BP cannot improve $\rv^{(t)}$ very much, even if many BP iterations are run. 
However, as $t$ increases, $\tau_t^2$ hopefully decreases, so BP is able to meaningfully improve the quality of the state estimates. 
Ideally, $\tau_t^2$ will eventually fall below the BP threshold of the outer LDPC code, at which point BP should be run until BP decoding succeeds.
In practice, conditions are not always ideal; nevertheless, increasing $N_t$ with $t$ seems to make sense.  
Though intuitive, we make no claims that the BP-N strategy is optimal, and we leave the optimal scheduling of BP iterations as an open problem for future work. 
Though we decide $N_t$ based solely on $t$, we note that the optimal $N_t^*$ may also depend on $\tau_t^2$. 

\begin{figure}
    \centering
    \begin{tikzpicture}

\definecolor{customred}{rgb}{0.63529,0.07843,0.18431} 
\definecolor{customblue}{rgb}{0.00000,0.44706,0.74118} 
\definecolor{customgreen}{rgb}{0.00000,0.49804,0.00000} 

\begin{semilogyaxis}[
    font=\small,
    width=7cm,
    height=5.5cm,
    scale only axis,
    every outer x axis line/.append style={white!15!black},
    every x tick label/.append style={font=\color{white!15!black}},
    xmin=1,
    xmax=2.75,
    xtick = {1.0, 1.25, 1.5, 1.75, 2.0, 2.25, 2.5, 2.75},
    xlabel={$E_b/N_0$ (dB)},
    xmajorgrids,
    every outer y axis line/.append style={white!15!black},
    every y tick label/.append style={font=\color{white!15!black}},
    ymin=0.0000008,
    ymax=0.5,
    ytick = {0.0000001, 0.000001, 0.00001, 0.0001, 0.001, 0.01, 0.1, 1.0},
    ylabel={BER},
    ymajorgrids,
    yminorgrids,
    legend style={at={(0,0)},anchor=south west, draw=black,fill=white,legend cell align=left}
]

\addplot [
    color=black,
    solid,
    line width=2.0pt,
    mark size=2.0pt,
    mark=square,
    mark options={solid}
]
table[row sep=crcr]{
    1.0 0.214110 \\
    1.25 0.191220 \\
    1.5 0.169673 \\
    1.75 0.125142 \\
    2.0 0.055629 \\
    2.25 0.004260 \\
    2.5 0.000039268 \\ 
    2.75 0.00000095788 \\ 
    3.00 0.000000771 \\ 
    3.25 0.0000003634 \\
    3.50 0.00000036005 \\
};
\addlegendentry{AMP + BP-N Denoiser};

\addplot [
    color=red,
    dashed,
    line width=2.0pt,
    mark size=3.5pt,
    mark=asterisk,
    mark options={solid}
]
table[row sep=crcr]{
    1.0 0.214850 \\
    1.25 0.194350 \\
    1.5 0.160370 \\
    1.75 0.116340 \\
    2.0 0.053313 \\
    2.25 0.004299 \\
    2.5 0.000020 \\
    2.75 0.000000091711 \\
};
\addlegendentry{AMP + BP-1-KG Denoiser};

\end{semilogyaxis}
\end{tikzpicture}
    \caption{This figure compares the BER performance of the SR-LDPC code when the BP-N and BP-$1$-KG denoisers are used. The BP-$1$-KG denoiser provides a significant reduction in decoding complexity while only minimally affecting the BER performance. }
    \label{fig:bpn_vs_bp1kg}
\end{figure}
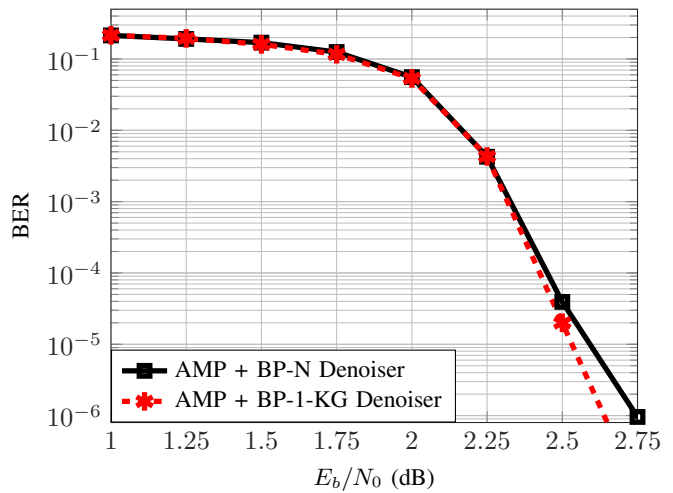

Furthermore, though the BP-N strategy is elegant, it requires $T(T+1)/2 = \mathcal{O}\left(T^2\right)$ total BP iterations, where $T$ is the number of AMP iterations to perform. 
For even moderate $T$, the complexity of this decoding strategy becomes significant. 
As an alternative approach, we propose the BP-$1$-KeepGraph (BP-1-KG) denoiser in which only one BP iteration is run per AMP iteration ($N_t = 1~\forall t$). 
However, instead of completely resetting all graph messages and local observations between AMP iterations, only the local observations are reset and the graph messages from the previous round are left to be incorporated into the current round's message passing.
Thus, this strategy requires only $\mathcal{O}(T)$ BP iterations.
Essentially, this approach uses noisy messages from previous rounds as information from further down the computation tree instead of passing fresh information across the entire computation tree during every AMP iteration. 
Figure~\ref{fig:bpn_vs_bp1kg} compares the performance of the SR-LDPC code using the BP-N and BP-$1$-KG denoisers. 
Demonstrably, there is minimal degradation in BER performance when the BP-$1$-KG denoiser is used at low SNRs and, somewhat surprisingly, a performance boost at high SNRs; thus, the BP-$1$-KG denoiser may be used as a pragmatic means to reduce decoding complexity.   

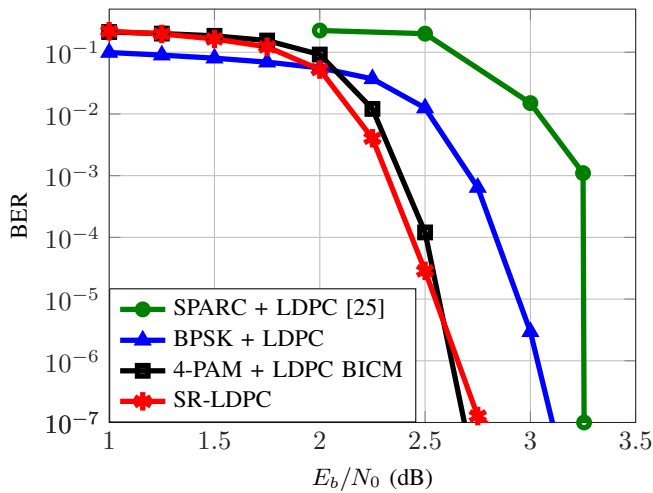
\begin{figure}[t!]
    \centering
    \begin{tikzpicture}

\definecolor{customred}{rgb}{0.63529,0.07843,0.18431} 
\definecolor{customblue}{rgb}{0.00000,0.44706,0.74118} 
\definecolor{customgreen}{rgb}{0.00000,0.49804,0.00000} 

\begin{semilogyaxis}[
    font=\small,
    width=7cm,
    height=5.5cm,
    scale only axis,
    every outer x axis line/.append style={white!15!black},
    every x tick label/.append style={font=\color{white!15!black}},
    xmin=1,
    xmax=3.5,
    xtick = {1.0, 1.5, 2.0, 2.5, 3.0, 3.5},
    xlabel={$E_b/N_0$ (dB)},
    xmajorgrids,
    every outer y axis line/.append style={white!15!black},
    every y tick label/.append style={font=\color{white!15!black}},
    ymin=0.0000001,
    ymax=0.5,
    ytick = { 0.00000001, 0.0000001, 0.0000001, 0.000001, 0.00001, 0.0001, 0.001, 0.01, 0.1, 1.0},
    ylabel={BER},
    ymajorgrids,
    yminorgrids,
    legend style={at={(0,0)},anchor=south west, draw=black,fill=white,legend cell align=left}
]

\addplot [
    color=customgreen,
    solid,
    line width=2.0pt,
    mark size=2.0pt,
    mark=o,
    mark options={solid}
]
table[row sep=crcr]{
    2.0 2.25e-1 \\
    2.5 2e-1 \\
    3.0 1.5e-2 \\
    3.25 1.1e-3 \\
    3.255 0.0000001 \\
    3.5 0 \\
    3.75 0 \\
};
\addlegendentry{SPARC + LDPC \cite{greig2017techniques}};

\addplot [
    color=blue,
    solid,
    line width=2.0pt,
    mark size=2.0pt,
    mark=triangle,
    mark options={solid}
]
table[row sep=crcr]{
    1.0 0.098983 \\
    1.25 0.089801 \\
    1.5 0.080139 \\
    1.75 0.069076 \\
    2.0 0.055831 \\
    2.25 0.037056 \\
    2.5 0.012483 \\
    2.75 0.00064230 \\
    3.0 0.0000029535 \\ 
    3.25 0.0000000009 \\ 
};
\addlegendentry{BPSK + LDPC};

\addplot [
    color=black,
    solid,
    line width=2.0pt,
    mark size=2.0pt,
    mark=square,
    mark options={solid}
]
table[row sep=crcr]{
    1.0 0.21277 \\
    1.25 0.20068 \\
    1.5 0.18443 \\
    1.75 0.15474 \\
    2.0 0.091709 \\
    2.25 0.012070 \\
    2.5 0.00012024 \\
    2.75 0.000000009 \\ 
};
\addlegendentry{4-PAM 
 + LDPC BICM};

\addplot [
    color=red,
    solid,
    line width=2.0pt,
    mark size=3.5pt,
    mark=asterisk,
    mark options={solid}
]
table[row sep=crcr]{
    1.0 0.21865 \\
    1.25 0.19773 \\
    1.5 0.16397 \\
    1.75 0.12211 \\
    2.0 0.052382 \\
    2.25 0.0040360 \\
    2.5 0.000028648 \\ 
    2.75 0.000000127 \\ 
    3.0 0.0000000244565 \\
};
\addlegendentry{SR-LDPC};

\end{semilogyaxis}
\end{tikzpicture}
    \caption{This figure compares the BER performance of an SR-LDPC code, a SPARC + LDPC concatenated code from \cite{greig2017techniques}, BPSK with binary NR LDPC coding, and a $4$-PAM with NR LDPC BICM scheme.}
    \label{fig:ber_vs_nr}
\end{figure}

\begin{figure}
    \centering
    \begin{tikzpicture}

\definecolor{customred}{rgb}{0.63529,0.07843,0.18431} 
\definecolor{customblue}{rgb}{0.00000,0.44706,0.74118} 
\definecolor{customgreen}{rgb}{0.00000,0.49804,0.00000} 

\begin{semilogyaxis}[
    font=\small,
    width=7cm,
    height=5.5cm,
    scale only axis,
    every outer x axis line/.append style={white!15!black},
    every x tick label/.append style={font=\color{white!15!black}},
    xmin=1,
    xmax=3.5,
    xtick = {1.0, 1.5, 2.0, 2.5, 3.0, 3.5},
    xlabel={$E_b/N_0$ (dB)},
    xmajorgrids,
    every outer y axis line/.append style={white!15!black},
    every y tick label/.append style={font=\color{white!15!black}},
    ymin=0.00001,
    ymax=1,
    ytick = {0.000001, 0.00001, 0.0001, 0.001, 0.01, 0.1},
    ylabel={CER},
    ymajorgrids,
    yminorgrids,
    legend style={at={(0,0)},anchor=south west, draw=black,fill=white,legend cell align=left}
]

\addplot [
    color=blue,
    solid,
    line width=2.0pt,
    mark size=1.4pt,
    mark=triangle,
    mark options={solid}
]
table[row sep=crcr]{
    1.0 1 \\
    1.25 1 \\
    1.5 1 \\
    1.75 1 \\
    2.0 1 \\
    2.25 0.9985 \\
    2.5 0.8395 \\
    2.75 0.15083 \\
    3.0 0.0019100 \\
    3.25 0.0000009 \\
};
\addlegendentry{BPSK + LDPC};

\addplot [
    color=black,
    solid,
    line width=2.0pt,
    mark size=2.0pt,
    mark=square,
    mark options={solid}
]
table[row sep=crcr]{
    1.0 1 \\
    1.25 1 \\
    1.5 1 \\
    1.75 0.999 \\
    2.0 0.8825 \\
    2.25 0.208 \\
    2.5 0.00346 \\
    2.75 0.000001 \\ 
};
\addlegendentry{4-PAM + LDPC BICM};

\addplot [
    color=red,
    solid,
    line width=2.0pt,
    mark size=3.5pt,
    mark=asterisk,
    mark options={solid}
]
table[row sep=crcr]{
    1.0 1.000000 \\
    1.25 1.000000 \\
    1.5 1.000000 \\
    1.75 0.9878 \\
    2.0 0.76042 \\
    2.25 0.077666 \\
    2.5 0.00054 \\ 
    2.75 0.000086957 \\
    3.0 0.0000159998 \\
};
\addlegendentry{SR-LDPC};

\end{semilogyaxis}
\end{tikzpicture}
    \vspace{-1.5mm}
    \caption{CER performance comparison of a SR-LDPC code, BPSK with NR LDPC coding, and a $4$-PAM with LDPC BICM scheme.
    The SR-LDPC code outperforms both benchmarks at a CER of $10^{-1}$, which is a common target for systems employing an ARQ outer loop.}
    \label{fig:cer_vs_nr}
\end{figure}
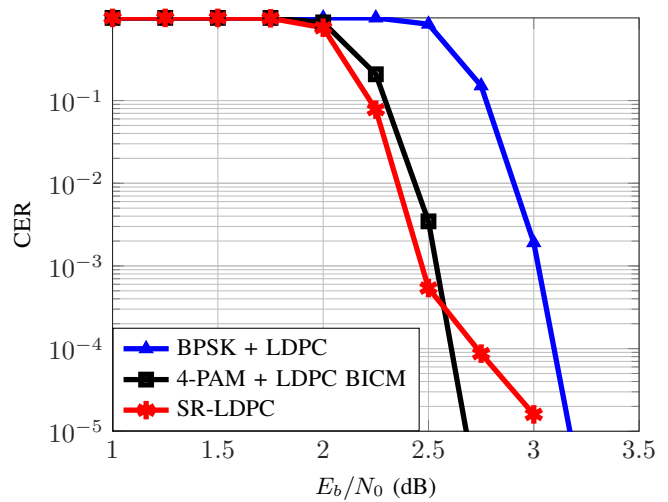

We now seek to compare the performance of the SR-LDPC code using the BP-1-KG denoiser to three pertinent benchmarks: a highly-optimized SPARC/LDPC construction from \cite{greig2017techniques}, a rate $R = 0.8$ NR binary LDPC code with BPSK signalling, and a rate $R = 0.4$ NR LDPC code with bit-interleaved coded modulation (BICM) using the $4$-PAM constellation.
As the scheme in \cite{greig2017techniques} was already described in Section~\ref{section:introduction}, its description will not be replicated here. 
We note that the latter two comparisons must be considered carefully as there exists a gap between the unconstrained and constrained capacities of the AWGN channel.
In the case of BPSK signalling, this gap is significant; however, this gap is negligible under $4$-PAM inputs at our chosen rate. 
A BER comparison of these three schemes is included in Fig.~\ref{fig:ber_vs_nr} and a corresponding CER comparison is included in Fig.~\ref{fig:cer_vs_nr}.
Note that the CER of the scheme from \cite{greig2017techniques} is not included as it was not provided in Greig and Venkataramanan's original paper. 

From these figures, it is clear that this SR-LDPC code provides an improvement of about $1$~dB at a BER of $10^{-3}$ over the SPARC/LDPC concatenated coding structure from \cite{greig2017techniques}. 
Additionally, the SR-LDPC code outperforms the BPSK + LDPC scheme by roughly $0.5$~dB and even outperforms the $4$-PAM with LDPC BICM by about $0.1$~dB at that same BER. 
In terms of CER, the SR-LDPC code provides a $0.5$~dB improvement over the BPSK + LDPC scheme and an improvement of under $0.1$~dB over the $4$-PAM + LDPC with BICM at a CER of $10^{-1}$, which is a common target CER when an ARQ outer loop is employed.
What makes these results even more impressive is the fact that this SR-LDPC code was generated randomly while the code in \cite{greig2017techniques} and the NR LDPC codes are both highly optimized. 
Thus, it is likely that further performance improvements are possible through the careful design of the SR-LDPC code. 

Recall that the entries of the sensing matrix $\Am$ are generated as i.i.d. $\mathcal{N}\left(0, \frac{1}{n}\right)$ random variables. 
As each SR-LDPC codeword is a linear combination of the columns of $\Am$, every channel input is therefore the realization of a $\mathcal{N}\left(0, \frac{L}{n}\right)$ Gaussian random variable. 
It is well-known that the capacity of the AWGN channel is achieved with a Gaussian input distribution, and that the performance of coded modulation schemes may be improved by shaping the constellation to be Gaussian-like~\cite{forney1998modulationcoding}. 
While much work has been done on forcing traditional constellations (e.g., M-QAM) to be Gaussian-like, SR-LDPC coding is a natural strategy to combine Gaussian signalling with traditional codes in a powerful way.  
Thus, we view SR-LDPC coding as a pragmatic strategy for obtaining shaping gains over the AWGN channel. 

\section{Conclusion}
\label{section:conclusion}

This article introduces sparse regression LDPC (SR-LDPC) codes and their decoding. 
SR-LDPC codes are formed by concatenating an inner sparse regression code with an outer non-binary LDPC code whose respective field size and section sizes are equal. 
Such codes can be efficiently decoded using AMP with a dynamic denoiser that runs BP on the factor graph of the outer LDPC code, thus allowing for soft information to be shared between inner and outer decoders.
It is shown that the proposed denoiser satisfies the conditions for state evolution to hold under non-separable pseudo-Lipschitz denoising functions. 
Furthermore, by exploiting the structure of SR-LDPC codes and the proposed decoding algorithm, a computationally efficient approximate state evolution recursion is presented that allows for rapid code optimization and hyperparameter tuning.

Numerical simulation results are presented to demonstrate that the proposed AMP-BP decoder, which jointly decodes inner and outer codes, significantly outperforms a traditional Forney-style decoding algorithm. 
Additionally, an SR-LDPC code is shown to significantly outperform a similar concatenated SPARC/LDPC code construction and outperform NR LDPC + $4$-PAM BICM.

A remarkable fact about the results presented in this article is that they were obtained for a randomly-generated SR-LDPC code. 
Thus, it is likely that further performance improvements may be obtained through proper optimization of the code structure. 
For example, it is known that the performance of uncoded SPARCs can be significantly improved via a nonuniform power allocation; yet, in this article, we employ a naive uniform power allocation. 
Thus, the optimal power allocation for SR-LDPC codes remains a promising open problem. 
Other open problems include the optimal design of the outer LDPC code and the optimal number of BP iterations to perform per AMP iteration. 

\bibliographystyle{IEEEbib}
\bibliography{IEEEabrv,journal}

\appendices

\section{Properties of BP Denoiser}
\label{appendix:properties_bp_denosier}

In this appendix, we prove that the BP-N denoiser is Lipschitz continuous under certain assumptions (Theorem~\ref{theorem:bp_denoiser_lipschitz}) and we derive the Onsager correction term associated with the BP denoiser (Proposition~\ref{proposition:onsager_correction_term}).

\subsection{Proof of Theorem~\ref{theorem:bp_denoiser_lipschitz}}

One of the conditions for the state evolution to hold for non-separable functions is that the denoiser must be pseudo-Lipschitz of a certain order~\cite{berthier2020state}.
For the problem at hand, the stronger Lipschitz condition is shown, which is sufficient.
Thus, the main objective of this section is to demonstrate that the denoiser introduced in Definition~\ref{definition:BPDenoiser} is Lipschitz continuous under Condition~\ref{condition:Sub-Girth-BP}.
To achieve this goal, our strategy is to demonstrate that the magnitudes of the entries in the Jacobian matrix of $\etav(\rv)$ with respect to $\rv$ are uniformly bounded.

\begin{figure}[t]
  \centering
  \begin{tikzpicture}
  [
  font=\small, line width=1pt, >=stealth', draw=black,
  check/.style={rectangle, minimum height=6mm, minimum width=6mm, draw=black, fill=gray!20},
  smallcheck/.style={rectangle, minimum height=3mm, minimum width=3mm, draw=black, fill=gray!20},
  trivialcheck/.style={rectangle, minimum height=3mm, minimum width=3mm, draw=black, fill=gray},
  section/.style={circle, minimum size=7mm, draw=black},
  emptysection/.style={circle, minimum size=5mm, draw=black}
  ]

\tikzset{
  connect/.pic={
    \draw[densely dotted, line width=0.75pt] (0,0) to (0,-0.3125);
    \draw[densely dotted, line width=0.75pt] (0,0) to (-0.1875,-0.25);
    \draw[densely dotted, line width=0.75pt] (0,0) to (0.1875,-0.25);
  }
}

\node[section] (v0) at (0.5,0.25) {$v_{\ell}$};
\node[trivialcheck] (t0) at (2,0.25) {}
  edge (v0);
\node[check] (c0a) at (-1,-1.25) {$c_p$};
\draw[->] (c0a) -- node[above,rotate=45]{$\boldsymbol{\mu}_{c_p \to v_{\ell}}$} (v0);
\node[smallcheck] (c0b) at (0.5,-0.75) {}
  edge[-] (v0);
\pic at (c0b.south) {connect};
\node[smallcheck] (c0c) at (1.5,-0.75) {}
  edge[-] (v0);
\pic at (c0c.south) {connect};

\node[emptysection] (v1a) at (-2,-2.25) {}
  edge[-] (c0a);
\pic at (v1a.south) {connect};
\node[emptysection] (v1b) at (-1,-2.25) {}
  edge[-] (c0a);
\pic at (v1b.south) {connect};
\node[section] (v1c) at (0,-2.5) {$v_k$}
  edge[-] (c0a);
\node[trivialcheck] (t1c) at (2,-2.5) {};
\draw[->] (t1c) -- node[above]{$\boldsymbol{\mu}_{c_0 \to v_k}$} (v1c);

\node[smallcheck] (c1a) at (-1,-3.5) {}
  edge[-] (v1c);
\pic at (c1a.south) {connect};
\node[smallcheck] (c1b) at (0,-3.5) {}
  edge[-] (v1c);
\pic at (c1b.south) {connect};
\node[check] (c1c) at (1,-3.75) {$c_{\rho}$}
  edge[-] (v1c);

\node[section] (v2a) at (-0.5,-5.25) {$v_j$};
\draw[->] (v2a) -- node[above,rotate=45]{$\boldsymbol{\mu}_{v_j \to c_{\rho}}$} (c1c);
\node[emptysection] (v2b) at (1,-4.75) {}
  edge[-] (c1c);
\node[emptysection] (v2c) at (2,-4.75) {}
  edge[-] (c1c);
\node[trivialcheck] (t1c) at (-2.5,-5.25) {};
\draw[->] (t1c) -- node[above]{$\boldsymbol{\mu}_{c_0 \to v_j}$} (v2a);

\end{tikzpicture}
  \caption{Computation tree for variable node $v_{\ell}$ obtained by taking the factor graph of the outer LDPC code, setting $v_{\ell}$ as the root, and retaining only the nodes involved in the computation of $\hat{\sv}_{\ell}\left(\rv, g\right)$. We use this data structure to compute the derivatives in \eqref{equation:JacobianMatrixEntries}. }
  \label{figure:computation_tree_illustration}
\end{figure}
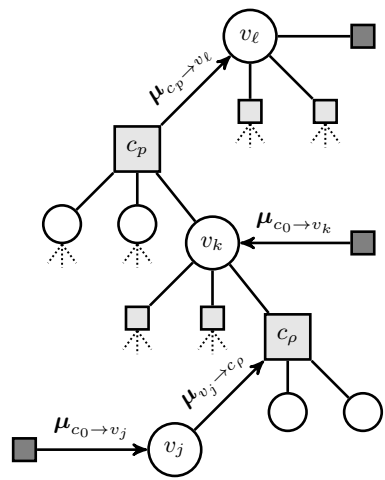

Recall that the denoiser assumes a sectional form, as described in Definition~\ref{definition:BPDenoiser}.
The vector estimate for section~$\ell$ becomes
\begin{equation*}
\hat{\sv}_{\ell} (\rv) = \sum_{g \in \mathbb{F}_q} \Pr \left( V_{\ell} = g \middle| \Rv_{\mathrm{tree}} = \rv_{\mathrm{tree}} \right) \ev_g ,
\end{equation*}
where $\Rv_{\mathrm{tree}}$ denotes the measurements associated with the computational tree of the LDPC code rooted at section~$\ell$.
The (realized) scaling factors found in \eqref{equation:BP-Estimate} are given by
\begin{equation}
\begin{split}
\hat{\sv}_{\ell} (\rv)
&= \frac{ \operatorname*{\bigcirc}_{c_p \in N_0(v_{\ell})} \boldsymbol{\mu}_{c_{p} \to v_{\ell}} }
{ \left\| \operatorname*{\bigcirc}_{c_p \in N_0(v_{\ell})} \boldsymbol{\mu}_{c_{p} \to v_{\ell}} \right\|_1 } .
\end{split}
\end{equation}
where $N_0(v_{\ell})$ denotes the neighborhood of $v_{\ell}$ including the local observation and $\boldsymbol{\mu}_{c_0 \to v_{\ell}} = \boldsymbol{\alpha}_{\ell}$.
We are ultimately interested in Jacobian entries of the form
\begin{equation} \label{equation:JacobianMatrixEntries}
\begin{split}
&\frac{\partial \hat{\sv}_{\ell} \left( \rv, g \right)}{\partial \rv_j(h)}
= \frac{\partial }{\partial \rv_j(h)}
\frac{ \prod_{c_p \in N_0(v_{\ell})} \boldsymbol{\mu}_{c_{p} \to v_{\ell}} (g) }
{ \left\| \operatorname*{\bigcirc}_{c_p \in N_0(v_{\ell})} \boldsymbol{\mu}_{c_{p} \to v_{\ell}} \right\|_1 } \\
&= \frac{\partial }{\partial \rv_j(h)}
\frac{ \boldsymbol{\alpha}_{\ell}(g) \prod_{c_p \in N(v_{\ell})} \boldsymbol{\mu}_{c_{p} \to v_{\ell}} (g) }
{ \sum_{k \in \mathbb{F}_q} \boldsymbol{\alpha}_{\ell}(k) \prod_{c_p \in N(v_{\ell})} \boldsymbol{\mu}_{c_{p} \to v_{\ell}} (k) }
\end{split}
\end{equation}
for $g, h \in \mathbb{F}_q$ and $\ell, j \in [L]$.
We adopt a divide-and-conquer approach to identify and bound these derivatives.
Specifically, we focus on the rooted tree obtained by taking the factor graph of the outer LDPC code, setting $v_{\ell}$ as the root, and retaining only the nodes involved in the computation of $\hat{\sv}_{\ell} ( \rv, g )$. 
Under Condition~\ref{condition:Sub-Girth-BP}, this sub-graph must form a proper tree; Fig.~\ref{figure:computation_tree_illustration} offers a notional diagram to illustrate the outcome of this process.

We seek to bound the magnitude of the derivatives in \eqref{equation:JacobianMatrixEntries} based on the distance between $v_{\ell}$ and $v_j$ in this rooted tree.
We begin with local observations.

\begin{proposition}[Local Observations] \label{proposition:DerivativeAlphaEll}
The partial derivatives of $\boldsymbol{\alpha}_j$ with respect to $\rv_k(h)$ are given by
\begin{equation}
\label{equation:DerivativeAlpha}
\frac{\partial \boldsymbol{\alpha}_j}{\partial \rv_k(h)} = 
\begin{cases}
\frac{1}{\tau^2} \boldsymbol{\alpha}_j (h) \left( \ev_h - \boldsymbol{\alpha}_j \right) & j = k \\
0 & j \neq k
\end{cases}
\end{equation}
for $h \in \mathbb{F}_{q}$ and where $\tau > 0$ is the standard deviation of the effective observation.
\end{proposition}
\begin{IEEEproof}
As defined in \eqref{equation:ConditionalProbability}, the vector $\boldsymbol{\alpha}_j$ is given by
\begin{equation*}
\boldsymbol{\alpha}_j (g)
= \frac{ e^{\frac{ \rv_j(g) }{\tau^2}} }{\sum_{k \in \mathbb{F}_q} e^{\frac{ \rv_j(k) }{\tau^2}}}
\qquad \forall g \in \mathbb{F}_q .
\end{equation*}
When $j \neq k$, it immediately follows that $\partial\boldsymbol{\alpha}_j/\partial\rv_k(h) = 0$ as $\boldsymbol{\alpha}_j$ does not depend on $\rv_k(h)$. 
We thus consider the case where $k = j$. 
When $g = h$, we have that
\begin{equation*}
\begin{split}
\frac{\partial \boldsymbol{\alpha}_j (h)}{\partial \rv_j(h)}
&= \frac{1}{\tau^2} \frac{ e^{\frac{ \rv_j(h) }{\tau^2}} }{\sum_{k \in \mathbb{F}_q} e^{\frac{ \rv_j(k) }{\tau^2}}}
- \frac{1}{\tau^2} \frac{ e^{\frac{ \rv_j(h) }{\tau^2}} e^{\frac{ \rv_j(h) }{\tau^2}} }{\left( \sum_{k \in \mathbb{F}_q} e^{\frac{ \rv_j(k) }{\tau^2}} \right)^2} \\
&= \frac{1}{\tau^2} \boldsymbol{\alpha}_j (h)
\left( 1 - \boldsymbol{\alpha}_j (h) \right) .
\end{split}
\end{equation*}
When $g \neq h$, we get
\begin{equation*}
\begin{split}
\frac{\partial \boldsymbol{\alpha}_j (g)}{\partial \rv_j(h)}
&= - \frac{1}{\tau^2} \frac{ e^{\frac{ \rv_j(g) }{\tau^2}} e^{\frac{ \rv_j(h) }{\tau^2}} }{\left( \sum_{\kappa \in \mathbb{F}_q} e^{\frac{ \rv_j(\kappa) }{\tau^2}} \right)^2}
= - \frac{1}{\tau^2} \boldsymbol{\alpha}_j (g) \boldsymbol{\alpha}_j (h) .
\end{split}
\end{equation*}
Collecting these findings and condensing them into a more compact form, we arrive at \eqref{equation:DerivativeAlpha}, which is the desired expression.
\end{IEEEproof}

\begin{corollary} \label{corollary:bounded_alpha_ell_derivatives}
The absolute value of the partial derivatives of $\boldsymbol{\alpha}_{j}$ with respect to $\rv_{k}(h)$ are bounded by
\begin{equation}
    \left|\frac{\partial\boldsymbol{\alpha}_{j}}{\partial\rv_{k}(h)}\right| \leq \frac{1}{4\tau^2}
\end{equation}
where $\tau > 0$ is the standard deviation of the effective observation. 
\end{corollary}

The proof of this corollary is trivial when $\boldsymbol{\alpha}_{j}$ is a valid probability vector, as is the case in this article. 
We also note that, based on the state evolution of AMP, $\tau^2 \geq \sigma^2$ at every iteration irrespective of the iteration number.
We can therefore establish a uniform bound across iterations.
We are now ready to show that the absolute value of \eqref{equation:JacobianMatrixEntries} is bounded whenever $\ell = j$, i.e., at the root level of the computation tree.

\begin{proposition}[Root Derivatives] \label{lemma:DerivativeEllBound}
The partial derivatives of $\hat{\sv}_{\ell} \left(\rv, g \right)$ with respect to $\rv_{\ell} (h)$ are given by
\begin{equation} \label{equation:DerivativeSell}
\frac{\partial \hat{\sv}_{\ell} \left( \rv \right)}{\partial \rv_{\ell} (h)}
= \frac{1}{\tau^2} \hat{\sv}_{\ell} \left( \rv, h \right) \left( \ev_h - \hat{\sv}_{\ell} \left( \rv \right) \right)
\qquad \forall h \in \mathbb{F}_{q}
\end{equation}
where $\tau > 0$ is the standard deviation of the effective observation.
\end{proposition}
\begin{IEEEproof}
Leveraging Proposition~\ref{proposition:DerivativeAlphaEll} and denoting the standard inner product by $\langle\cdot, \cdot\rangle$, we have
\begin{equation*}
\begin{split}
\frac{\partial \hat{\sv}_{\ell} \left( \rv \right)}{\partial \rv_{\ell} (h)}
&= \frac{\partial}{\partial \rv_{\ell} (h)} \frac{ \boldsymbol{\alpha}_{\ell} \circ \operatorname*{\bigcirc}_{c_p \in N(v_{\ell})} \boldsymbol{\mu}_{c_p \to v_{\ell}} }
{ \left\| \boldsymbol{\alpha}_{\ell} \circ \operatorname*{\bigcirc}_{c_p \in N(v_{\ell})} \boldsymbol{\mu}_{c_p \to v_{\ell}} \right\|_1 } \\
&= \frac{ \frac{\partial \boldsymbol{\alpha}_{\ell}}{\partial \rv_{\ell} (h)}
\circ \left( \operatorname*{\bigcirc}_{c_p \in \mathrm{N}(v_{\ell})} \boldsymbol{\mu}_{c_p \to v_{\ell}} \right) }
{ \left\| \boldsymbol{\alpha}_{\ell} \circ \left( \operatorname*{\bigcirc}_{c_p \in \mathrm{N}(v_{\ell})} \boldsymbol{\mu}_{c_p \to v_{\ell}} \right) \right\|_1 } \\
&- \hat{\sv}_{\ell} \left( \rv \right) \frac{ \left\langle \frac{\partial \boldsymbol{\alpha}_{\ell}}{\partial \rv_{\ell} (h)},
\operatorname*{\bigcirc}_{c_p \in \mathrm{N}(v_{\ell})} \boldsymbol{\mu}_{c_p \to v_{\ell}} \right\rangle }
{ \left\| \boldsymbol{\alpha}_{\ell} \circ \left( \operatorname*{\bigcirc}_{c_p \in \mathrm{N}(v_{\ell})} \boldsymbol{\mu}_{c_p \to v_{\ell}} \right) \right\|_1 } \\
&= \frac{\boldsymbol{\alpha}_{\ell} (h)}{\tau^2} \frac{ \left( \ev_h - \boldsymbol{\alpha}_{\ell} \right)
\circ \left( \operatorname*{\bigcirc}_{c_p \in \mathrm{N}(v_{\ell})} \boldsymbol{\mu}_{c_p \to v_{\ell}} \right) }
{ \left\| \boldsymbol{\alpha}_{\ell} \circ \left( \operatorname*{\bigcirc}_{c_p \in \mathrm{N}(v_{\ell})} \boldsymbol{\mu}_{c_p \to v_{\ell}} \right) \right\|_1 } \\
&- \frac{\boldsymbol{\alpha}_{\ell} (h)}{\tau^2} \hat{\sv}_{\ell} \left( \rv \right) \frac{ \left\langle \ev_h - \boldsymbol{\alpha}_{\ell},
\operatorname*{\bigcirc}_{c_p \in \mathrm{N}(v_{\ell})} \boldsymbol{\mu}_{c_p \to v_{\ell}} \right\rangle }
{ \left\| \boldsymbol{\alpha}_{\ell} \circ \left( \operatorname*{\bigcirc}_{c_p \in \mathrm{N}(v_{\ell})} \boldsymbol{\mu}_{c_p \to v_{\ell}} \right) \right\|_1 } \\
&= \frac{\boldsymbol{\alpha}_{\ell} (h)}{\tau^2}
\frac{ \ev_h \circ \left( \operatorname*{\bigcirc}_{c_p \in \mathrm{N}(v_{\ell})} \boldsymbol{\mu}_{c_p \to v_{\ell}} \right) }
{ \left\| \boldsymbol{\alpha}_{\ell} \circ \left( \operatorname*{\bigcirc}_{c_p \in \mathrm{N}(v_{\ell})} \boldsymbol{\mu}_{c_p \to v_{\ell}} \right) \right\|_1 } \\
&- \frac{\boldsymbol{\alpha}_{\ell} (h)}{\tau^2} \hat{\sv}_{\ell} \left( \rv \right)
\frac{ \left\langle \ev_h, \operatorname*{\bigcirc}_{c_p \in \mathrm{N}(v_{\ell})} \boldsymbol{\mu}_{c_p \to v_{\ell}} \right\rangle }
{ \left\| \boldsymbol{\alpha}_{\ell} \circ \left( \operatorname*{\bigcirc}_{c_p \in \mathrm{N}(v_{\ell})} \boldsymbol{\mu}_{c_p \to v_{\ell}} \right) \right\|_1 } \\
&= \frac{1}{\tau^2} \hat{\sv}_{\ell} \left( \rv, h \right) \left( \ev_h - \hat{\sv}_{\ell} \left( \rv \right) \right),
\end{split}
\end{equation*}
which is the desired expression. 
\end{IEEEproof}
\begin{corollary} \label{corollary:bounded_root_derivatives}
The absolute value of the partial derivatives of $\hat{\sv}_{\ell}\left(\rv\right)$ with respect to $\rv_{\ell}(h)$ are bounded by
\begin{equation}
    \left|\frac{\partial\hat{\sv}_{\ell}\left(\rv\right)}{\partial\rv_{\ell}(h)}\right| \leq \frac{1}{4\tau^2}
\end{equation}
\end{corollary}
The proof of this corollary follows that of Corollary \ref{corollary:bounded_alpha_ell_derivatives} because, like $\boldsymbol{\alpha}_{j}$, $\hat{\sv}_{\ell}\left(\rv\right)$ forms a valid probability vector. 

Proposition~\ref{lemma:DerivativeEllBound} offers a blueprint for the general result we wish to establish.
Yet, the situation gets more complicated when $\ell \neq j$ because we have to involve the message passing rules.
In doing so, we obtain a key intermediate result using mathematical induction.
We start with the variable node closest to the root node, and then progress outward step by step.

To circumvent a notational nightmare, we restrict the proof to cases where all edge weights are equal to $1 \in \mathbb{F}_{q}$.
Conceptually, the edge can be interpreted as permutations on the belief vectors.
From the point of view of bounding partial derivatives, this is a benign operation, yet the accounting that comes with permutations is dreadful, hence our focus on the simpler case.
Moving forward, we assume the following condition.

\begin{condition} \label{condition:Edge1}
All edge weights within the factor graph of the LDPC outer code are equal to $1 \in \mathbb{F}_{q}$.
\end{condition}

The extension of the following propositions to the case with arbitrary edge weights (i.e., beyond Condition~\ref{condition:Edge1}) is conceptually straightforward.

\begin{proposition} \label{proposition:DerivativeJGammaBound}
Suppose Condition~\ref{condition:Sub-Girth-BP} holds and let $v_j$ be a descendant of root node $v_{\ell}$ in the computation tree.
Moreover, let $c_p \in N(v_{\ell})$ be the unique check neighbor of $v_{\ell}$ on the path from $v_{\ell}$ to $v_j$ within the tree.
Then, there exists vector $\boldsymbol{\nu}$, with $\zerov \preceq \boldsymbol{\nu} \preceq \boldsymbol{\mu}_{c_p \to v_{\ell}}$, such that the partial derivative of $\boldsymbol{\mu}_{c_p \to v_{\ell}}$ with respect to $\rv_j (h)$ is given by
\begin{equation} \label{equation:DerivativeMuJ}
\frac{\partial \boldsymbol{\mu}_{c_p \to v_{\ell}}}{\partial \rv_j (h)}
= \frac{1}{\tau^2} \left( \boldsymbol{\nu} - \left\| \boldsymbol{\nu} \right\|_1
\boldsymbol{\mu}_{c_p \to v_{\ell}} \right),
\end{equation}
where $\tau > 0$ is the standard deviation of the effective observation. 
Here, $\preceq$ denotes elementwise comparison of the entries in the vector. 
\end{proposition}
\begin{IEEEproof}
Under Condition~\ref{condition:Sub-Girth-BP}, we know that $v_j$ appears at most once within the computation tree rooted at $v_{\ell}$. 
Thus, we establish \eqref{equation:DerivativeMuJ} via mathematical induction on the distance between $v_{\ell}$ and its descendant $v_j$ on the computation tree.
The distance that we are interested in only considers the number of variable nodes between $v_{\ell}$ and $v_j$. 
Before beginning, we point out that if $v_j$ is not a descendant of $v_{\ell}$, then the corresponding partial derivatives vanish and the claim is immediate.

We begin with generic results that are useful for both the base case and the inductive step.
Let $v_j$ be a descendant of $v_{\ell}$ and let $v_k$ be the variable child of $v_{\ell}$ on the path from $v_{\ell}$ to $v_j$.
Let $c_p$ be the unique check node in $N(v_{\ell}) \cap N(v_k)$ and let $c_{\varrho}$ be the unique check node child of $v_k$ on the path from $v_k$ to $v_j$; if $v_k = v_j$, let $\varrho = 0$. 
Then, we have that
\begin{equation} \label{equation:DerivativeMukp}
\begin{split}
&\frac{\partial \boldsymbol{\mu}_{v_k \to c_p}}{\partial \rv_j (h)}
= \frac{\partial}{\partial \rv_j (h)}
\frac{ \operatorname*{\bigcirc}_{c_{\xi} \in N_0(v_k) \setminus c_{p}} \boldsymbol{\mu}_{c_{\xi} \to v_k} }
{ \left\| \operatorname*{\bigcirc}_{c_{\xi} \in N_0(v_k) \setminus c_{p}}  \boldsymbol{\mu}_{c_{\xi} \to v_k} \right\|_1 } .
\end{split}
\end{equation}
We can also examine the partial derivatives of the probability vector $\boldsymbol{\mu}_{c_{\varrho} \to v_k}$.
Suppose $v_k \neq v_j$ and let $v_o$ be the unique variable child of $v_k$ on the path from $v_{\ell}$ to $v_j$.
Using the $\mathbb{F}_q$ convolution, we have
\begin{equation} \label{equation:DerivativeMupl}
\begin{split}
\frac{\partial \boldsymbol{\mu}_{c_{\varrho} \to v_k}}{\partial \rv_j (h)}
&= \frac{\partial}{\partial \rv_j (h)} \left( \bigodot_{v_l \in N(c_{\varrho}) \setminus v_k} \mkern-18mu \boldsymbol{\mu}_{v_l \to c_{\varrho}} \right) \\
&= \frac{\partial \boldsymbol{\mu}_{v_o \to c_{\varrho}}}{\partial \rv_j (h)} \odot
\left( \bigodot_{v_l \in N(c_{\varrho}) \setminus \{ v_k, v_o \}} \mkern-36mu \boldsymbol{\mu}_{v_l \to c_{\varrho}} \right) \\
&= \frac{\partial \boldsymbol{\mu}_{v_o \to c_{\varrho}}}{\partial \rv_j (h)} \odot
\boldsymbol{\nu}_{c_{\varrho} \setminus v_k, v_o} .
\end{split}
\end{equation}
We emphasize that $\boldsymbol{\nu}_{c_{\varrho} \setminus v_k, v_o}$, as defined implicitly above, is a probability distribution.

Having established these results, we turn our attention to the base case where $v_j$ is a variable child of $v_{\ell}$ ($v_k = v_j$, $\varrho = 0$). 
Applying \eqref{equation:DerivativeMukp} and Proposition~\ref{proposition:DerivativeAlphaEll}, we obtain
\begin{equation}
\begin{split}
\frac{\partial \boldsymbol{\mu}_{v_j \to c_p}}{\partial \rv_j (h)}
&= \frac{\partial}{\partial \rv_j (h)}
\frac{ \boldsymbol{\alpha}_{j}\circ\left(\operatorname*{\bigcirc}_{c_{\xi} \in N(v_j) \setminus c_{p}} \boldsymbol{\mu}_{c_{\xi} \to v_j} \right)}
{ \left\| \boldsymbol{\alpha}_{j}\circ\left(\operatorname*{\bigcirc}_{c_{\xi} \in N(v_j) \setminus c_{p}}  \boldsymbol{\mu}_{c_{\xi} \to v_j}\right) \right\|_1 } \\
&= \frac{ \frac{\partial \boldsymbol{\alpha}_j}{\partial \rv_j (h)} \circ
\left( \operatorname*{\bigcirc}_{c_{\xi} \in N(v_j) \setminus c_p} \boldsymbol{\mu}_{c_{\xi} \to v_j} \right) }
{ \left\langle \boldsymbol{\alpha}_j, \operatorname*{\bigcirc}_{c_{\xi} \in N(v_j) \setminus c_p} \boldsymbol{\mu}_{c_{\xi} \to v_j} \right\rangle } \\
&- \boldsymbol{\mu}_{v_j \to c_p} \frac{ \left\langle \frac{\partial \boldsymbol{\alpha}_j}{\partial \rv_j (h)},
\operatorname*{\bigcirc}_{c_{\xi} \in N(v_j) \setminus c_p} \boldsymbol{\mu}_{c_{\xi} \to v_j} \right\rangle }
{ \left\langle \boldsymbol{\alpha}_j, \operatorname*{\bigcirc}_{c_{\xi} \in N(v_j) \setminus c_p} \boldsymbol{\mu}_{c_{\xi} \to v_j} \right\rangle } \\
&= \frac{1}{\tau^2} \frac{ \boldsymbol{\alpha}_j (h) \ev_h \circ
\left( \operatorname*{\bigcirc}_{c_{\xi} \in N(v_j) \setminus c_p} \boldsymbol{\mu}_{c_{\xi} \to v_j} \right) }
{ \left\langle \boldsymbol{\alpha}_j, \operatorname*{\bigcirc}_{c_{\xi} \in N(v_j) \setminus c_p} \boldsymbol{\mu}_{c_{\xi} \to v_j} \right\rangle } \\
&- \frac{1}{\tau^2} \boldsymbol{\mu}_{v_j \to c_p}
\frac{ \left\langle \boldsymbol{\alpha}_j (h) \ev_h ,
\operatorname*{\bigcirc}_{c_{\xi} \in N(v_j) \setminus c_p} \boldsymbol{\mu}_{c_{\xi} \to v_j} \right\rangle }
{ \left\langle \boldsymbol{\alpha}_j, \operatorname*{\bigcirc}_{c_{\xi} \in N(v_j) \setminus c_p} \boldsymbol{\mu}_{c_{\xi} \to v_j} \right\rangle } \\
&= \frac{1}{\tau^2} \left( \boldsymbol{\nu}_{v_j \to c_p}
- \left\| \boldsymbol{\nu}_{v_j \to c_p} \right\|_1 \boldsymbol{\mu}_{v_j \to c_p} \right),
\end{split}
\end{equation}
where
\begin{equation*}
\boldsymbol{\nu}_{v_j \to c_p} = \frac{ \boldsymbol{\alpha}_j (h) \ev_h \circ
\left( \operatorname*{\bigcirc}_{c_{\xi} \in N(v_j) \setminus c_p} \boldsymbol{\mu}_{c_{\xi} \to v_j} \right) }
{ \left\langle \boldsymbol{\alpha}_j, \operatorname*{\bigcirc}_{c_{\xi} \in N(v_j) \setminus c_p} \boldsymbol{\mu}_{c_{\xi} \to v_j} \right\rangle } .
\end{equation*}
By construction, we have $\zerov \preceq \boldsymbol{\nu}_{v_j \to c_p} \preceq \boldsymbol{\mu}_{v_j \to c_p}$.
We turn to the second graph operation and apply \eqref{equation:DerivativeMupl}, which yields
\begin{equation} \label{equation:DerivativeMu0jh}
\begin{split}
&\frac{\partial \boldsymbol{\mu}_{c_p \to v_{\ell}}}{\partial \rv_j(h)}
= \frac{\partial \boldsymbol{\mu}_{v_j \to c_p}}{\partial \rv_j (h)}
\odot \boldsymbol{\nu}_{c_p \setminus v_{\ell}, v_j} \\
&= \frac{1}{\tau^2} \left( \boldsymbol{\nu}_{v_j \to c_p} - \left\| \boldsymbol{\nu}_{v_j \to c_p} \right\|_1 \boldsymbol{\mu}_{v_j \to c_p} \right) 
\odot \boldsymbol{\nu}_{c_p \setminus v_{\ell}, v_j} \\
&= \frac{1}{\tau^2} \left( \boldsymbol{\nu}_{v_j \to c_p} \odot \boldsymbol{\nu}_{c_p \setminus v_{\ell}, v_j}
- \left\| \boldsymbol{\nu}_{v_j \to c_p} \right\|_1 \boldsymbol{\mu}_{c_p \to v_{\ell}} \right) .
\end{split}
\end{equation}
Thus, in this case, we take $\boldsymbol{\nu} = \boldsymbol{\nu}_{v_j \to c_p} \odot \boldsymbol{\nu}_{c_p \setminus v_{\ell}, v_j}$ as a suitable vector.
Based on the fact that $\boldsymbol{\nu}_{c_p \setminus v_{\ell}, v_j}$ is a probability vector, together with the aforementioned component-wise ordering, we gather that
\begin{equation*}
\begin{split}
\zerov &\preceq \boldsymbol{\nu} = \boldsymbol{\nu}_{v_j \to c_p} \odot \boldsymbol{\nu}_{c_p \setminus v_{\ell}, v_j} \\
&\preceq \boldsymbol{\mu}_{v_j \to c_p} \odot \boldsymbol{\nu}_{c_p \setminus v_{\ell}, v_j}
= \boldsymbol{\mu}_{c_p \to v_{\ell}} .
\end{split}
\end{equation*}
Moreover, leveraging the properties of the $\mathbb{F}_q$ convolution for non-negative vectors, we get
\begin{equation*}
\left\| \boldsymbol{\nu} \right\|_1
= \left\| \boldsymbol{\nu}_{v_j \to c_p} \right\|_1 \left\| \boldsymbol{\nu}_{c_p \setminus v_{\ell}, v_j} \right\|_1
= \left\| \boldsymbol{\nu}_{v_j \to c_p} \right\|_1 .
\end{equation*}
Thus, for this choice of $\boldsymbol{\nu}$, we arrive at
\begin{equation} \label{equation:DerivativeMu0jhNu}
\frac{\partial \boldsymbol{\mu}_{c_p \to v_{\ell}}}{\partial \rv_j(h)}
= \frac{1}{\tau^2} \left( \boldsymbol{\nu} - \left\| \boldsymbol{\nu} \right\|_1
\boldsymbol{\mu}_{c_p \to v_{\ell}} \right) ,
\end{equation}
as claimed.
That is, the base case conforms to the structure of Proposition~\ref{proposition:DerivativeJGammaBound}.

We now consider the inductive step in our proof. 
As our hypothesis, we assume that \eqref{equation:DerivativeMuJ} holds for all computation trees wherein the distance between the root node and $v_j$ is less than or equal to $\gamma \in \mathbb{N}$.
Consider a rooted computation tree and suppose the distance between $v_{\ell}$ and its descendant $v_j$ in the tree is exactly $\gamma + 1$.
Under Condition~\ref{condition:Sub-Girth-BP}, there is a unique path from $v_{\ell}$ to node $v_j$.
Let $v_k$ be the variable child of $v_{\ell}$ that is also an ascendant of $v_j$, and denote the unique check node that connects the two by $c_p \in N(v_{\ell}) \cap N(v_k)$.
Furthermore, let $c_{\varrho} \in N(v_k)$ be the unique check node within this neighborhood that is an ascendant of $v_j$ on the computation tree.
Finally, let $v_o \in N(c_{\varrho})$ be the unique variable child of $v_k$ that is also an ascendant of $v_j$ (or, perhaps, $v_j$ itself).

The sub-tree starting at $v_k$ can be viewed as a rooted tree containing $v_j$; the graph distance between these two variable nodes within the sub-tree is exactly $\gamma$.
As such, our inductive hypothesis applies.
That is, there exists vector $\boldsymbol{\nu}$ such that $\zerov \preceq \boldsymbol{\nu} \preceq \boldsymbol{\mu}_{c_{\varrho} \to v_k}$ where the partial derivative of $\boldsymbol{\mu}_{c_{\varrho} \to v_k}$ with respect to $\rv_j (h)$ is equal to
\begin{equation}
\frac{\partial \boldsymbol{\mu}_{c_{\varrho} \to v_k}}{\partial \rv_j (h)}
= \frac{1}{\tau^2} \left( \boldsymbol{\nu} - \left\| \boldsymbol{\nu} \right\|_1
\boldsymbol{\mu}_{c_{\varrho} \to v_k} \right) .
\end{equation}

Applying \eqref{equation:DerivativeMukp} and our inductive hypothesis, we have that
\begin{equation} \label{equation:paritalMukp}
\begin{split}
\frac{\partial \boldsymbol{\mu}_{v_k \to c_p}}{\partial \rv_j (h)}
&= \frac{\partial}{\partial \rv_j (h)}
\frac{\operatorname*{\bigcirc}_{c_{\xi} \in N_0(v_k) \setminus c_{p}} \boldsymbol{\mu}_{c_{\xi} \to v_k} }
{ \left\| \operatorname*{\bigcirc}_{c_{\xi} \in N_0(v_k) \setminus c_{p}}  \boldsymbol{\mu}_{c_{\xi} \to v_k} \right\|_1 } \\
&= \frac{ \frac{\partial \boldsymbol{\mu}_{c_{\varrho} \to v_k}}{\partial \rv_j (h)} \circ
\left( \operatorname*{\bigcirc}_{c_{\xi} \in N_0(v_k) \setminus c_p, c_{\varrho}} \boldsymbol{\mu}_{c_{\xi} \to v_k} \right) }
{ \left\langle \boldsymbol{\mu}_{c_{\varrho} \to v_k}, \operatorname*{\bigcirc}_{c_{\xi} \in N_0(v_k) \setminus c_p, c_{\varrho}} \boldsymbol{\mu}_{c_{\xi} \to v_k} \right\rangle } \\
&- \boldsymbol{\mu}_{v_k \to c_p} \frac{ \left\langle \frac{\partial \boldsymbol{\mu}_{c_{\varrho} \to v_k}}{\partial \rv_j (h)},
\operatorname*{\bigcirc}_{c_{\xi} \in N_0(v_k) \setminus c_p, c_{\varrho}} \boldsymbol{\mu}_{c_{\xi} \to v_k} \right\rangle }
{ \left\langle \boldsymbol{\mu}_{c_{\varrho} \to v_k}, \operatorname*{\bigcirc}_{c_{\xi} \in N_0(v_k) \setminus c_p, c_{\varrho}} \boldsymbol{\mu}_{c_{\xi} \to v_k} \right\rangle } \\
&= \frac{1}{\tau^2} \frac{ \boldsymbol{\nu} \circ
\left( \operatorname*{\bigcirc}_{c_{\xi} \in N_0(v_k) \setminus c_p, c_{\varrho}} \boldsymbol{\mu}_{c_{\xi} \to v_k} \right) }
{ \left\langle \boldsymbol{\mu}_{c_{\varrho} \to v_k}, \operatorname*{\bigcirc}_{c_{\xi} \in N_0(v_k) \setminus c_p, c_{\varrho}} \boldsymbol{\mu}_{c_{\xi} \to v_k} \right\rangle } \\
&- \frac{1}{\tau^2} \boldsymbol{\mu}_{v_k \to c_p}
\frac{ \left\langle \boldsymbol{\nu},
\operatorname*{\bigcirc}_{c_{\xi} \in N_0(v_k) \setminus c_p, c_{\varrho}} \boldsymbol{\mu}_{c_{\xi} \to v_k} \right\rangle }
{ \left\langle \boldsymbol{\mu}_{c_{\varrho} \to v_k}, \operatorname*{\bigcirc}_{c_{\xi} \in N_0(v_k) \setminus c_p, c_{\varrho}} \boldsymbol{\mu}_{c_{\xi} \to v_k} \right\rangle } \\
&= \frac{1}{\tau^2} \left( \boldsymbol{\nu}_{v_k \to c_p}
- \left\| \boldsymbol{\nu}_{v_k \to c_p} \right\|_1 \boldsymbol{\mu}_{v_k \to c_p} \right) ,
\end{split}
\end{equation}
where we have utilized the shorthand notation
\begin{equation*}
\boldsymbol{\nu}_{v_k \to c_p} = \frac{ \boldsymbol{\nu} \circ
\left( \operatorname*{\bigcirc}_{c_{\xi} \in N_0(v_k) \setminus c_p, c_{\varrho}} \boldsymbol{\mu}_{c_{\xi} \to v_k} \right) }
{ \left\langle \boldsymbol{\mu}_{c_{\varrho} \to v_k}, \operatorname*{\bigcirc}_{c_{\xi} \in N_0(v_k) \setminus c_p, c_{\varrho}} \boldsymbol{\mu}_{c_{\xi} \to v_k} \right\rangle } .
\end{equation*}
We emphasize that two of the terms in the derivation above cancel out, as before.
Furthermore, by construction, we immediately obtain $\zerov \preceq \boldsymbol{\nu}_{v_k \to c_p} \preceq \boldsymbol{\mu}_{v_k \to c_p}$.
These observations closely parallel the description for the base case.

The derivation of the second graph operation for the inductive step is in complete analogy with the base case, except for labeling.
Specifically, we apply \eqref{equation:DerivativeMupl} and obtain
\begin{equation}
\begin{split}
&\frac{\partial \boldsymbol{\mu}_{c_p \to v_{\ell}}}{\partial \rv_j(h)}
= \frac{\partial \boldsymbol{\mu}_{v_k \to c_p}}{\partial \rv_j (h)}
\odot \boldsymbol{\nu}_{c_p \setminus v_{\ell}, v_k} \\
&= \frac{1}{\tau^2} \left( \boldsymbol{\nu}_{v_k \to c_p} - \left\| \boldsymbol{\nu}_{v_k \to c_p} \right\|_1 \boldsymbol{\mu}_{v_k \to c_p} \right) 
\odot \boldsymbol{\nu}_{c_p \setminus v_{\ell}, v_k} \\
&= \frac{1}{\tau^2} \left( \boldsymbol{\nu}_{v_k \to c_p} \odot \boldsymbol{\nu}_{c_p \setminus v_{\ell}, v_k}
- \left\| \boldsymbol{\nu}_{v_k \to c_p} \right\|_1 \boldsymbol{\mu}_{c_p \to v_{\ell}} \right) .
\end{split}
\end{equation}
For the inductive step, we define $\boldsymbol{\nu}' = \boldsymbol{\nu}_{v_k \to c_p} \odot \boldsymbol{\nu}_{c_p \setminus v_{\ell}, v_k}$ as the candidate vector.
Based on component-wise ordering, we can write
\begin{equation*}
\begin{split}
\zerov &\preceq \boldsymbol{\nu}' = \boldsymbol{\nu}_{v_k \to c_p} \odot \boldsymbol{\nu}_{c_p \setminus v_{\ell}, v_k} \\
&\preceq \boldsymbol{\mu}_{v_k \to c_p} \odot \boldsymbol{\nu}_{c_p \setminus v_{\ell}, v_k}
= \boldsymbol{\mu}_{c_p \to v_{\ell}} .
\end{split}
\end{equation*}
As before, we have that
\begin{equation*}
\left\| \boldsymbol{\nu}' \right\|_1
= \left\| \boldsymbol{\nu}_{v_k \to c_p} \right\|_1 \left\| \boldsymbol{\nu}_{c_p \setminus v_{\ell}, v_k} \right\|_1
= \left\| \boldsymbol{\nu}_{v_k \to c_p} \right\|_1 .
\end{equation*}
Hence, candidate vector $\boldsymbol{\nu}'$ is such that $\zerov \preceq \boldsymbol{\nu}' \preceq \boldsymbol{\mu}_{c_p \to v_{\ell}}$ and
\begin{equation}
\frac{\partial \boldsymbol{\mu}_{c_p \to v_{\ell}}}{\partial \rv_j(h)}
= \frac{1}{\tau^2} \left( \boldsymbol{\nu}' - \left\| \boldsymbol{\nu}' \right\|_1
\boldsymbol{\mu}_{c_p \to v_{\ell}} \right) .
\end{equation}
This completes the proof for Proposition~\ref{proposition:DerivativeJGammaBound}.
\end{IEEEproof}

We have nearly attained out goal of showing that the magnitudes of the entries in the Jacobian matrix of $\etav(\rv)$ with respect to $\rv$ are uniformly bounded.
To achieve the desired result, it sufficies to connect the partial derivative of the incoming message with the partial derivative of state estimate $\hat{\sv}_{\ell} \left( \rv, g \right)$.
This is accomplished below.

\begin{proposition} \label{proposition:DerivativeJBound}
Under Condition~\ref{condition:Sub-Girth-BP}, the absolute value of the entries in the Jacobian are bounded by
\begin{equation} \label{equation:DerivativeJBound}
\left| \frac{\partial \hat{\sv}_{\ell} \left( \rv, g \right)}{\partial \rv_{j} (h)} \right|
\leq \frac{1}{\tau^2} \quad \forall g, h \in \mathbb{F}_q, \ell, j \in [L] ,
\end{equation}
where $\tau > 0$ represents the standard deviation of the effective observation.
\end{proposition}
\begin{IEEEproof}
When $v_j$ does not appear in the rooted tree of $v_{\ell}$, the partial derivative is equal to zero and the result immediately follows. 
Furthermore, when $v_{\ell} = v_j$, the result follows from Corollary~\ref{corollary:bounded_root_derivatives}.
Thus, we can focus on the scenario wherein $v_j$ is a descendant of $v_{\ell}$.

Let $c_p$ be the unique check node in $N(v_{\ell})$ that lies on the path between $v_{\ell}$ and $v_j$.
By Proposition~\ref{proposition:DerivativeJGammaBound}, there exists vector $\boldsymbol{\nu}$, with $\zerov \preceq \boldsymbol{\nu} \preceq \boldsymbol{\mu}_{c_p \to v_{\ell}}$, such that the partial derivative of $\boldsymbol{\mu}_{c_p \to v_{\ell}}$ with respect to $\rv_j (h)$ is given by
\begin{equation}
\frac{\partial \boldsymbol{\mu}_{c_p \to v_{\ell}}}{\partial \rv_j (h)}
= \frac{1}{\tau^2} \left( \boldsymbol{\nu} - \left\| \boldsymbol{\nu} \right\|_1
\boldsymbol{\mu}_{c_p \to v_{\ell}} \right) .
\end{equation}
Drawing an analogy to \eqref{equation:paritalMukp}, we have
\begin{equation}
\begin{split}
\frac{\partial \hat{\sv}_{\ell} \left( \rv \right)}{\partial \rv_j (h)}
&= \frac{\partial}{\partial \rv_j (h)}
\frac{ \operatorname*{\bigcirc}_{c_{\xi} \in N_0(v_{\ell})} \boldsymbol{\mu}_{c_{\xi} \to v_{\ell}} }
{ \left\| \operatorname*{\bigcirc}_{c_{\xi} \in N_0(v_{\ell})}  \boldsymbol{\mu}_{c_{\xi} \to v_{\ell}} \right\|_1 } \\
&= \frac{1}{\tau^2} \frac{ \boldsymbol{\nu} \circ
\left( \operatorname*{\bigcirc}_{c_{\xi} \in N_0(v_{\ell}) \setminus c_p} \boldsymbol{\mu}_{c_{\xi} \to v_{\ell}} \right) }
{ \left\langle \boldsymbol{\mu}_{c_p \to v_{\ell}}, \operatorname*{\bigcirc}_{c_{\xi} \in N_0(v_{\ell}) \setminus c_p} \boldsymbol{\mu}_{c_{\xi} \to v_{\ell}} \right\rangle } \\
&- \frac{1}{\tau^2} \hat{\sv}_{\ell} \left( \rv \right)
\frac{ \left\langle \boldsymbol{\nu},
\operatorname*{\bigcirc}_{c_{\xi} \in N_0(v_{\ell}) \setminus c_p} \boldsymbol{\mu}_{c_{\xi} \to v_{\ell}} \right\rangle }
{ \left\langle \boldsymbol{\mu}_{c_p \to v_{\ell}}, \operatorname*{\bigcirc}_{c_{\xi} \in N_0(v_{\ell}) \setminus c_p} \boldsymbol{\mu}_{c_{\xi} \to v_{\ell}} \right\rangle } \\
&= \frac{1}{\tau^2} \left( \boldsymbol{\nu}_{v_{\ell}}
- \left\| \boldsymbol{\nu}_{v_{\ell}} \right\|_1 \hat{\sv}_{\ell} \left( \rv \right) \right) ,
\end{split}
\end{equation}
where we have implicitly defined
\begin{equation*}
\boldsymbol{\nu}_{v_{\ell}} = \frac{ \boldsymbol{\nu} \circ
\left( \operatorname*{\bigcirc}_{c_{\xi} \in N_0(v_{\ell}) \setminus c_p} \boldsymbol{\mu}_{c_{\xi} \to v_{\ell}} \right) }
{ \left\langle \boldsymbol{\mu}_{c_{p} \to v_{\ell}}, \operatorname*{\bigcirc}_{c_{\xi} \in N_0(v_{\ell}) \setminus c_p} \boldsymbol{\mu}_{c_{\xi} \to v_{\ell}} \right\rangle } .
\end{equation*}
We note that $\zerov \preceq \boldsymbol{\nu}_{v_{\ell}} \preceq \hat{\sv}_{\ell} \left( \rv \right)$.
Thus, we have that:
\begin{equation} \label{eq:thm_positive}
\frac{\partial \hat{\sv}_{\ell} \left( \rv \right)}{\partial \rv_j (h)}
\leq \frac{1}{\tau^2} \boldsymbol{\nu}_{v_{\ell}}
\leq \frac{1}{\tau^2} \hat{\sv}_{\ell} \left( \rv \right) .
\end{equation}
Since we are interested in bounding the absolute value of the partial derivatives, we also consider a lower bound. 
\begin{equation}
\frac{\partial \hat{\sv}_{\ell} \left( \rv \right)}{\partial \rv_j (h)}
\geq - \frac{1}{\tau^2} \left\| \boldsymbol{\nu}_{v_{\ell}} \right\|_1 \hat{\sv}_{\ell} \left( \rv \right)
\geq - \frac{1}{\tau^2} \hat{\sv}_{\ell} \left( \rv \right) .
\end{equation}
Combining these two observations with the properties of probability vectors, we obtain the desired expression.
\end{IEEEproof}

The proof of Theorem~\ref{theorem:bp_denoiser_lipschitz} thus follows immediately from this result because, since the magnitudes of the entries of the Jacobian matrix of $\etav(\rv)$ with respect to $\rv$ are uniformly bounded, the BP denoiser is Lipschitz continuous.

\subsection{Proof of Proposition~\ref{proposition:onsager_correction_term}}
\label{section:derivation_onsager_term}

Intuitively, the role of the Onsager term is to (asymptotically) cancel the first-order correlations between $\Am^{\mathrm{T}} \zv^{(t)}$ and $\sv^{(t)}$ and thereby maintain a structure conducive to prompt convergence and analysis.
This factor, emblematic of AMP algorithms, appears in \eqref{equation:AMP-Residual} and is given by
\begin{equation}
\label{equation:ot_def}
\begin{split}
\frac{\zv^{(t-1)}}{n} \operatorname{div} &\etav_{t-1} \left( \Am^{\mathrm{T}} \zv^{(t-1)} + \sv^{(t-1)} \right)\\
&=\frac{\zv^{(t-1)}}{n} \operatorname{div} \etav_{t-1} \left( \rv^{(t-1)} \right)
\end{split}
\end{equation}
where the $\operatorname{div}$ operator can be expanded into
\begin{equation} \label{equation:DivOperator}
\operatorname{div} \etav \left( \rv \right)
= \sum_{\ell \in [L]} \operatorname{div} \hat{\sv}_{\ell} ( \rv, \tau )
= \sum_{\ell \in [L]} \sum_{g \in \mathbb{F}_q} \frac{\partial \hat{s}_{\ell} \left( g, \rv, \tau \right)}{\partial \rv_{\ell}(g)} .
\end{equation}
Thus, as an intermediate step, we must calculate the partial derivative of $\hat{s}_{\ell} \left( g, \rv, \tau \right)$ with respect to $\rv_{\ell}(g)$.
This computation is rendered much simpler under Condition~\ref{condition:Sub-Girth-BP}, which ensures that the message passing operations employed during denoising yield valid computation trees without cycles.

\begin{lemma} \label{lemma:BP-PartialSestimate}
Under Condition~\ref{condition:Sub-Girth-BP}, the partial derivative of $\hat{s}_{\ell} \left( g, \rv, \tau \right)$ with respect to $\rv_{\ell}(g)$ is equal to
\begin{equation} \label{equation:BP-PartialSestimate}
\frac{\partial \hat{s}_{\ell} \left( g, \rv, \tau \right)}{\partial \rv_{\ell}(g)}
= \frac{1}{\tau^2} \hat{s}_{\ell} \left( g, \rv, \tau \right)
\left( 1 - \hat{s}_{\ell} \left( g, \rv, \tau \right) \right)
\end{equation}
where $g \in \mathbb{F}_q$.
\end{lemma}
\begin{IEEEproof}
Recall that the output of the BP denoiser defined in \eqref{equation:BP-Estimate} can be expressed as
\begin{equation}
\begin{split}
\hat{s}_{\ell} \left( g, \rv, \tau \right)
&= \frac{ \boldsymbol{\alpha}_{\ell} (g) \prod_{c_p \in N(v_{\ell})} \boldsymbol{\mu}_{c_{p} \to v_{\ell}}(g) }
{ \sum_{h \in \mathbb{F}_q} \boldsymbol{\alpha}_{\ell} (h) \prod_{c_p \in N(v_{\ell})} \boldsymbol{\mu}_{c_{p} \to v_{\ell}}(h) } \\
&= \frac{ \boldsymbol{\alpha}_{\ell} (g) \boldsymbol{\mu}_{v_{\ell} \to c_0}(g) }
{ \sum_{h \in \mathbb{F}_q} \boldsymbol{\alpha}_{\ell} (h) \boldsymbol{\mu}_{v_{\ell} \to c_0}(h) } .
\end{split}
\end{equation}
Under Condition~\ref{condition:Sub-Girth-BP}, belief vector $\boldsymbol{\mu}_{v_{\ell} \to c_0}$ is based solely on extrinsic information and, hence, it is determined based on $\left\{ \rv_j : j \in [L] \setminus \{ \ell \} \right\}$.
Consequence, we gather that
\begin{equation*}
\frac{\partial \boldsymbol{\mu}_{v_{\ell} \to c_0}(g)}{\partial \rv_{\ell}(g)} = 0 .
\end{equation*}
Under such circumstances, we can calculate the desired derivative in a straightforward manner, with
\begin{equation*}
\begin{split}
\frac{\partial \hat{s}_{\ell} \left( g, \rv, \tau \right)}{\partial \rv_{\ell}(g)}
&= \frac{\partial}{\partial \rv_{\ell}(g)}
\frac{ e^{\frac{ \rv_{\ell}(g)}{\tau^2}} \boldsymbol{\mu}_{v_{\ell} \to c_0}(g) }
{ \sum_{h \in \mathbb{F}_q} e^{\frac{ \rv_{\ell}(h)}{\tau^2}} \boldsymbol{\mu}_{v_{\ell} \to c_0}(h) } \\
&= \frac{1}{\tau^2} \frac{ e^{\frac{ \rv_{\ell}(g)}{\tau^2}} \boldsymbol{\mu}_{v_{\ell} \to c_0}(g) }{\sum_{h \in \mathbb{F}_{q}} e^{\frac{ \rv_{\ell}(g)}{\tau^2}} \boldsymbol{\mu}_{v_{\ell} \to c_0}(g) } \\
&\quad- \frac{1}{\tau^2}\frac{ \left( e^{\frac{ \rv_{\ell}(g)}{\tau^2}} \boldsymbol{\mu}_{v_{\ell} \to c_0}(g) \right)^2 }{\left( \sum_{h \in \mathbb{F}_{q}} e^{\frac{ \rv_{\ell}(g)}{\tau^2}} \boldsymbol{\mu}_{v_{\ell} \to c_0}(g) \right)^2} \\
&= \frac{1}{\tau^2} \hat{s}_{\ell} \left( g, \rv, \tau \right)
\left( 1 - \hat{s}_{\ell} \left( g, \rv, \tau \right) \right) .
\end{split}
\end{equation*}
This last line corresponds to the statement of the lemma.
\end{IEEEproof}

It is worth emphasizing that the derivative in \eqref{equation:BP-PartialSestimate} remains unchanged irrespective of the number of BP rounds computed on the factor graph, so long as Condition~\ref{condition:Sub-Girth-BP} is satisfied.
The divergence of \eqref{equation:DivOperator} assumes the same simple form under such circumstances.

\begin{proposition} \label{proposition:DivComputationBP}
The divergence of $\etav \left( \rv \right)$ with respect to $\rv$ is equal to
\begin{equation} \label{equation:BP-OnsagerCorrection2}
\begin{split}
&\operatorname{div} \etav \left( \rv \right)
= \frac{1}{\tau^2} \left( \left\| \etav \left( \rv \right) \right\|_1 - \left\| \etav \left( \rv \right) \right\|^2 \right) .
\end{split}
\end{equation}
\end{proposition}
\begin{IEEEproof}
We expand the $\operatorname{div}$ operator as
\begin{equation} \label{equation:OnsagerDerivation}
\begin{split}
\operatorname{div} \etav \left( \rv \right)
&= \sum_{\ell \in [L]} \operatorname{div} \hat{\sv}_{\ell} \left( \rv, \tau \right)
= \sum_{\ell \in [L]} \sum_{g \in \mathbb{F}_q} \frac{\partial \hat{s}_{\ell} \left( g, \rv, \tau \right)}{\partial \rv_{\ell}(g)} \\
&= \sum_{\ell \in [L]} \sum_{g \in \mathbb{F}_q} \frac{1}{\tau^2} \hat{s}_{\ell} \left( g, \rv, \tau \right)
\left( 1 - \hat{s}_{\ell} \left( g, \rv, \tau \right) \right) \\
&= \frac{1}{\tau^2} \left( \left\| \etav \left( \rv \right) \right\|_1
- \left\| \etav \left( \rv \right) \right\|^2 \right) .
\end{split}
\end{equation}
The last equality follows from the fact that, since $\hat{s}_{\ell} \left( g, \rv, \tau \right)$ lies between zero and one, the corresponding partial derivative with respect to $\rv_{\ell}(g)$ found in Lemma~\ref{lemma:BP-PartialSestimate} is always non-negative.
\end{IEEEproof}
The proof of Proposition~\ref{proposition:onsager_correction_term} follows immediately from the definition of the Onsager term \eqref{equation:ot_def} and Proposition~\ref{proposition:DivComputationBP}. 

\section{State Evolution}
\label{appendix:state_evolution}

In this appendix, we provide proofs for the propositions, lemmas, corollaries, and theorems from Section~\ref{section:state_evolution}.

\subsection{Proofs from Section~\ref{subsection:geometric_uniformity}}

\noindent \textbf{Proof of Proposition~\ref{proposition:geometric_uniformity}.}
\begin{IEEEproof}
Fix two points $\sv, \sv' \in \mathcal{S}$.
Since every point in $\mathcal{S}$ is obtained by indexing an LDPC codeword, there exists $\vv, \vv'$ in LDPC codebook $\mathcal{V}$ such that $\vv$ maps to $\sv$ and, similarly, $\vv'$ maps to $\sv'$.
Furthermore, since the outer LDPC code is a linear code, we have $\uv = \vv' - \vv \in \mathcal{V}$;
that is, $\uv$ is also a valid codeword.

Consider the invertible translation $\boldsymbol{\upsilon} \mapsto \boldsymbol{\upsilon} \oplus \uv$ in $\mathbb{F}_q^L$.
Then, for any codeword $\boldsymbol{\upsilon} \in \mathcal{V}$, we get $\boldsymbol{\upsilon} \oplus \uv \in \mathcal{V}$ because codebook $\mathcal{V}$ is closed under addition.
Focusing on every section individually, we have the mapping $\upsilon_{\ell} \mapsto \upsilon_{\ell} \oplus u_{\ell}$.
This action induces a bijection acting on $\mathbb{F}_q$.
Likewise, this produces a permutation of basis vectors with
\begin{equation*}
\ev_{v_{\ell}} \mapsto \ev_{\pi(v_{\ell})} = \ev_{v_{\ell} \oplus u_{\ell}} .
\end{equation*}
Based on this correspondence, we can define a permutation matrix $\Pi^{(\ell)}$ that acts on $\mathbb{R}^q$ for each $\ell \in [L]$.
Then, aggregating these $L$ permutation matrices, we get the isometry $\boldsymbol{\Pi} = \operatorname{diag} \left( \Pi^{(1)}, \ldots, \Pi^{(L)} \right)$ which, by construction, maps $\sv$ to $\sv'$ while leaving $\mathcal{S}$ invariant.
Since $\sv$ and $\sv'$ are arbitrary, we conclude that $\mathcal{S}$ is geometrically uniform, as claimed in the proposition.
\end{IEEEproof}

\noindent\textbf{Proof of Proposition~\ref{proposition:SectionPermuationInvariance}.}
\begin{IEEEproof}
We already know that the $2$-norm of $\rv_{\ell} - \ev_j$ is invariant to permutations in its vector argument.
Then, under Condition~\ref{condition:AsymptoticCharacterization}, we have
\begin{equation} \label{equation:ConditionalGaussianDistribution}
\begin{split}
f_{\Rv_{\ell} | \Sv_{\ell}} \left( \rv_{\ell} | \ev_g \right)
&= \frac{1}{(2 \pi)^{\frac{q}{2}} \tau^q }
\exp \left( - \frac{\left\| \rv_{\ell} - \ev_g \right\|^2}{2 \tau^2} \right) \\
&= \frac{1}{(2 \pi)^{\frac{q}{2}} \tau^q }
\exp \left( - \frac{\left\| \Pi \rv_{\ell} - \Pi \ev_{g} \right\|^2}{2 \tau^2} \right) \\
&= f_{\Rv_{\ell} | \mathbf{S}_{\ell}} \left( \Pi \rv_{\ell} \middle| \Pi \ev_{g} \right) .
\end{split}
\end{equation}
Leveraging the underlying order on field elements, we can write $\Pi \ev_{g} = \ev_{\pi(g)}$, where $\pi (\cdot)$ is the permutation on $\mathbb{F}_q$ induced by matrix $\Pi$.
\end{IEEEproof}

\noindent\textbf{Proof of Corollary~\ref{corollary:edge_label_effect_on_distribution}.}
\begin{IEEEproof}
To begin, recall that $\Rv_{\ell}^{\times \omega}$ corresponds to the operator introduced in Definition~\ref{defintion:TimesOperator}.
This operator reorders the entries of a vector based on the mapping $g \mapsto \omega \otimes g$, together with the bijection of Remark~\ref{remark:NotationOverload}.
This action therefore creates a permutation matrix $\Pi$ on the vector entries of its argument.
However, the zeroth element of $\Rv$ necessarily remains in its original location under this mapping because $\omega \otimes 0 = 0$.
Thus, we can apply Proposition~\ref{proposition:SectionPermuationInvariance} and get
\begin{equation}
\begin{split}
f_{\Rv_{\ell} | \Sv_{\ell}} \left( \rv_{\ell} | \ev_0 \right)
&= f_{\Rv_{\ell} | \mathbf{S}_{\ell}} \left( \Pi \rv_{\ell} \middle| \Pi \ev_0 \right) \\
&= f_{\Rv_{\ell} | \mathbf{S}_{\ell}} \left(\rv_{\ell}^{\times \omega} \middle| \ev_0 \right) .
\end{split}
\end{equation}
Consequently, the conditional distributions of $\Rv_{\ell}$ and $\Rv_{\ell}^{\times \omega}$ are identical, given vector input $\ev_0$.
\end{IEEEproof}

\subsection{Proofs from Section~\ref{subsection:properties_bp_messages}}

\noindent \textbf{Proof of Lemma~\ref{lemma:fq_conv_dominant_gs_lv_rv}.}
\begin{IEEEproof}
This lemma is best proved by induction; however, in the interest of space, we only provide a proof of the base case.
Let $\Lv$ and $\Mv$ be two independent group-symmetric likelihood-vector random variables.
Define $\Nv = \Lv \odot \Mv$.
Then, the components of the convolution are given by
\begin{equation*}
N_g = \left( \Lv \odot \Mv \right)_g = \sum_{h \in \mathbb{F}_q} L_h M_{g - h} .
\end{equation*}
As a function of two random variables with non-negative entries, it immediately follows that $\Nv$ is a likelihood vector random variable. 
What remains is to show that $\Nv$ is dominant and  group-symmetric.
For any $i \in \mathbb{F}_q \setminus \{ 0 \}$, we have
\begin{equation}
\begin{split}
&\left( \Nv^{\times i} \right)_g
= \left( \left( \Lv \odot \Mv \right)^{\times i} \right)_g
= \left( \Lv \odot \Mv \right)_{g \otimes i} \\
&= \sum_{h \in \mathbb{F}_q} L_h M_{g \otimes i - h}
= \sum_{h \in \mathbb{F}_q} L_{h \otimes i^{-1} \otimes i} M_{(g - h \otimes i^{-1}) \otimes i} \\
&= \sum_{j \in \mathbb{F}_q} L_{j \otimes i} M_{(g - j) \otimes i}
= \sum_{j \in \mathbb{F}_q} \left( \Lv^{\times i} \right)_{j} \left( \Mv^{\times i} \right)_{(g - j)} \\
&= \left( \Lv^{\times i} \odot \Mv^{\times i} \right)_g
\end{split}
\end{equation}
where $j = h \otimes i^{-1}$.
By definition, $\Lv \overset{d}{=} \Lv^{\times i}$ and $\Mv \overset{d}{=} \Mv^{\times i}$, where $\overset{d}{=}$ denotes equality in distribution.
Thus, we have that 
\begin{equation}
    \begin{split}
        \Nv = \left(\Lv \odot \Mv\right) &\overset{d}{=} \left(\Lv^{\times i} \odot \Mv^{\times i}\right) = \Nv^{\times i}. \\
    \end{split}
\end{equation}
Thus, $\Nv$ is a group symmetric likelihood vector random variable. 
Now, note that 
\begin{equation*}
\begin{split}
&\mathbb{E} [N_0] = \mathbb{E} \left[ \sum_{h \in \mathbb{F}_q} L_h M_{-h} \right]
= \sum_{h \in \mathbb{F}_q} \mathbb{E}[L_h] \mathbb{E}[M_{-h}] \\
&= \mathbb{E}[L_0] \mathbb{E}[M_0] + (q-1) \mathbb{E}[L_{\bullet}] \mathbb{E}[M_{\bullet}] \\
&\geq \mathbb{E}[L_0] \mathbb{E}[M_{\bullet}] + \mathbb{E}[L_{\bullet}] \mathbb{E}[M_0]
+ (q-2) \mathbb{E}[L_{\bullet}] \mathbb{E}[M_{\bullet}] \\
&= \mathbb{E}[N_{\bullet}] .
\end{split}
\end{equation*}
The inequality above is a consequence of the fact that, by assumption, $\Lv$ and $\Mv$ are dominant symmetric.
Hence, $\mathbb{E}[L_0] \geq \mathbb{E}[L_{\bullet}]$ and $\mathbb{E}[M_0] \geq \mathbb{E}[M_{\bullet}]$.
Thus, $\Nv$ is a dominant group-symmetric likelihood vector random variable. 

The extension of these findings to the convolution of multiple independent vectors follows from a straightforward induction argument.
\end{IEEEproof}

\noindent\textbf{Proof of Lemma~\ref{lemma:ConvolutionNormL1}.}
\begin{IEEEproof}
Again, in the interest of space, we only prove the base case when $n = 2$. 
The inductive case for $n > 2$ follows directly. 
Let $\lv$ and $\mv$ be likelihood-vectors over $\mathbb{F}_q$, and define $\nv = \lv \odot \mv$.
Since $\mathbb{F}_q$ is a finite set and likelihood-vectors have non-negative entries, we get
\begin{equation}
\begin{split}
\left\| \nv \right\|_1
&= \sum_{g \in \mathbb{F}_q} \nv_g
= \sum_{g \in \mathbb{F}_q} \left( \lv \odot \mv \right)_g \\
&= \sum_{g \in \mathbb{F}_q} \sum_{h \in \mathbb{F}_q} \ell_h m_{g - h}
= \sum_{h \in \mathbb{F}_q} \ell_h \sum_{g \in \mathbb{F}_q} m_{g - h} \\
&= \sum_{h \in \mathbb{F}_q} \ell_h \left\| \mv \right\|_1
= \left\| \lv \right\|_1 \left\| \mv \right\|_1 .
\end{split}
\end{equation}
That is, the one-norm of $\nv$ is equal to the product of the one-norms of $\lv$ and $\mv$.
\end{IEEEproof}

\noindent\textbf{Proof of Corollary~\ref{corollary:OneNormConvolution}.}
\begin{IEEEproof}
Let $\omega \in \Omega$ denote an outcome within the underlying sample space.
For any such realization, Lemma~\ref{lemma:ConvolutionNormL1} states that
\begin{equation*}
\left\| \Nv (\omega) \right\|_1 = \prod_{p \in [n]} \left\| \Lv^{(p)} (\omega) \right\|_1 .
\end{equation*}
Taking expectations with respect to the corresponding probability law, we get
\begin{equation}
\begin{split}
\mathbb{E} \left[ \left\| \Nv (\omega) \right\|_1 \right]
&= \mathbb{E} \left[ \prod_{p \in [n]} \left\| \Lv^{(p)} (\omega) \right\|_1 \right] \\
&= \prod_{p \in [n]} \mathbb{E} \left[ \left\| \Lv^{(p)} (\omega) \right\|_1 \right]
\end{split}
\end{equation}
where the last step is a consequence of the likelihood-vector random variables being independent.
Hence, the expectations decouple and the result follows.
\end{IEEEproof}

\noindent\textbf{Proof of Proposition~\ref{proposition:ConvolutionGroupSymmetricLikelihoodRV}.}
\begin{IEEEproof}
Consider the case of two independent dominant group-symmetric likelihood-vector random variables, $\Lv$ and $\Mv$, with $\Nv = \Lv \odot \Mv$.
Examining the zeroth component of $\Nv$, we write
\begin{equation*}
\mathbb{E} \left[ N_0 \right]
= \mathbb{E} [L_0] \mathbb{E} [M_0] + (q-1) \mathbb{E} [L_{\bullet}] \mathbb{E} [M_{\bullet}] .
\end{equation*}
Similarly, the expected value of any other component within random vector $\Nv$ is of the form
\begin{equation*}
\mathbb{E} \left[ N_{\bullet} \right] = \mathbb{E} [L_0] \mathbb{E} [M_{\bullet}] + \mathbb{E} [L_{\bullet}] \mathbb{E} [M_0] + (q-2) \mathbb{E} [L_{\bullet}] \mathbb{E} [M_{\bullet}] .
\end{equation*}
Combining these two equations, we arrive at the expression
\begin{equation} \label{equation:DifferenceNZeroBullet}
\begin{split}
\mathbb{E}  \left[ N_0 \right] - \mathbb{E} \left[ N_{\bullet} \right] &= 
\mathbb{E} [L_0] \mathbb{E} [M_0] - \mathbb{E} [L_0] \mathbb{E} [M_{\bullet}] \\
&\hspace{5mm} - \mathbb{E} [L_{\bullet}] \mathbb{E} [M_0] + \mathbb{E} [L_{\bullet}] \mathbb{E} [M_{\bullet}] \\
&= \left( \mathbb{E} [L_0] - \mathbb{E} [L_{\bullet}] \right)
\left( \mathbb{E} [M_0] - \mathbb{E} [M_{\bullet}] \right) ,
\end{split}
\end{equation}
which seems propitious for the application of mathematical induction.

To proceed with the induction argument, we introduce a collection of convolved vectors: $\Nv^{(n)} = \bigodot_{p \in [n]} \Lv^{(p)}$.
The hypothesis can be formulated as
\begin{equation} \label{equation:InductionNDiffence}
\mathbb{E} \left[ N_0^{(n)} \right] - \mathbb{E} \left[ N_{\bullet}^{(n)} \right]
= \prod_{p \in [n]} \left( \mathbb{E} \left[ L_0^{(p)} \right] - \mathbb{E} \left[ L_{\bullet}^{(p)} \right] \right) .
\end{equation}
The base case of $n = 1$, is immediate by construction.
For the inductive step, assume that \eqref{equation:InductionNDiffence} holds for fixed $n \in \mathbb{N}$, $n >0$.
Note that
\begin{equation*}
\Nv^{(n+1)}
= \Lv^{(n+1)} \odot \Nv^{(n)} .
\end{equation*}
Also, note that $\Lv^{(n+1)}$ and $\Nv^{(n)}$ are independent because $\left\{ \Lv^{(p)} \right\}$ is a set of independent likelihood-vector random variables and, consequently, $\Lv^{(n+1)}$ and $\left\{ \Lv^{(p)} : p \in [n] \right\}$ form independent collections.
Hence, \eqref{equation:DifferenceNZeroBullet} applies and
\begin{equation*}
\begin{split}
&\mathbb{E} \left[ N_0^{(n+1)} \right] - \mathbb{E} \left[ N_{\bullet}^{(n+1)} \right] \\
&= \left( \mathbb{E} \left[L_0^{(n+1)}\right] - \mathbb{E} \left[L_{\bullet}^{(n+1)}\right] \right)
\left( \mathbb{E} \left[ N_0^{(n)} \right] - \mathbb{E} \left[ N_{\bullet}^{(n)} \right] \right) \\
&= \left( \mathbb{E} \left[L_0^{(n+1)}\right] - \mathbb{E} \left[L_{\bullet}^{(n+1)}\right] \right)
\prod_{p \in [n]} \left( \mathbb{E} \left[ L_0^{(p)} \right] - \mathbb{E} \left[ L_{\bullet}^{(p)} \right] \right) \\
&= \prod_{p \in [n+1]} \left( \mathbb{E} \left[ L_0^{(p)} \right] - \mathbb{E} \left[ L_{\bullet}^{(p)} \right] \right) ,
\end{split}
\end{equation*}
where the penultimate equality follows from our inductive hypothesis.
This completes the mathematical induction.

Lemma~\ref{lemma:fq_conv_dominant_gs_lv_rv} asserts that $\Nv = \bigodot_{p \in [n]} \Lv^{(p)}$ is a dominant group-symmetric likelihood-vector random variable.
Thus,
\begin{equation} \label{equation:MeanNOne}
\begin{split}
\mathbb{E} \left[ \| \Nv \|_1 \right] &= \mathbb{E} \left[ N_0 \right] + (q-1) \mathbb{E} \left[ N_{\bullet} \right] \\
&= \left( \mathbb{E} \left[ N_0 \right] - \mathbb{E} \left[ N_{\bullet} \right] \right) + q \mathbb{E} \left[ N_{\bullet} \right] \\
&= q \mathbb{E} \left[ N_0 \right] - (q-1) \left( \mathbb{E} \left[ N_0 \right] - \mathbb{E} \left[ N_{\bullet} \right] \right) .
\end{split}
\end{equation}
Furthermore, combining Corollary~\ref{corollary:OneNormConvolution} and the fact that the random vectors in $\left\{ \Lv^{(p)} \right\}$ are independent, we get
\begin{equation} \label{equation:MeanNOneNorm}
\mathbb{E} \left[ \left\| \Nv \right\|_1 \right]
= \prod_{p \in [n]} \left( \mathbb{E} \left[ L_0^{(p)} \right] + (q-1) \mathbb{E} \left[ L_{\bullet}^{(p)} \right] \right) .
\end{equation}
Isolating $\mathbb{E} \left[ N_0 \right]$ in \eqref{equation:MeanNOne} and then applying \eqref{equation:InductionNDiffence} \& \eqref{equation:MeanNOneNorm}, we get
\begin{equation*}
\begin{split}
\mathbb{E} \left[ N_0 \right]
&= \frac{1}{q} \mathbb{E} \left[ \| \Nv \|_1 \right] 
+ \frac{(q-1)}{q} \left( \mathbb{E} \left[ N_0 \right] - \mathbb{E} \left[ N_{\bullet} \right] \right) \\
&= \frac{1}{q} \prod_{p \in [n]} \left( \mathbb{E} \left[ L_0^{(p)} \right] + (q-1) \mathbb{E} \left[ L_{\bullet}^{(p)} \right] \right) \\
&\qquad + \left(\frac{q-1}{q}\right) \prod_{p \in [n]} \left( \mathbb{E} \left[ L_0^{(p)} \right] - \mathbb{E} \left[ L_{\bullet}^{(p)} \right] \right).
\end{split}
\end{equation*}
Similarly, isolating $\mathbb{E} \left[ N_{\bullet} \right]$, we arrive at
\begin{equation*}
\begin{split}
\mathbb{E} [N_{\bullet}] &= \frac{1}{q} \mathbb{E} \left[ \left\| \Nv \right\|_1 \right] - \frac{1}{q} \left(\mathbb{E}[N_0] - \mathbb{E}[N_{\bullet}]\right) \\
&= \frac{1}{q} \prod_{p \in [n]} \left( \mathbb{E} \left[ L_0^{(p)} \right] + (q-1) \mathbb{E} \left[ L_{\bullet}^{(p)} \right] \right) \\
&\qquad - \frac{1}{q} \prod_{p \in [n]} \left( \mathbb{E} \left[ L_0^{(p)} \right] - \mathbb{E} \left[ L_{\bullet}^{(p)} \right] \right) .
\end{split}
\end{equation*}
This completes the proof.
\end{IEEEproof}

\noindent\textbf{Proof of Corollary~\ref{corollary:ConvolutionGroupSymmetricProbabilityRV}.}
\begin{IEEEproof}
First, we stress that a probability-vector random variable, as described in Definition~\ref{definition:SymmetricProbVectorRV}, is also a likelihood-vector random variable, albeit with additional structure.
Thus, some of the results derived above for likelihood-vector random variables readily apply in the current scenario.
For instance, in view of Lemma~\ref{lemma:ConvolutionNormL1}, we have
\begin{equation*}
\left\| \bigodot_{p \in [n]} \bar{\Lv}^{(p)} \right\|_1
= \prod_{p \in [n]} \left\| \bar{\Lv}^{(p)} \right\|_1 = 1 .
\end{equation*}
Thus, $\left\| \bar{\Nv} \right\|_1 = 1$ and, consequently, $\bar{\Nv}$ is a valid probability-vector random variable as suggested by our notation.
Second, we emphasize that the expected value of all its components, except for the zeroth entry, are equal,
Given that a probability-vector takes on values in the simplex, we can therefore write
\begin{equation*}
\mathbb{E} \left[ \bar{L}_{\bullet} \right] = \frac{1 - \mathbb{E} \left[ \bar{L}_0 \right]}{q-1} .
\end{equation*}
The difference between $\mathbb{E} \left[ \bar{L}_0 \right]$ and $\mathbb{E} \left[ \bar{L}_{\bullet} \right]$ is subject to
\begin{equation*}
\begin{split}
\mathbb{E} \left[ \bar{L}_0 \right] - \mathbb{E} \left[ \bar{L}_{\bullet} \right]
&= \mathbb{E} \left[ \bar{L}_0 \right] - \frac{1 - \mathbb{E} \left[ \bar{L}_0 \right]}{q-1} \\
&= \frac{q}{q-1} \left( \mathbb{E} \left[ \bar{L}_0 \right] - \frac{1}{q} \right) .
\end{split}
\end{equation*}
Collecting these findings and substituting the equivalent forms into Proposition~\ref{proposition:ConvolutionGroupSymmetricLikelihoodRV}, we arrive at the claimed expressions.
\end{IEEEproof}

\noindent\textbf{Proof of Lemma~\ref{lemma:SymmetricGaussianProbabilityRV}.}
\begin{IEEEproof}
Recall that, for a dominant permutation-symmetric Gaussian probability-vector random variable, we can write the individual components of $\bar{\Lv}$ as
\begin{equation} \label{equation:BarLg}
\begin{split}
\bar{L}_g &= \frac{f_{\Rv_{\ell} | \Sv_{\ell}} \left( \Rv_{\ell} | \ev_g \right)}
{\sum_{h \in \mathbb{F}_q} f_{\Rv_{\ell} | \Sv_{\ell}} \left( \Rv_{\ell} | \ev_h \right)}
\end{split}
\end{equation}
where $f_{\Rv_{\ell} | \Sv_{\ell}} \left( \cdot | \ev_g \right)$ is specified in \eqref{equation:ConditionalGaussianDistribution}.
Interestingly, we can express the second moment of $\bar{L}_g$ as
\begin{equation*}
\begin{split}
&\mathbb{E} \left[ \bar{L}_g^2 \right]
= \mathbb{E}_{0} \left[ \left( \frac{f_{\Rv_{\ell} | \Sv_{\ell}} \left( \Rv_{\ell} | \ev_g \right)}
{\sum_{h \in \mathbb{F}_q} f_{\Rv_{\ell} | \Sv_{\ell}} \left( \Rv_{\ell} | \ev_h \right)} \right)^2 \right] \\
&= \int_{\mathbb{R}^q}
\frac{f_{\Rv_{\ell} | \Sv_{\ell}} \left( \rv_{\ell} | \ev_g \right) f_{\Rv_{\ell} | \Sv_{\ell}} \left( \rv_{\ell} | \ev_g \right)}
{\left( \sum_{h \in \mathbb{F}_q} f_{\Rv_{\ell} | \Sv_{\ell}} \left( \rv_{\ell} | \ev_h \right) \right)^2 }
f_{\Rv_{\ell} | \Sv_{\ell}} \left( \rv_{\ell} | \ev_0 \right) d\rv_{\ell} \\
&= \int_{\mathbb{R}^q}
\frac{f_{\Rv_{\ell} | \Sv_{\ell}} \left( \rv_{\ell} | \ev_g \right) f_{\Rv_{\ell} | \Sv_{\ell}} \left( \rv_{\ell} | \ev_0 \right)}
{\left( \sum_{h \in \mathbb{F}_q} f_{\Rv_{\ell} | \Sv_{\ell}} \left( \rv_{\ell} | \ev_h \right) \right)^2 }
f_{\Rv_{\ell} | \Sv_{\ell}} \left( \rv_{\ell} | \ev_g \right) d\rv_{\ell} \\
&= \int_{\mathbb{R}^q}
\frac{f_{\Rv_{\ell} | \Sv_{\ell}} \left( \rv_{\ell} | \ev_0 \right) f_{\Rv_{\ell} | \Sv_{\ell}} \left( \rv_{\ell} | \ev_g \right)}
{\left( \sum_{h \in \mathbb{F}_q} f_{\Rv_{\ell} | \Sv_{\ell}} \left( \rv_{\ell} | \ev_h \right) \right)^2 }
f_{\Rv_{\ell} | \Sv_{\ell}} \left( \rv_{\ell} | \ev_0 \right) d\rv_{\ell} \\
&= \mathbb{E} \left[ \bar{L}_0 \bar{L}_g \right] .
\end{split}
\end{equation*}
In the third equality, we leverage the invariance in the problem structure established in Proposition~\ref{proposition:SectionPermuationInvariance}.
This fact, together with the symmetry in the region of integration, enables us to permute the indices.
With this relation, we can rewrite the two-norm of $\bar{\Lv}$ as
\begin{equation} \label{equation:DerivationSymmetricGaussianProbabilityRV}
\begin{split}
\mathbb{E} \left[ \left\| \bar{\Lv} \right\|_2^2 \right]
&= \mathbb{E} \left[ \sum_{g \in \mathbb{F}_q} \bar{L}_g^2 \right]
= \mathbb{E} \left[ \sum_{g \in \mathbb{F}_{q}} \bar{L}_0\bar{L}_g\right] \\
&= \mathbb{E} \left[ \bar{L}_0 \left( \sum_{g \in \mathbb{F}_q} \bar{L}_g \right) \right]
= \mathbb{E} \left[ \bar{L}_0 \left\| \bar{\Lv} \right\|_1 \right] \\
&= \mathbb{E} \left[ \bar{L}_0 \right] .
\end{split}
\end{equation}
This chain of equalities reveals the intricate relation between $\mathbb{E} \left[ \left\| \bar{\Lv} \right\|_2^2 \right]$ and $\mathbb{E} \left[ \bar{L}_0 \right]$.
\end{IEEEproof}

\noindent\textbf{Proof of Proposition~\ref{proposition:BalancedProbabilityRV}.}
\begin{IEEEproof}
First, suppose \eqref{equation:BalancedProbabilityRV} holds.
Then, we can write
\begin{equation*}
\begin{split}
\mathbb{E} \left[ \left\| \bar{\Lv} \right\|_2^2 \right]
&= \mathbb{E} \left[ \bar{L}_0 \right]
= \mathbb{E} \left[ \bar{L}_0 \left\| \bar{\Lv} \right\|_1 \right] \\
&= \mathbb{E} \left[ \bar{L}_0 \left( \bar{L}_0 + (q-1) \bar{L}_{\bullet} \right) \right] \\
&= \mathbb{E} \left[ \bar{L}_0^2 \right] + (q-1) \mathbb{E} \left[ \bar{L}_0 \bar{L}_{\bullet} \right] .
\end{split}
\end{equation*}
At the same time, by definition, we have
\begin{equation*}
\mathbb{E} \left[ \left\| \bar{\Lv} \right\|_2^2 \right]
= \mathbb{E} \left[ \bar{L}_0^2 \right] + (q-1) \mathbb{E} \left[ \bar{L}_{\bullet}^2 \right] .
\end{equation*}
Equating both expressions for the two-norm of $\bar{\Lv}$, we deduce that $\mathbb{E} \left[ \bar{L}_0 \bar{L}_{\bullet} \right] = \mathbb{E} \left[ \bar{L}_{\bullet}^2 \right]$.
To get the converse, we assume that $\bar{\Lv}$ is balanced and then parallel the progression in \eqref{equation:DerivationSymmetricGaussianProbabilityRV}, which yields
\begin{equation*}
\begin{split}
&\mathbb{E} \left[ \left\| \bar{\Lv} \right\|_2^2 \right]
= \mathbb{E} \left[ \bar{L}_0^2 \right] + (q-1) \mathbb{E} \left[ \bar{L}_{\bullet}^2 \right] \\
&= \mathbb{E} \left[ \bar{L}_0^2 \right] + (q-1) \mathbb{E} \left[ \bar{L}_0 \bar{L}_{\bullet} \right] \\
&= \mathbb{E} \left[ \bar{L}_0 \left( \bar{L}_0 + (q-1) \bar{L}_{\bullet} \right) \right]
= \mathbb{E} \left[ \bar{L}_0 \right] .
\end{split}
\end{equation*}
We emphasize that in the two instances above, we have leveraged the fact that, for any realization of $\bar{\Lv}$,
\begin{equation*}
\left\| \bar{\Lv} \right\|_1
= \bar{L}_0 + (q-1) \bar{L}_{\bullet} = 1 .
\end{equation*}
Combining these two results, we get the desired logical equivalence.
\end{IEEEproof}

\noindent\textbf{Proof of Corollary~\ref{corollary:mse_of_structured_rv}.}
\begin{IEEEproof}
Let $\Bar{\Lv}$ be a balanced dominant permutation-symmetric probability-vector random variable. 
Then, it follows that
\begin{equation*}
    \begin{split}
        \mathbb{E}\left[\|\Bar{\Lv} - \ev_0\|_2^2\right] &= \sum_{g\in\mathbb{F}_q} \mathbb{E}\left[\left(\Bar{\Lv}_g - \ev_0(g)\right)^2\right] \\
        &= \mathbb{E}\left[1 - 2\Bar{\Lv}_0 + \Bar{\Lv}_0^2\right] + \sum_{g\in\mathbb{F}_q\setminus0} \mathbb{E}\left[\Bar{\Lv}_g^2\right] \\
        &= 1 - 2\mathbb{E}\left[\Bar{\Lv}_0\right] + \mathbb{E}\left[\|\Bar{\Lv}\|_2^2\right] \\
        &= 1 - \mathbb{E}\left[\|\Bar{\Lv}\|_2^2\right],
    \end{split}
\end{equation*}
where the last line uses the fact that $\mathbb{E}\left[\Bar{\Lv}_0\right] = \mathbb{E}\left[\|\Bar{\Lv}\|_2^2\right]$.
\end{IEEEproof}

\noindent\textbf{Proof of Theorem~\ref{theorem:ConvolutionGaussianSymmetricMSE}}
\begin{IEEEproof}
We show this result via mathematical induction.
To facilitate the proof, we need to expand our notation slightly with
\begin{equation*}
\bar{\Nv}^{(n)} = \bigodot_{p \in [n]} \bar{\Lv}^{(p)} .
\end{equation*}
In this context, the base case is immediate.
When $n=1$, we have $\bar{\Nv}^{(1)} = \bar{\Lv}^{(1)}$ and, based on our assumptions, we can write
\begin{equation*}
\begin{split}
\mathbb{E} \left[ \left\| \bar{\Nv}^{(1)} \right\|_2^2 \right]
= \mathbb{E} \left[ \left\| \bar{\Lv}^{(1)} \right\|_2^2 \right] 
= \mathbb{E} \left[ \bar{L}_0^{(1)} \right]
= \mathbb{E} \left[ \bar{N}_0^{(1)} \right] .
\end{split}
\end{equation*}
For the inductive step, assume that \eqref{equation:ConvolutionGaussianSymmetricMSE} holds for $n$ fixed.
Then, consider the case where
\begin{equation*}
\bar{\Nv}^{(n+1)} = \bar{\Lv}^{(n+1)} \odot \bar{\Nv}^{(n)} .
\end{equation*}
Within this part of the proof, we use the abridged notation $\bar{\Nv} = \bar{\Nv}^{(n+1)}$, $\bar{\Lv} = \bar{\Lv}^{(n+1)}$, and $\bar{\Mv} = \bar{\Nv}^{(n)}$ to lighten the exposition.
We stress that, under our inductive hypothesis, $\bar{\Lv}$ and $\bar{\Mv}$ are both balanced dominant permutation-symmetric probability-vector random variables.
As a first step, we seek a convenient expression for the square of the two-norm of $\bar{\Nv}$,
\begin{equation} \label{equation:NormNintoSummands}
\begin{split}
&\mathbb{E} \left[ \left\| \bar{\Nv} \right\|^2 \right]
= \sum_{g \in \mathbb{F}_q} \mathbb{E} \left[ \bar{N}_g^2 \right] \\
&= \sum_{g \in \mathbb{F}_q} \mathbb{E} \left[ \left( \sum_{h \in \mathbb{F}_q} \bar{L}_{g-h} \bar{M}_h \right)
\left( \sum_{\iota \in \mathbb{F}_q} \bar{L}_{g-\iota} \bar{M}_{\iota} \right) \right] \\
&= \sum_{g \in \mathbb{F}_q} \sum_{h \in \mathbb{F}_q} \sum_{\iota \in \mathbb{F}_q} 
\mathbb{E} \left[ \bar{L}_{g-h} \bar{L}_{g-\iota} \right] \mathbb{E} \left[ \bar{M}_h \bar{M}_{\iota} \right] \\
&= \sum_{h \in \mathbb{F}_q} \mathbb{E} \left[ \bar{M}_h \bar{M}_h \right]
\sum_{g \in \mathbb{F}_q} \mathbb{E} \left[ \bar{L}_{g-h} \bar{L}_{g-h} \right] \\
&+ \sum_{h \in \mathbb{F}_q} \sum_{\iota \in \mathbb{F}_q \setminus h}
\mathbb{E} \left[ \bar{M}_h \bar{M}_{\iota} \right]
\sum_{g \in \mathbb{F}_q} \mathbb{E} \left[ \bar{L}_{g-h} \bar{L}_{g-\iota} \right] .
\end{split}
\end{equation}
We emphasize that $\bar{\Lv}$ and $\bar{\Mv}$ are independent and, as such, we can split the expectations.
Focusing on the first summand, we have
\begin{equation*}
\begin{split}
&\sum_{h \in \mathbb{F}_q} \mathbb{E} \left[ \bar{M}_h \bar{M}_h \right]
\sum_{g \in \mathbb{F}_q} \mathbb{E} \left[ \bar{L}_{g-h} \bar{L}_{g-h} \right] \\
&= \sum_{h \in \mathbb{F}_q} \mathbb{E} \left[ \bar{M}_h^2 \right]
\sum_{g \in \mathbb{F}_q} \mathbb{E} \left[ \bar{L}_{g-h}^2 \right]
= \sum_{h \in \mathbb{F}_q} \mathbb{E} \left[ \bar{M}_h^2 \right]
\mathbb{E} \left[ \left\| \bar{\Lv} \right\|_2^2 \right] \\
&= \mathbb{E} \left[ \left\| \bar{\Lv} \right\|_2^2 \right]
\mathbb{E} \left[ \left\| \bar{\Mv} \right\|_2^2 \right] .
\end{split}
\end{equation*}
Turning to the second summand, we get
\begin{equation*}
\begin{split}
&\sum_{h \in \mathbb{F}_q} \sum_{\iota \in \mathbb{F}_q \setminus h}
\mathbb{E} \left[ \bar{M}_h \bar{M}_{\iota} \right]
\sum_{g \in \mathbb{F}_q} \mathbb{E} \left[ \bar{L}_{g-h} \bar{L}_{g-\iota} \right] \\
&= \sum_{h \in \mathbb{F}_q} \sum_{\iota \in \mathbb{F}_q \setminus h}
\mathbb{E} \left[ \bar{M}_h \bar{M}_{\iota} \right]
\left( 2 \mathbb{E} \left[ \bar{L}_0 \bar{L}_{\bullet} \right]
+ (q-2) \mathbb{E} \left[ \bar{L}_{\bullet} \bar{L}_{\star} \right] \right) \\
&= \left( 1 - \mathbb{E} \left[ \left\| \bar{\Mv} \right\|_2^2 \right] \right)
\left( 2 \mathbb{E} \left[ \bar{L}_0 \bar{L}_{\bullet} \right]
+ (q-2) \mathbb{E} \left[ \bar{L}_{\bullet} \bar{L}_{\star} \right] \right) \\
&= \left( 1 - \mathbb{E} \left[ \left\| \bar{\Mv} \right\|_2^2 \right] \right)
\left( 2 \mathbb{E} \left[ \bar{L}_0 \bar{L}_{\bullet} \right]
+ \mathbb{E} \left[ \bar{L}_{\bullet} \right] - 2 \mathbb{E} \left[ \bar{L}_{\bullet}^2 \right] \right) \\
&= \frac{1}{q-1} \left( 1 - \mathbb{E} \left[ \left\| \bar{\Lv} \right\|_2^2 \right] \right)
\left( 1 - \mathbb{E} \left[ \left\| \bar{\Mv} \right\|_2^2 \right] \right) .
\end{split}
\end{equation*}
The subscript notation $\bar{L}_{\bullet} \bar{L}_{\star}$ refers to any two distinct, non-zero elements in $\mathbb{F}_q$.
In the second equality, we have utilized the fact that $\bar{\Mv}$ is normalized with $\left\| \bar{\Mv} \right\|_1 = 1$ and, hence,
\begin{equation*}
1 - \left\| \bar{\Mv} \right\|_2^2
= \left\| \bar{\Mv} \right\|_1^2 - \left\| \bar{\Mv} \right\|_2^2
= \sum_{h \in \mathbb{F}_q} \sum_{\iota \in \mathbb{F}_q \setminus h} \bar{M}_h \bar{M}_{\iota} .
\end{equation*}
The third equality relies on the identity
\begin{equation*}
\begin{split}
&\mathbb{E} \left[ \bar{L}_{\bullet} \bar{L}_{\star} \right]
= \mathbb{E} \left[ \mathbb{E} \left[ \bar{L}_{\bullet} \bar{L}_{\star} \middle| \bar{L}_0, \bar{L}_{\bullet} \right] \right] \\
&= \mathbb{E} \left[ \bar{L}_{\bullet} \mathbb{E} \left[ \bar{L}_{\star} \middle| \bar{L}_0, \bar{L}_{\bullet} \right] \right]
= \mathbb{E} \left[ \bar{L}_{\bullet} \left( \frac{1 - \bar{L}_0 - \bar{L}_{\bullet}}{q-2} \right) \right] \\
&= \frac{\mathbb{E} \left[ \bar{L}_{\bullet} \right] - \mathbb{E} \left[ \bar{L}_0 \bar{L}_{\bullet} \right] - \mathbb{E} \left[ \bar{L}_{\bullet}^2 \right]}{q-2}
= \frac{\mathbb{E} \left[ \bar{L}_{\bullet} \right] - 2 \mathbb{E} \left[ \bar{L}_{\bullet}^2 \right]}{q-2} .
\end{split}
\end{equation*}
The last equality makes use of the relations $\mathbb{E} \left[ \bar{L}_0 \bar{L}_{\bullet} \right] = \mathbb{E} \left[ \bar{L}_{\bullet}^2 \right]$ and $\mathbb{E}\left[\bar{L}_0\right] = \mathbb{E}\left[\|\bar{\Lv}\|_2^2\right]$, which hold for balanced dominant permutation-symmetric probability-vector random variables.
Combining our findings for the constituent sums in \eqref{equation:NormNintoSummands}, we arrive at
\begin{equation*}
\begin{split}
&\mathbb{E} \left[ \left\| \bar{\Nv} \right\|_2^2 \right] \\
&= \mathbb{E} \left[ \left\| \bar{\Lv} \right\|_2^2 \right]
\mathbb{E} \left[ \left\| \bar{\Mv} \right\|_2^2 \right] \\
&\qquad + \frac{1}{q-1} \left( 1 - \mathbb{E} \left[ \left\| \bar{\Lv} \right\|_2^2 \right] \right)
\left( 1 - \mathbb{E} \left[ \left\| \bar{\Mv} \right\|_2^2 \right] \right) \\
&= \frac{q}{q-1} \left( \mathbb{E} \left[ \left\| \bar{\Lv} \right\|_2^2 \right] - \frac{1}{q} \right)
\left( \mathbb{E} \left[ \left\| \bar{\Mv} \right\|_2^2 \right] - \frac{1}{q} \right) + \frac{1}{q} .
\end{split}
\end{equation*}
This demonstrates that the inductive step is valid.
Corollary~\ref{corollary:ConvolutionGroupSymmetricProbabilityRV} connects the expression for $\mathbb{E} \left[ \left\| \bar{\Nv} \right\|_2^2 \right]$ to the mean of $\bar{N}_0$, given the assumed condition
\begin{equation*}
\mathbb{E} \left[ \left\| \bar{\Lv}^{(p)} \right\|_2^2 \right] = \mathbb{E} \left[ \bar{L}_0^{(p)} \right]
\qquad \forall p \in [n] .
\end{equation*}
Finally, Proposition~\ref{proposition:BalancedProbabilityRV} ensures that $\bar{\Nv}$ is balanced.
This completes the proof.
\end{IEEEproof}

\subsection{Proofs from Section~\ref{subsection:computing_state_evolution}}

\noindent\textbf{Proof of Proposition~\ref{proposition:check_to_var_mse}.}
\begin{IEEEproof}
This result is obtained immediately by combining Theorem~\ref{theorem:ConvolutionGaussianSymmetricMSE} and Corollary~\ref{corollary:mse_of_structured_rv}. 
\end{IEEEproof}

\noindent \textbf{Proof of Proposition~\ref{proposition:var_to_chk_approximation}.}
\begin{IEEEproof}

The operation at the variable node yields
\begin{equation*}
\begin{split}
\tilde{\boldsymbol{\mu}}_{v_{\ell} \to c_p}
&= \frac{ \boldsymbol{\alpha}_{\ell} \circ \left( \operatorname*{\bigcirc}_{c_{\xi} \in N(v_{\ell}) \setminus c_p} \tilde{\boldsymbol{\mu}}_{c_{\xi} \to v_{\ell}} \right) }
{ \left\| \boldsymbol{\alpha}_{\ell} \circ \left( \operatorname*{\bigcirc}_{c_{\xi} \in N(v_{\ell}) \setminus c_p} \tilde{\boldsymbol{\mu}}_{c_{\xi} \to v_{\ell}} \right) \right\|_1 } \\
\end{split}
\end{equation*}
or, component-wise,
\begin{equation*}
\begin{split}
\tilde{\boldsymbol{\mu}}_{v_{\ell} \to c_p} (g)
&= \frac{e^{\frac{\rv_{\ell}(g)}{\tau^2}}
\prod_{c_{\xi} \in N(v_{\ell}) \setminus c_p} \exp \left(  \frac{\rv_{\xi,\ell}(g)}{\tau_{\xi,\ell}^2} \right)}
{\sum_{h \in \mathbb{F}_q} \left( e^{\frac{\rv_{\ell}(h)}{\tau^2}}
\prod_{c_{\xi} \in N(v_{\ell}) \setminus c_p}
\exp \left( \frac{\rv_{\xi,\ell}(h)}{\tau_{\xi,\ell}^2} \right) \right)} \\
&= \frac{\exp \left( \frac{\rv_{\ell}(g)}{\tau^2} +
\sum_{c_{\xi} \in N(v_{\ell}) \setminus c_p} \frac{\rv_{\xi,\ell}(g)}{\tau_{\xi,\ell}^2} \right)}
{\sum_{h \in \mathbb{F}_q} \exp \left( \frac{\rv_{\ell}(h)}{\tau^2} +
\sum_{c_{\xi} \in N(v_{\ell}) \setminus c_p} \frac{\rv_{\xi,\ell}(h)}{\tau_{\xi,\ell}^2} \right)} .
\end{split}
\end{equation*}
We emphasize that the argument of the exponential is a Gaussian random variable with mean
\begin{equation*}
\begin{split}
&\mathbb{E} \left[ \frac{\rv_{\ell}(g)}{\tau^2} +
\sum_{\substack{c_{\xi} \in N(v_{\ell}) \\ c_{\xi} \neq c_p}} \frac{\rv_{\xi,\ell}(g)}{\tau_{\xi,\ell}^2} \right] = \sv_{\ell}(g) \left( \frac{1}{\tau^2} +
\sum_{\substack{c_{\xi} \in N(v_{\ell}) \\ c_{\xi} \neq c_p}} \frac{1}{\tau_{\xi,\ell}^2} \right)
\end{split}
\end{equation*}
and variance
\begin{equation*}
\begin{split}
&\operatorname{Var} \left[ \frac{\rv_{\ell}(g)}{\tau^2} +
\sum_{\substack{c_{\xi} \in N(v_{\ell}) \\ c_{\xi} \neq c_p}} \frac{\rv_{\xi,\ell}(g)}{\tau_{\xi,\ell}^2} \right] = \left( \frac{1}{\tau^2} +
\sum_{\substack{c_{\xi} \in N(v_{\ell})  \\ c_{\xi} \neq c_p}} \frac{1}{\tau_{\xi,\ell}^2} \right) .
\end{split}
\end{equation*}
Due to normalization, this becomes statistically equivalent to observing
\begin{equation*}
\tilde{\rv}_{\ell,p} = \sv + \nv_{\ell,p},
\end{equation*}
where $\nv_{\ell,p}$ is i.i.d.\ with Gaussian entries
\begin{equation}
\mathcal{N} \left( 0, \frac{1}{\frac{1}{\tau^2} +
\sum_{c_{\xi} \in N(v_{\ell}) \setminus c_p} \frac{1}{\tau_{\xi,\ell}^2}} \right).
\end{equation}
Denoting the variance of $\nv_{\ell, p}$ by $\Tilde{\tau}_{\ell, p}^2$, it follows that 
\begin{equation}
    \mathbb{E}\left[\Tilde{\boldsymbol{\mu}}_{v_{\ell} \rightarrow c_p}(0)\right] = \Psi\left(\Tilde{\tau}^2_{\ell, p}\right).
\end{equation}
Since $\Tilde{\boldsymbol{\mu}}_{v_{\ell} \rightarrow c_p}$ is a balanced dominant permutation-symmetric probability-vector random variable, the corresponding MSE may be obtained by inspection using Corollary~\ref{corollary:mse_of_structured_rv}.
\end{IEEEproof}

\end{document}